\documentclass{IEEEtran}

\makeatletter
\usepackage{mathtools,amssymb,yhmath}\interdisplaylinepenalty=2500
	\DeclareMathOperator*\lcm{lcm}
	\DeclareMathOperator*\depth{depth}

\usepackage{amsthm}
	\newtheorem{thm}{Theorem}
	\newtheorem{lem}[thm]{Lemma}
	\newtheorem{cor}[thm]{Corollary}
	\newtheorem{prop}[thm]{Proposition}
	\newtheorem{conj}[thm]{Claim}

\usepackage{tikz}
	\definecolor{encolor}{HTML}{BBEECC}
	\definecolor{decolor}{HTML}{DEC0DE}
	\definecolor{recycol}{HTML}{167629}
	\tikzset{
		every picture/.style={
			cap=round,join=round,
			baseline={([yshift=-.5ex]current bounding box)},
		},
		rp/.style={remember picture},o/.style={overlay},
		hl/.style={help lines},
		<^/.style={above left},^/.style={above},^>/.style={above right},
		<</.style={      left}                 ,>>/.style={      right},
		<_/.style={below left},_/.style={below},_>/.style={below right},
		dot/.pic={\fill circle(1pt);},
	}\tikzset{
		bb/.style={/utils/exec=\bbsetupcs,nodes={inner ysep=1,inner xsep=0}},
		G/.pic={\def\coor{coordinate}\path
			(-  \bbwire,   \bbgape)\coor(1L)(   \bbwire,   \bbgape)\coor(1R)
			(-  \bbwire,-  \bbgape)\coor(0L)(   \bbwire,-  \bbgape)\coor(0R);},
		F/.pic={\draw[pic actions]pic{G}(1L)--(1R)(0L)--(0R);},
		E/.pic={\pic[thick]{F};\draw[thick,fill=encolor]
			(-  \bbsize,-  \bbsize)rectangle(   \bbsize,   \bbsize)
			(-  \bbgape,-  \bbgape)       --(   \bbgape,   \bbgape);},
		D/.pic={\pic[thick]{F};\draw[thick,fill=decolor]
			(-  \bbsize,-  \bbsize)rectangle(   \bbsize,   \bbsize)
			(-  \bbgape,   \bbgape)       --(   \bbgape,-  \bbgape)..controls
			( .5\bbgape,-.5\bbgape)    and  ( .5\bbgape, .5\bbgape)..
			(   \bbgape,   \bbgape)       --(-  \bbgape,-  \bbgape);},
	}
	\newdimen\bbsize\bbsize8pt  
	\newdimen\bbwire\bbwire12pt 
	\newdimen\bbgape\bbgape4pt  
	\pgfmathdeclarefunction{xoffset}{1}{\pgfmathparse{
		((2^#1-1)*\bbwire-2*#1\bbgape)/3}}
	\def\bbsetupcs{
		\xdef\bbxmax{0}
		\def\pgfpointxyz##1##2##3{
			\ifnum##2>\bbxmax\xdef\bbxmax{##2}\fi
			\pgfpoint
				{(2\bbwire*##2+xoffset(\bbxmax)-xoffset(\bbxmax+1-##2))*##11}
				{0b##3*2\bbwire}
		}
	}
	\tikzset{
		ut/.style={
			/utils/exec=\utsetupcs,
			nodes={inner sep=1,minimum size=2,scale=2.5^(\utdepth)/3^(\utdepth)},
			baseline=.5\utupper-.5ex
		},
		s/.style={scale=5/6},
		h/.style={fill=yellow},
	}
	\def\utdepth{0}
	\newdimen\utupper\utupper8\bbwire
	\newdimen\utmiddle
	\newdimen\utlower\utlower0cm
	\def\utnode$#1$#2$#3#4{
		\pgfmathsetlength\utmiddle{.5\utupper+.5\utlower}
		\draw(\utdepth,\utmiddle)node(\utdepth){$#1$}
		\ifnum\utdepth>0(\the\numexpr\utdepth-1)--(\utdepth)\fi;
		\edef\utdepth{\the\numexpr\utdepth+1}
		{\utlower\utmiddle#3}{\utupper\utmiddle#4}
		\draw[o](\utdepth-.333,\utmiddle)node{$#2$};
	}
	\def\utsetupcs{
		\def\pgfpointxy##1##2{\pgfpoint{2\bbwire*(1.2^(##1)*5-5)}{0pt}}
	}
	\def\vtnode#1$#2$#3#4{
		\pgfmathsetlength\utmiddle{.5\utupper+.5\utlower}
		\ifx\utnotfirsthild#2\relax\else
			\draw(\utdepth,\utmiddle)node#1(\utdepth){$#2$}
			\ifnum\utdepth>0(\the\numexpr\utdepth-1)--(\utdepth)\fi;
		\fi
		\pgfmathsetlength\utmiddle{((#3-1)*\utupper+\utlower)/#3}
		{
			\edef\utdepth{\the\numexpr\utdepth+1}
			\utlower\utmiddle
			#4
		}
		\let\utnext\relax
		\ifnum#3>1
			\utupper\utmiddle
			\def\utnext{\vtnode$\utnotfirsthild${\numexpr#3-1\relax}}
		\fi
		\utnext
	}
	\tikzset{
		sd/.style={
			every path/.style={line width=\sdwidth,cap=butt,draw opacity=.2},
			nodes={scale={min(1,\sdwidth/.5cm)},o,opacity=1}
		}
	}
	\newdimen\sdwidth\sdwidth1cm
	\newdimen\sdlepht\sdlepht.5cm
	\newdimen\sdright\sdright.5cm
	\newdimen\sddepth\sddepth0cm
	\def\sddive(#1)#2;{
		\draw(\sddepth,        \sdlepht)..controls(\sddepth+#1cm*.4,\sdlepht)
		  and(\sddepth+#1cm*.6,\sdright)..        (\sddepth+#1cm,   \sdright)#2;
		\advance\sddepth by#1cm\sdlepht\sdright
	}
	\def\sdcutby.#1#2#3{
		{
			\pgfmathsetlength\sdright{\sdlepht+(1-.#1)*\sdwidth}
			\pgfmathsetlength\sdlepht{\sdlepht+(1-.#1)*\sdwidth/2}
			\pgfmathsetlength\sdwidth{.#1*\sdwidth}
			#2
		}{
			\pgfmathsetlength\sdright{\sdlepht-.#1*\sdwidth}
			\pgfmathsetlength\sdlepht{\sdlepht-.#1*\sdwidth/2}
			\pgfmathsetlength\sdwidth{(1-.#1)*\sdwidth}
			#3
		}
	}
	\def\sdsave[#1]{
		\expandafter\xdef\csname sdsaved#1width\endcsname{\the\sdwidth}
		\expandafter\xdef\csname sdsaved#1lepht\endcsname{\the\sdlepht}
		\expandafter\xdef\csname sdsaved#1right\endcsname{\the\sdright}
		\expandafter\xdef\csname sdsaved#1depth\endcsname{\the\sddepth}
	}
	\def\sdload[#1](#2)#3;{
		{
			\sdwidth\csname sdsaved#1width\endcsname
			\sdlepht\csname sdsaved#1lepht\endcsname
			\sdright\csname sdsaved#1right\endcsname
			\pgfmathsetlength\sdright{\sdright-\sdwidth/2}
			\sddepth\csname sdsaved#1depth\endcsname
			\sddive(#2);
		}
		{
			\pgfmathsetlength\sdright{\sdright+\sdwidth/2}
			\sddive(#2)#3;
		}
		\advance\sdwidth by\csname sdsaved#1width\endcsname
		\advance\sddepth by#2cm
		\pgfmathsetlength\sdright{\sdright+\sdwidth/2}
		\sdlepht\sdright
	}
	\tikzset{
		tri/.style={
			scale=#1,
			xshift=-.5cm,
			execute at end picture={
				\global\pgf@picminx-\pgf@picmaxx
			}
		},
		tri/.default=1,
		Tri/.style={tri=2},
		TRi/.style={tri=4},
		TRI/.style={tri=8,
			execute at end picture={
				\global\pgf@picminx-4cm
				\global\pgf@picmaxx36mm
			}
		}
	}
	\pgfmathdeclarefunction{H}{1}{\pgfmathparse{-#1*log2(#1)-(1-#1)*log2(1-#1)}}
	\pgfmathdeclarefunction{h2}{1}{\pgfmathparse{-#1*log2(#1)-(1-#1)*log2(1-#1)}}
	\pgfmathdeclarefunction{mu}{0}{\pgfmathparse{3.627}}
	\pgfmathdeclarefunction{G}{1}{\pgfmathparse{1/(mu+1-mu*h2(#1))}}
	\pgfdeclarelayer{foreground}
	\pgfsetlayers{main,foreground}
	\def\onforeground#1;{\begin{pgfonlayer}{foreground}#1;\end{pgfonlayer}}

\usepackage[breakable,skins]{tcolorbox}
	\newenvironment{rtr}{
		\begin{tcolorbox}[breakable,enhanced jigsaw,size=fbox]
			\IEEEdlabelindent0pt\raggedright
			\begin{description}[\IEEEsetlabelwidth{Recruit}]
	}{
			\end{description}
		\end{tcolorbox}
	}

\usepackage[utf8]{inputenc}\let\DUC\DeclareUnicodeCharacter
	\DUC{03B1}\alpha     
	\DUC{03B2}\beta      
	\DUC{03B3}\gamma     
	\DUC{03B4}\delta     
	\DUC{03F5}\epsilon   
	\DUC{03B6}\zeta      
	\DUC{03B7}\eta       
	\DUC{03B8}\theta     
	\DUC{03B9}\iota      
	\DUC{03BA}\kappa     
	\DUC{03BB}\lambda    
	\DUC{03BC}\mu        
	\DUC{03BD}\nu        
	\DUC{03BE}\xi        
	\DUC{03C0}\pi        
	\DUC{03C1}\rho       
	\DUC{03C3}\sigma     
	\DUC{03C4}\tau       
	\DUC{03C5}\upsilon   
	\DUC{03D5}\phi       
	\DUC{03C7}\chi       
	\DUC{03C8}\psi       
	\DUC{03C9}\omega     
	\DUC{0393}\Gamma     
	\DUC{0394}\Delta     
	\DUC{03B5}\varepsilon
	\DUC{0398}\Theta     
	\DUC{03D1}\vartheta  
	\DUC{039B}\Lambda    
	\DUC{039E}\Xi        
	\DUC{03A0}\Pi        
	\DUC{03D6}\varpi     
	\DUC{03F1}\varrho    
	\DUC{03A3}\Sigma     
	\DUC{03C2}\varsigma  
	\DUC{03A5}\Upsilon   
	\DUC{03A6}\Phi       
	\DUC{03C6}\varphi    
	\DUC{03A8}\Psi       
	\DUC{03A9}\Omega     
	
	\DUC{1D53C}{\mathbb E}  
	\DUC{1D53D}{\mathbb F}  
	\DUC{1D540}{\mathbb I}  
	\DUC{2115}{\mathbb N}   
	\DUC{2119}{\mathbb P}   
	\DUC{211A}{\mathbb Q}   
	\DUC{211D}{\mathbb R}   
	\DUC{1D49C}{\mathcal A} 
	\DUC{1D49E}{\mathcal C} 
	\DUC{1D49F}{\mathcal D} 
	\DUC{1D4AF}{\mathcal T} 
	\DUC{1D4B3}{\mathcal X} 
	\DUC{1D4B4}{\mathcal Y} 
	\DUC{2113}{\ell}        
	
	\DUC{00B7}\cdot     
	\DUC{00D7}\times    
	\DUC{2020}\dagger   
	\DUC{2022}\bullet   
	\DUC{2192}\to       
	\DUC{21A6}\mapsto   
	\DUC{2202}\partial  
	\DUC{2208}\in       
	\DUC{2211}\sum      
	\DUC{221A}\sqrt     
	\DUC{221E}\infty    
	\DUC{2227}\wedge    
	\DUC{222A}\cup      
	\DUC{222B}\int      
	\DUC{2248}\approx   
	\DUC{2254}\coloneqq 
	\DUC{2264}\le       
	\DUC{2265}\ge       
	\DUC{2282}\subset   
	\DUC{2297}\otimes   
	\DUC{27F6}\longrightarrow 
	\DUC{27FC}\longmapsto 
	
	\def\({\bigl(}
	\def\){\bigr)}
	\DUC{FF08}{\Bigl(} 
	\DUC{FF09}{\Bigr)} 
	\DUC{FF3B}{\bigl[} 
	\DUC{FF5C}{\bigm|} 
	\DUC{FF3D}{\bigr]} 
	\DUC{FF5B}{\Bigl\{} 
	\DUC{FF5D}{\Bigr\}} 
	
	\DUC{266D}\smashflat 
	\DUC{266F}\smashsharp 
	\def\smashflat{{\smash\flat}}
	\def\smashsharp{{\smash\sharp}}

\usepackage[ampersand]{easylist}

\usepackage{soul}

\usepackage[pdfusetitle,unicode]{hyperref}
	\def\TP$#1$#2{\texorpdfstring{$#1$}{#2}}

	\def\G(#1){pic(#1){G}}
	\def\F(#1){pic(#1){F}}
	\def\E(#1){pic(#1){E}}
	\def\D(#1){pic(#1){D}}
	\def\latin{\emph}
	\def\caution{}
	\def\[{\begin{equation}}
	\def\]{\end{equation}}
	\def\Capacity{\textnormal{Capacity}}
	\def\Cramer{Cram\'er}
	\def\Bha{Bhattacharyya}
	\def\Arikan{Ar\i kan}
	\def\Ari{\text{Ar\i}}
	\def\KSU{\text{K\smash{\c S}U}}
	\def\GBLB{\text{GBLB}}

	\def\RS#1{{\text{RS}#1}}

	\def\fek{_\subset^k}
	\def\fe#1{_\subset^#1}
	\def\CF{\Lambda^*}
	\def\rat{_\text{rat}}
	\def\err{_\text{err}}
	\def\ub{^\text{upper}_\text{bound}}
	\def\lb{^\text{lower}_\text{bound}}
	\def\pf{^{\text{per}}_\text{fect}}
	\def\Y{\underline Y}
	\def\absT{\lvert T\rvert}

\begin{document}
	\title{Polar-like Codes and Asymptotic Tradeoff among \\
	       Block Length, Code Rate, and Error Probability}
	\author{Hsin-Po Wang and Iwan Duursma \\
	        University of Illinois at Urbana--Champaign \\
	        \{hpwang2, duursma\}@illinois.edu}
\maketitle

\begin{abstract}
	%
	A general framework is proposed that includes
	polar codes over arbitrary channels with arbitrary kernels.
	The asymptotic tradeoff among
	block length $N$, code rate $R$, and error probability $P$ is analyzed.
	
	Given a tradeoff between $N,P$ and a tradeoff between $N,R$,
	we return an interpolating tradeoff among $N,R,P$
	(Theorem~\ref{thm:disposable}).
	Quantitatively,
	if $P=\exp(-N^{β^*})$ is possible for some $β^*$
	and if $R=\Capacity-N^{1/μ^*}$ is possible for some $1/μ^*$,
	then $(P,R)=\(\exp(-N^{β'}),\Capacity-N^{-1/μ'}\)$ is possible
	for some pair $(β',1/μ')$
	determined by $β^*$, $1/μ^*$, and auxiliary information.
	In fancy words,
	an error exponent regime tradeoff plus a scaling exponent regime tradeoff
	implies a moderate deviations regime tradeoff.
	
	The current world records are:
	\cite{GX13,MHU16,WD18} analyzing \Arikan's codes over BEC;
	\cite{FT17} analyzing \Arikan's codes over AWGN; and
	\cite{BGNRS18,BGS18} analyzing general codes over general channels.
	An attempt is made to generalize all at once.
	(Section~\ref{sec:bigtriangle}.)
	
	As a corollary, a grafted variant of polar coding
	almost catches up the code rate and error probability of random codes
	with complexity slightly larger than $N\log N$ over BEC.
	In particular,
	$(P,R)=\(\exp(-N^{.33}),\Capacity-N^{-.33}\)$ is possible
	(Corollary~\ref{cor:hypotenuse}).
	In fact, all points in this triangle are possible $(β',1/μ')$-pairs.
	
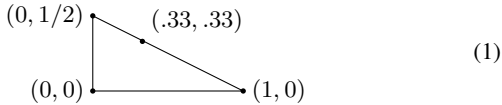
\[\tikz[Tri]\draw
		(0,0 )pic{dot}node[<<]{$(0,0)$}--
		(0,.5)pic{dot}node[<<]{$(0,1/2)$}--
		(1,0 )pic{dot}node[>>]{$(1,0)$}--cycle
		(.33,.33)pic{dot}node[^>]{$(.33,.33)$}
	;\]
\end{abstract}



\section{Introduction}

\IEEEPARstart{I}{n the}
	theory of two-terminal error correcting codes,
	three of the most important parameters of block codes are
	block length $N$, code rate $R$, and error probability $P$.
	Though we want codes with small $N$, higher $R$, and lower $P$,
	these goals contradict each other.
	Thus it becomes essential to quantify the tradeoffs.
	
	Given a memoryless channel $W$ with symmetric capacity $I(W)$,
	there exists polar codes with
	\begin{align}
		\log(-\log P) &∈ Θ(\log N) &\text{as }N\to\infty.
		\shortintertext{It is also shown that there exist polar codes with}
		-\log\(I(W)-R\) &∈ Θ(\log N) &\text{as }N\to\infty.
	\end{align}
	This work aims to characterize the pairs of ratios
	\[（
		\liminf_{N\to\infty}\frac{\log(-\log P)}{\log N},
		\liminf_{N\to\infty}\frac{-\log\(I(W)-R\)}{\log N}
	）\]
	that are realized by polar codes.
	
	It has been shown before that the pair of ratios for block codes lies in
	
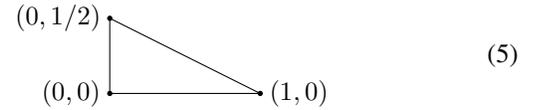
\[\tikz[Tri]\draw
		(0,0 )pic{dot}node[<<]{$(0,0)$}--
		(0,.5)pic{dot}node[<<]{$(0,1/2)$}--
		(1,0 )pic{dot}node[>>]{$(1,0)$}--cycle
	;\]
	and random codes achieve the hypotenuse.
	This motivates two questions:
	whether polar codes can achieve the hypotenuse (yes for BEC)
	and what price we pay in terms of complexity (slightly more than $N\log N$).
	
	See Section~\ref{sec:bigtriangle} for big pictures.

\subsection{Channel polarization}

	Channel polarization \cite{Arikan09}
	is a method to synthesize some channels to form
	some extremely-unreliable channels and some extremely-reliable channels.
	The users then can transmit uncoded messages through extremely-reliable ones
	while transmitting predictable symbols through extremely-unreliable ones.
	
	We summarize channel polarization as follows.
	Say we are going to communicate over this BEC
	\[\tikz[bb]{\draw[o]
		(-,1,0)\G(E)(+,1,0)\G(D);
		\draw(E1R)--node[above]{$W$}(D1L)
	}.\]
	We have two magic devices
	\begin{gather}
		\tikz[bb]\pic{E};
		\shortintertext{and}
		\tikz[bb]\pic{D};
	\end{gather}
	such that if we wire two i.i.d.\ instances of $W$ as follows
	\[\tikz[bb]\draw
		(-,1,0)\E(E)(+,1,0)\D(D)
		(E1R)--node[^>]{$W$}(D1L)
		(E0R)--node[<_]{$W$}(D0L)
		(E1L)node[<<]{$A$}(D1R)node[>>]{$B$}
		(E0L)node[<<]{$C$}(D0R)node[>>]{$D$}
	;, \label{eq:cir2}\]
	then pin $A$ to pin $B$ forms a less reliable synthetic channel $W^♭$,
	while pin $C$ to pin $D$ forms a more reliable synthetic channel $W^♯$.
	Graphically, Formula~(\ref{eq:cir2}) is equivalent to
	\[\tikz[bb]\draw
		(E1L)--node[^>]{$W^♭$}(D1R)
		(E0L)--node[<_]{$W^♯$}(D0R)
	;.\]
	
	Formula~(\ref{eq:cir2}) being the base step,
	the next step is to duplicate Formula~(\ref{eq:cir2}) and wire them as
	\[\tikz[bb]{\draw[o]
		(-,2,1)\G(1)(-,1,1)\E(2)(+,1,1)\D(3)(+,2,1)\G(4)
		(-,2,0)\G(5)(-,1,0)\E(6)(+,1,0)\D(7)(+,2,0)\G(8)
	;\draw
		(11R)--(21L)(21R)--node[^>]{$W$}(31L)(31R)--(41L)
		(51R)--(20L)(20R)--node[<_]{$W$}(30L)(30R)--(81L)
		(10R)--(61L)(61R)--node[^>]{$W$}(71L)(71R)--(40L)
		(50R)--(60L)(60R)--node[<_]{$W$}(70L)(70R)--(80L)
	;}, \label{eq:xpo2x}\]
	which is equivalent to four synthetic channels as
	
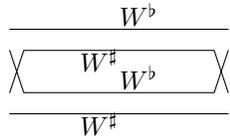
\[\tikz[bb]{\draw
		(11R)--(21L)--node[^>]{$W^♭$}(31R)--(41L)
		(51R)--(20L)--node[<_]{$W^♯$}(30R)--(81L)
		(10R)--(61L)--node[^>]{$W^♭$}(71R)--(40L)
		(50R)--(60L)--node[<_]{$W^♯$}(70R)--(80L)
	;}\]
	or simply
	
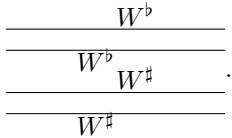
\[\tikz[bb]\draw
		(11R)--node[^>]{$W^♭$}(41L)
		(10R)--node[<_]{$W^♭$}(40L)
		(51R)--node[^>]{$W^♯$}(81L)
		(50R)--node[<_]{$W^♯$}(80L)
	;.\]
	Further wire Formula~(\ref{eq:xpo2x}) as
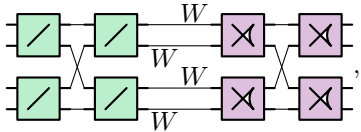
	\[\tikz[bb]\draw
		(-,2,1)\E(1)(-,1,1)\E(2)(+,1,1)\D(3)(+,2,1)\D(4)
		(-,2,0)\E(5)(-,1,0)\E(6)(+,1,0)\D(7)(+,2,0)\D(8)
		(11R)--(21L)(21R)--node[^>]{$W$}(31L)(31R)--(41L)
		(10R)--(61L)(20R)--node[<_]{$W$}(30L)(30R)--(81L)
		(51R)--(20L)(61R)--node[^>]{$W$}(71L)(71R)--(40L)
		(50R)--(60L)(60R)--node[<_]{$W$}(70L)(70R)--(80L)
	;, \label{eq:cir4}\]
	which is equivalent to
	
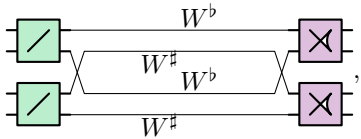
\[\tikz[bb]\draw
		(-,2,1)\E(1)(+,2,1)\D(4)
		(-,2,0)\E(5)(+,2,0)\D(8)
		(11R)--(21L)--node[^>]{$W^♭$}(31R)--(41L)
		(51R)--(20L)--node[<_]{$W^♯$}(30R)--(81L)
		(10R)--(61L)--node[^>]{$W^♭$}(71R)--(40L)
		(50R)--(60L)--node[<_]{$W^♯$}(70R)--(80L)
	;,\]
	to
	
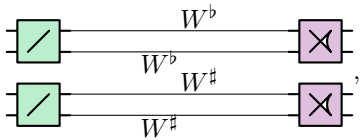
\[\tikz[bb]\draw
		(-,2,1)\E(1)(+,2,1)\D(4)
		(-,2,0)\E(5)(+,2,0)\D(8)
		(11R)--node[^>]{$W^♭$}(41L)
		(10R)--node[<_]{$W^♭$}(40L)
		(51R)--node[^>]{$W^♯$}(81L)
		(50R)--node[<_]{$W^♯$}(80L)
	;,\]
	and to
	
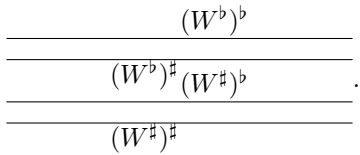
\[\tikz[bb]\draw
		(11L)--node[^>]{$(W^♭)^♭$}(41R)
		(10L)--node[<_]{$(W^♭)^♯$}(40R)
		(51L)--node[^>]{$(W^♯)^♭$}(81R)
		(50L)--node[<_]{$(W^♯)^♯$}(80R)
	;.\]
	Here
	$(W^♭)^♭$ is a synthetic channel less reliable than $W^♭$;
	synthetic channel $(W^♭)^♯$ is  more reliable than $W^♭$;
	synthetic channel $(W^♯)^♭$ is less reliable than $W^♯$; and
	synthetic channel $(W^♯)^♯$ is more reliable than $W^♯$.
	
	After Formula~(\ref{eq:cir4}), the next, larger construction is
	two copies of Formula~(\ref{eq:cir4}) plus four more pairs of magic devices
	\[\tikz[bb]{\draw
		foreach\J in{00,01,10,11}{
			foreach\l in{3,2,1}{
				(-,\l,\J)\E(\l E\J)(+,\l,\J)\D(\l D\J)}
			(1E\J1R)--node[^>]{$W$}(1D\J1L)
			(1E\J0R)--node[<_]{$W$}(1D\J0L)};
		\def\DRAW#1#2{
			\draw foreach\J in{000,001,010,011,100,101,110,111}{
				(#2E\K R)--(#1E\J L)(#1D\J R)--(#2D\K L)};}
		\def\K{\expandafter\k\J}
		\def\k#1#2#3{#3#1#2}\DRAW12
		\def\k#1#2#3{#1#3#2}\DRAW23
	}. \label{eq:cir8}\]
	It is equivalent to
	
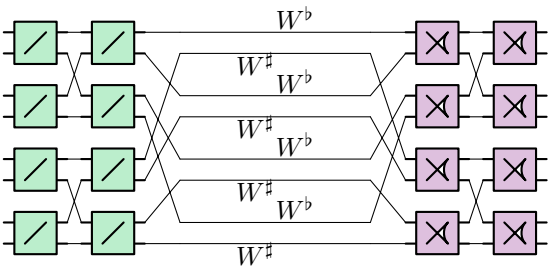
\[\tikz[bb]{\draw
		foreach\J in{00,01,10,11}{
			foreach\l in{3,2}{
				(-,\l,\J)\E(\l E\J)(+,\l,\J)\D(\l D\J)}
			(1E\J1L)--node[^>]{$W^♭$}(1D\J1R)
			(1E\J0L)--node[<_]{$W^♯$}(1D\J0R)};
		\def\DRAW#1#2{
			\draw foreach\J in{000,001,010,011,100,101,110,111}{
				(#2E\K R)--(#1E\J L)(#1D\J R)--(#2D\K L)};}
		\def\K{\expandafter\k\J}
		\def\k#1#2#3{#3#1#2}\DRAW12
		\def\k#1#2#3{#1#3#2}\DRAW23
	},\]
	to
	
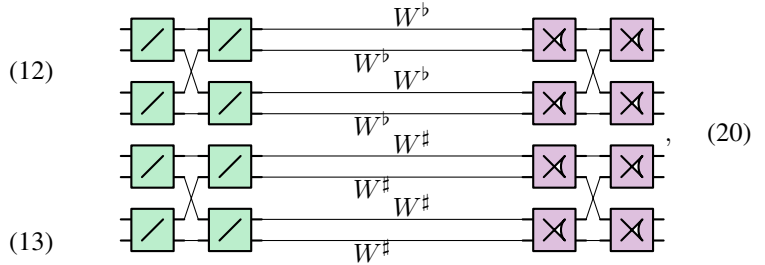
\[\tikz[bb]{\draw
		foreach \J in{00,01,10,11}{
			foreach\l in{3,2}{
				(-,\l,\J)\E(\l E\J)(+,\l,\J)\D(\l D\J)}}
		foreach\J in{0,1}{
			(2E1\J1R)--node[^>]{$W^♭$}(2D1\J1L)
			(2E1\J0R)--node[<_]{$W^♭$}(2D1\J0L)}
		foreach\J in{0,1}{
			(2E0\J1R)--node[^>]{$W^♯$}(2D0\J1L)
			(2E0\J0R)--node[<_]{$W^♯$}(2D0\J0L)};
		\def\DRAW#1#2{
			\draw foreach\J in{000,001,010,011,100,101,110,111}{
				(#2E\K R)--(#1E\J L)(#1D\J R)--(#2D\K L)};}
		\def\K{\expandafter\k\J}
		\def\k#1#2#3{#1#3#2}\DRAW23
	},\]
	to
	
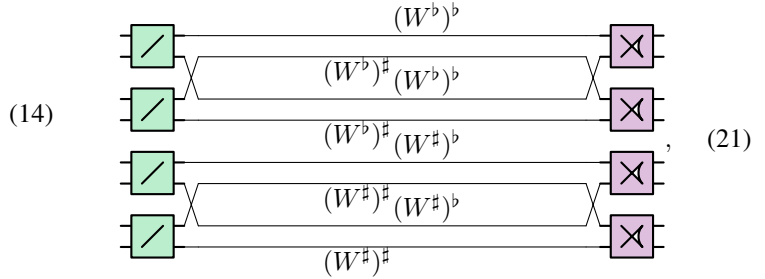
\[\tikz[bb]{\draw
		foreach\J in{00,01,10,11}{
			foreach\l in{3}{
				(-,\l,\J)\E(\l E\J)(+,\l,\J)\D(\l D\J)}}
		foreach\J in{0,1}{
			(2E1\J1L)--node[^>]{$(W^♭)^♭$}(2D1\J1R)
			(2E1\J0L)--node[<_]{$(W^♭)^♯$}(2D1\J0R)}
		foreach\J in{0,1}{
			(2E0\J1L)--node[^>]{$(W^♯)^♭$}(2D0\J1R)
			(2E0\J0L)--node[<_]{$(W^♯)^♯$}(2D0\J0R)};
		\def\DRAW#1#2{
			\draw foreach\J in{000,001,010,011,100,101,110,111}{
				(#2E\K R)--(#1E\J L)(#1D\J R)--(#2D\K L)};}
		\def\K{\expandafter\k\J}
		\def\k#1#2#3{#1#3#2}\DRAW23
	},\]
	to
	
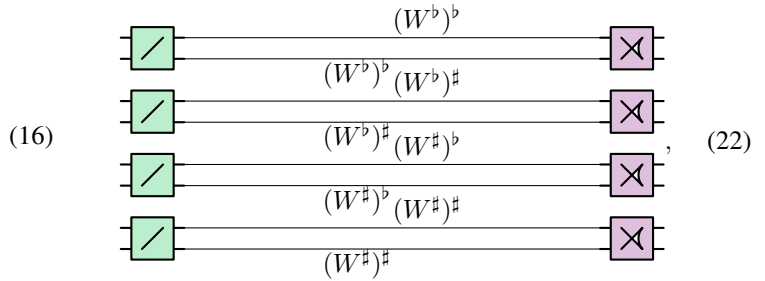
\[\tikz[bb]{\draw
		foreach\J in{00,01,10,11}{
			foreach\l in{3}{
				(-,\l,\J)\E(\l E\J)(+,\l,\J)\D(\l D\J)}}
		(3E111R)--node[^>]{$(W^♭)^♭$}(3D111L)
		(3E110R)--node[<_]{$(W^♭)^♭$}(3D110L)
		(3E101R)--node[^>]{$(W^♭)^♯$}(3D101L)
		(3E100R)--node[<_]{$(W^♭)^♯$}(3D100L)
		(3E011R)--node[^>]{$(W^♯)^♭$}(3D011L)
		(3E010R)--node[<_]{$(W^♯)^♭$}(3D010L)
		(3E001R)--node[^>]{$(W^♯)^♯$}(3D001L)
		(3E000R)--node[<_]{$(W^♯)^♯$}(3D000L);
	},\]
	and finally to
	\[\tikz[bb]{\draw
		(3E111L)--node[^>]{$((W^♭)^♭)^♭$}(3D111R)
		(3E110L)--node[<_]{$((W^♭)^♭)^♯$}(3D110R)
		(3E101L)--node[^>]{$((W^♭)^♯)^♭$}(3D101R)
		(3E100L)--node[<_]{$((W^♭)^♯)^♯$}(3D100R)
		(3E011L)--node[^>]{$((W^♯)^♭)^♭$}(3D011R)
		(3E010L)--node[<_]{$((W^♯)^♭)^♯$}(3D010R)
		(3E001L)--node[^>]{$((W^♯)^♯)^♭$}(3D001R)
		(3E000L)--node[<_]{$((W^♯)^♯)^♯$}(3D000R);
	}.\]
	Here $((W^♭)^♭)^♭$ is a synthetic channel less reliable than $(W^♭)^♭$; etc.
	
	After Formula~(\ref{eq:cir8}), the next, larger construction
	is going to be two copies of Formula~(\ref{eq:cir8})
	plus one extra layer of magic devices.
	
	The game goes on endlessly.
	\Arikan{} then observes that synthetic channels generated in this way
	tend to be either extremely reliable  or extremely unreliable.
	That is to say, they \emph{polarize}.

\subsection{Channel polarization in Tree Notation}

	Draw
	
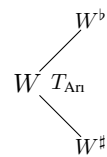
\[\tikz[ut]{
		\utnode$W$T_\Ari${
			\utnode$W^♭$${}{} }{
			\utnode$W^♯$${}{} }
	} \label{eq:tre2}\]
	to capture the fact that Formula~(\ref{eq:cir2})
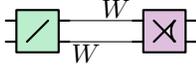
	\begin{equation*}\tikz[bb]\draw
		(-,1,0)\E(E)(+,1,0)\D(D)
		(E1R)--node[^>]{$W$}(D1L)
		(E0R)--node[<_]{$W$}(D0L)
	;\end{equation*}
	transforms two instances of $W$ into a $W^♭$ and a $W^♯$.
	We will later call this tree $𝒯\pf(W,T_\Ari,1)$ (guess why).
	
	Similarly, draw
	
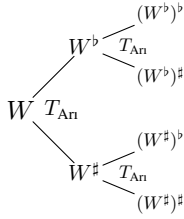
\[\tikz[ut]{
		\utnode$W$T_\Ari${
			\utnode$W^♭$T_\Ari${
				\utnode$(W^♭)^♭$${}{} }{
				\utnode$(W^♭)^♯$${}{} } }{
			\utnode$W^♯$T_\Ari${
				\utnode$(W^♯)^♭$${}{} }{
				\utnode$(W^♯)^♯$${}{} } }
	} \label{eq:tre4}\]
	to capture the fact that Formula~(\ref{eq:cir4})
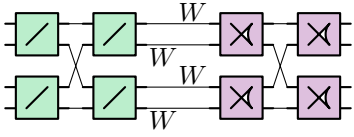
	\begin{equation*}\tikz[bb]\draw
		(-,2,1)\E(1)(-,1,1)\E(2)(+,1,1)\D(3)(+,2,1)\D(4)
		(-,2,0)\E(5)(-,1,0)\E(6)(+,1,0)\D(7)(+,2,0)\D(8)
		(11R)--(21L)(21R)--node[^>]{$W$}(31L)(31R)--(41L)
		(10R)--(61L)(20R)--node[<_]{$W$}(30L)(30R)--(81L)
		(51R)--(20L)(61R)--node[^>]{$W$}(71L)(71R)--(40L)
		(50R)--(60L)(60R)--node[<_]{$W$}(70L)(70R)--(80L)
	;\end{equation*}
	transforms four instances of $W$ into two pairs of $W^♭$ and $W^♯$.
	Two $W^♭$ are then transformed into a $(W^♭)^♭$ and a $(W^♭)^♯$;
	two $W^♯$ are then transformed into a $(W^♯)^♭$ and a $(W^♯)^♯$.
	We will later call this tree $𝒯\pf(W,T_\Ari,2)$ (guess why).
	
	Similarly, draw
	
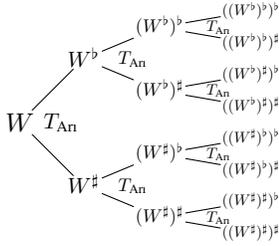
\[\tikz[ut]{
		\utnode$W$T_\Ari${
			\utnode$W^♭$T_\Ari${
				\utnode$(W^♭)^♭$T_\Ari${
					\utnode$((W^♭)^♭)^♭$${}{} }{
					\utnode$((W^♭)^♭)^♯$${}{} } }{
				\utnode$(W^♭)^♯$T_\Ari${
					\utnode$((W^♭)^♯)^♭$${}{} }{
					\utnode$((W^♭)^♯)^♯$${}{} } } }{
			\utnode$W^♯$T_\Ari${
				\utnode$(W^♯)^♭$T_\Ari${
					\utnode$((W^♯)^♭)^♭$${}{} }{
					\utnode$((W^♯)^♭)^♯$${}{} } }{
				\utnode$(W^♯)^♯$T_\Ari${
					\utnode$((W^♯)^♯)^♭$${}{} }{
					\utnode$((W^♯)^♯)^♯$${}{} } } }
	} \label{eq:tre8}\]
	to capture Formula~(\ref{eq:cir8})
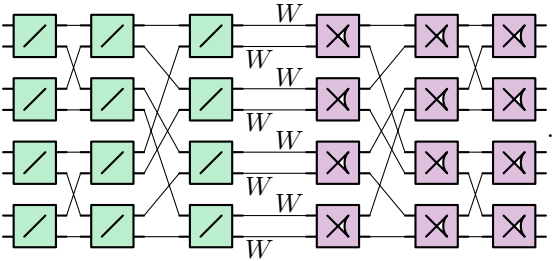
	\begin{equation*}\tikz[bb]{
		\foreach\J in{00,01,10,11}{
			\draw foreach\l in{3,2,1}{
					(-,\l,\J)\E(\l E\J)(+,\l,\J)\D(\l D\J)}
				(1E\J1R)--node[^>]{$W$}(1D\J1L)
				(1E\J0R)--node[<_]{$W$}(1D\J0L);}
		\def\DRAW#1#2{
			\draw foreach\J in{000,001,010,011,100,101,110,111}{
				(#2E\K R)--(#1E\J L)(#1D\J R)--(#2D\K L)};}
		\def\K{\expandafter\k\J}
		\def\k#1#2#3{#3#1#2}\DRAW12
		\def\k#1#2#3{#1#3#2}\DRAW23
	}.\end{equation*}
	That is, eight instances of $W$ are transformed into four pairs of $W^♭,W^♯$,
	into two quadruples of $(W^♭)^♭$, $(W^♭)^♯$, $(W^♯)^♭$, $(W^♯)^♯$,
	and finally into
	$((W^♭)^♭)^♭$, $((W^♭)^♭)^♯$, $((W^♭)^♯)^♭$, $((W^♭)^♯)^♯$,
	$((W^♯)^♭)^♭$, $((W^♯)^♭)^♯$, $((W^♯)^♯)^♭$, $((W^♯)^♯)^♯$.
	We will later call this tree $𝒯\pf(W,T_\Ari,3)$ (guess why).
	
	It is not hard to imagine that the next construction will
	transform sixteen instances of $W$ to ``some intermediate things'',
	and finally to $(((W^♭)^♭)^♭)^♭$ to $(((W^♯)^♯)^♯)^♯$.

\subsection{Generalize the Tree Notation}

	The tree notation comes with generalizations.

\subsubsection{Arbitrary Polar Kernels}

	\cite{KSU10}
	Given, say, an $ℓ$-by-$ℓ$ matrix $G_\KSU$ as a polar kernel,
	it induces a transformation $T_\KSU$.
	We may draw an $ℓ$-ary tree, starting from
	
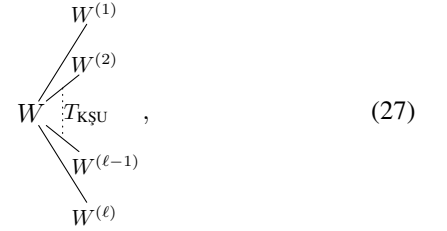
\[\tikz[ut]{\draw
		(0,.5\utupper)node(0){$W$}
		       +(.5,0)node(T)[s,>>]{$T_\KSU$}
		(1,.9\utupper)node(1)[s]{$W^{(1)}$}(0)--(1)
		(1,.7\utupper)node(1)[s]{$W^{(2)}$}(0)--coordinate(2)(1)
		(1,.3\utupper)node(1)[s]{$\phantom{W^{(ℓ)}}$}(0)--coordinate(3)(1)
		      (1.west)node[s,>>]{$W^{(ℓ-1)}$}
		(1,.1\utupper)node(1)[s]{$W^{(ℓ)}$}(0)--(1)
		(2)edge[dotted](3)
	}, \label{eq:trel}\]
	instead of a binary tree.
	This, when $ℓ=7$, translates into the circuit setup
	
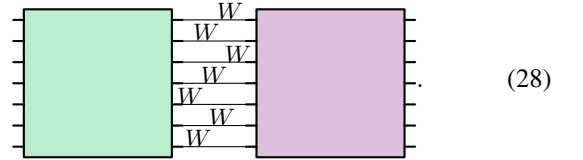
\[\tikz[domain=0:360]{
		\draw[thick,fill=encolor](-16pt,4pt)rectangle+(-56pt,56pt);
		\draw[thick,fill=decolor]( 16pt,4pt)rectangle+( 56pt,56pt);
		\foreach\i in{1,...,7}{
			\draw[thick](-72pt,\i*8pt)--+(-4pt,0)(-16pt,\i*8pt)--+(4pt,0)
			            ( 16pt,\i*8pt)--+(-4pt,0)( 72pt,\i*8pt)--+(4pt,0);
			\draw       (-16pt,\i*8pt)--+(32pt,0)
				({mod(5+3*\i,7)*3-9pt},\i*8pt)node[^,inner sep=0]{\small$W$};
		}
	}.\]
	Here the top pair of pins forms $W^{(1)}$,
	and the bottom pair of pins forms $W^{(7)}$.
	We will later call this tree $𝒯\pf(W,T_\KSU,1)$ (guess why).

\subsubsection{Unbalanced Tree}

	This is motivated by attempts of optimization of polar codes.
	The generalization comes in two perspectives.
	
	First perspective \cite{AYK11,SG13,SGVTG14,ZZWZP15,ZZPYG14}: 
	in a tree like Formula~(\ref{eq:tre8}) or a larger tree,
	it could be the case some synthetic channel, say $(W^♭)^♭$,
	is so bad that applying further transformations sounds useless.
	If so, we may remove children of $(W^♭)^♭$ to get
	
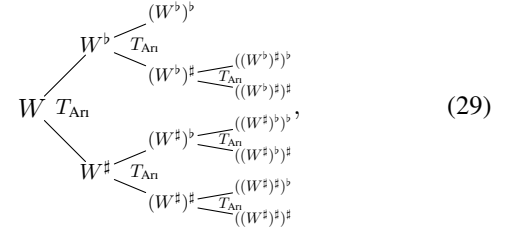
\[\tikz[ut]{
		\utnode$W$T_\Ari${
			\utnode$W^♭$T_\Ari${
				\utnode$(W^♭)^♭$${}{} }{
				\utnode$(W^♭)^♯$T_\Ari${
					\utnode$((W^♭)^♯)^♭$${}{} }{
					\utnode$((W^♭)^♯)^♯$${}{} } } }{
			\utnode$W^♯$T_\Ari${
				\utnode$(W^♯)^♭$T_\Ari${
					\utnode$((W^♯)^♭)^♭$${}{} }{
					\utnode$((W^♯)^♭)^♯$${}{} } }{
				\utnode$(W^♯)^♯$T_\Ari${
					\utnode$((W^♯)^♯)^♭$${}{} }{
					\utnode$((W^♯)^♯)^♯$${}{} } } }
	}, \label{eq:tre7}\]
	which translates into the circuit
	
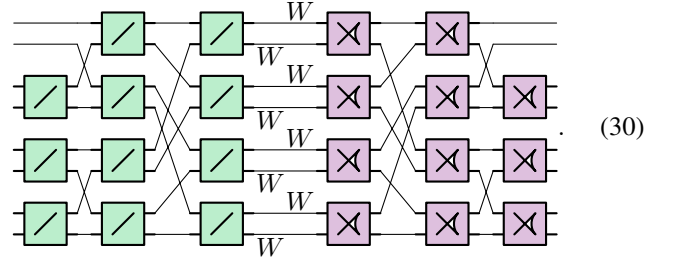
\[\tikz[bb]{
		\foreach\J in{00,01,10,11}{
			\draw foreach\l in{3,2,1}{\ifnum\J\l=113
					(-,\l,\J)\F(\l E\J)(+,\l,\J)\F(\l D\J)\else
					(-,\l,\J)\E(\l E\J)(+,\l,\J)\D(\l D\J)\fi}
				(1E\J1R)--node[^>]{$W$}(1D\J1L)
				(1E\J0R)--node[<_]{$W$}(1D\J0L);}
		\def\DRAW#1#2{
			\draw foreach\J in{000,001,010,011,100,101,110,111}{
				(#2E\K R)--(#1E\J L)(#1D\J R)--(#2D\K L)};}
		\def\K{\expandafter\k\J}
		\def\k#1#2#3{#3#1#2}\DRAW12
		\def\k#1#2#3{#1#3#2}\DRAW23
	}. \label{eq:cir7}\]
	That is, eight instances of $W$ are transformed into four pairs of $W^♭,W^♯$,
	into two quadruples of $(W^♭)^♭$, $(W^♭)^♯$, $(W^♯)^♭$, $(W^♯)^♯$,
	and, \caution{notice the difference},
	while keeping two $(W^♭)^♭$,
	the other six are transformed into $((W^♭)^♯)^♭$, $((W^♭)^♯)^♯$,
	$((W^♯)^♭)^♭$, $((W^♯)^♭)^♯$, $((W^♯)^♯)^♭$, $((W^♯)^♯)^♯$.
	
	Second perspective \cite{EKMFLK15,WLZZ15,EECB17,EKMFLK17,WYY18,WYXY18}:
	in a tree like Formula~(\ref{eq:tre4}),
	it could be that some synthetic channel, say $(W^♭)^♯$,
	might not polarize enough,
	i.e., it is neither extremely good nor extremely bad.
	thus we further polarize it by applying additional $T_\Ari$ as follows:
	\[\tikz[ut]{
		\utnode$W$T_\Ari${
			\utnode$W^♭$T_\Ari${
				\utnode$(W^♭)^♭$${}{} }{
				\utnode$(W^♭)^♯$T_\Ari${
					\utnode$((W^♭)^♯)^♭$${}{} }{
					\utnode$((W^♭)^♯)^♯$${}{} } } }{
			\utnode$W^♯$T_\Ari${
				\utnode$(W^♯)^♭$${}{} }{
				\utnode$(W^♯)^♯$${}{} } }
	}, \label{eq:tre5}\]
	which translates into the circuit
	\[\tikz[bb]{
		\foreach\J in{00,01,10,11}{
			\draw foreach\l in{3,2,1}{
				\ifnum\l=3
					\ifnum\J=10
						(-,\l,\J)\E(\l E\J)(+,\l,\J)\D(\l D\J)
					\else
						(-,\l,\J)\F(\l E\J)(+,\l,\J)\F(\l D\J)
					\fi
				\else
					(-,\l,\J)\E(\l E\J)(+,\l,\J)\D(\l D\J)
				\fi}
				(1E\J1R)--node[^>]{$W$}(1D\J1L)
				(1E\J0R)--node[<_]{$W$}(1D\J0L);}
		\def\DRAW#1#2{
			\draw foreach\J in{000,001,010,011,100,101,110,111}{
				(#2E\K R)--(#1E\J L)(#1D\J R)--(#2D\K L)};}
		\def\K{\expandafter\k\J}
		\def\k#1#2#3{#3#1#2}\DRAW12
		\def\k#1#2#3{#1#3#2}\DRAW23
	}. \label{eq:cir5}\]
	That is, eight instances of $W$ are transformed into four pairs of $W^♭,W^♯$,
	into two quadruples of $(W^♭)^♭$, $(W^♭)^♯$, $(W^♯)^♭$, $(W^♯)^♯$,
	and, \caution{notice another difference},
	only the two $(W^♭)^♯$ are transformed into $((W^♭)^♯)^♭$, $((W^♭)^♯)^♯$.

\subsubsection{Multi-Kernel}

	\hskip0ptplus3em\cite{BBGL17,BGLB17,GBLB17,BCL18}
	The Transformation $T_\Ari$ generates codes
	whose block lengths are powers of $2$.
	A transformation $T_\KSU$ induced by an $ℓ$-by-$ℓ$ matrix
	generates codes whose block lengths are powers of $ℓ$.
	For something in between, say length-$72$,
	one can apply $T_\Ari$ three times and then apply
	$T_\GBLB$, a transformation induced by a $3$-by-$3$ matrix, two times.
	
	Here is a small (length-$6$) example.
	First $T_\Ari$ is applied, and then $T_\GBLB$ is applied.
	\[\tikz[ut]{
		\draw(0,.5\utupper)node(0){$W$}(.666,.5\utupper)node[s]{$T_\Ari$};
		\foreach\i in{1,2}{
			\pgfmathsetlength\utmiddle{(\i*\utlower+(2-\i)*\utupper)/2}
			\pgfmathsetlength\utupper{\utmiddle+(\utupper-\utlower)/2}
			\pgfmathsetlength\utlower{\utmiddle}
			\pgfmathsetlength\utmiddle{.5\utlower+.5\utupper}
			\def\W{W\ifcase\i\or^♭\or^♯\fi}
			\draw(1,\utmiddle)node[s](1){$\W$}
				(1.4,\utmiddle)node[s,s,_>]{$T_\GBLB$}(1)--(0);
		{\foreach\j in{1,2,3}{
			\pgfmathsetlength\utmiddle{(\j*\utlower+(3-\j)*\utupper)/3}
			\pgfmathsetlength\utupper{\utmiddle+(\utupper-\utlower)/3}
			\pgfmathsetlength\utlower{\utmiddle}
			\pgfmathsetlength\utmiddle{.5\utlower+.5\utupper}
			\def\X{(\W)^{(\j)}}
			\draw(2,\utmiddle)node[s,s](2){$\X$}(2)--(1);
		}}}
	} \label{eq:tre6}\]
	This translates into the circuit drawn below.
	\newdimen\bbbwire
	\pgfmathsetlength\bbbwire{\bbwire+\bbgape}
	\tikzset{
		G3/.pic={\def\coor{coordinate}\draw[thick]
			(-\bbbwire, 2\bbgape)\coor(0L)--(\bbbwire, 2\bbgape)\coor(0R)
			(-\bbbwire, 0\bbgape)\coor(1L)--(\bbbwire, 0\bbgape)\coor(1R)
			(-\bbbwire,-2\bbgape)\coor(2L)--(\bbbwire,-2\bbgape)\coor(2R)
		;},
		E3/.pic={\pic{G3};\draw[thick,fill=encolor]
			(-\bbsize-\bbgape,-\bbsize-\bbgape)rectangle
			( \bbsize+\bbgape, \bbsize+\bbgape)
		;},
		D3/.pic={\pic{G3};\draw[thick,fill=decolor]
			(-\bbsize-\bbgape,-\bbsize-\bbgape)rectangle
			( \bbsize+\bbgape, \bbsize+\bbgape)
		;}
	}
	\newdimen\bbbpose
	\pgfmathsetlength\bbbpose{4\bbwire+\bbgape+10pt}
	\[\tikz[bb]\draw
		                   (-,1,10)\E(0)(+,1,10)\D(1)
		(-\bbbpose,3.5\bbwire)pic(2){E3}(\bbbpose,3.5\bbwire)pic(3){D3}
		                   (-,1,01)\E(4)(+,1,01)\D(5)
		(-\bbbpose, .5\bbwire)pic(6){E3}(\bbbpose, .5\bbwire)pic(7){D3}
		                   (-,1,00)\E(8)(+,1,00)\D(9)
		(20R)--(01L)(01R)--node[^>]{$W$}(11L)(11R)--(30L)
		(60R)--(00L)(00R)--node[<_]{$W$}(10L)(10R)--(70L)
		(21R)--(41L)(41R)--node[^>]{$W$}(51L)(51R)--(31L)
		(61R)--(40L)(40R)--node[<_]{$W$}(50L)(50R)--(71L)
		(22R)--(81L)(81R)--node[^>]{$W$}(91L)(91R)--(32L)
		(62R)--(80L)(80R)--node[<_]{$W$}(90L)(90R)--(72L)
	; \label{eq:cir6}\]

\subsubsection{Alphabet Extension}

	\cite{PSL11,PSL16}
	There is a special type of channel transformations
	corresponding to field extensions $𝔽_q⊂𝔽_{q^k}$ for any $q,k$.
	That is to say, $k$ independent copies of a $q$-ary erasure channel
	can transmit a $q^k$-bit.
	We claim an erasure if any of $k$ symbols in the ground field misses.
	Denote the transformation by $T\fek$.
	Draw
	\[\tikz[ut]\draw
		(0,.5\utupper)node(0){$W$}+(1,0)node[s](1){$W^k$}
		(0)--node[above,s]{$T\fek$}(1)
		(0)--(1)
	;\]
	for $W^k$ the $k$-th power of the channel $W$.
	Here is a $k=5$ translation.
	
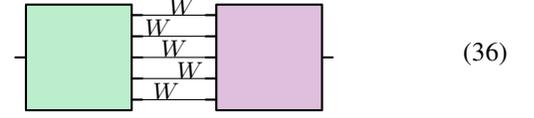
\[\tikz[domain=0:32,smooth]{
		\foreach\i in{1,...,5}{
			\draw[thick](-16pt,\i*8pt)--+( 4pt,0)( 16pt,\i*8pt)--+(-4pt,0);
			\draw       (-16pt,\i*8pt)--+(32pt,0)
				({mod(3+3*\i,5)*3-6pt},\i*8pt)node[^,inner sep=0]{\small$W$};
		}
			\draw[thick]( 56pt,24pt)--+( 4pt,0);
			\draw[thick](-56pt,24pt)--+(-4pt,0);
		\draw[thick,fill=encolor](-16pt,4pt)rectangle+(-40pt,40pt);
		\draw[thick,fill=decolor]( 16pt,4pt)rectangle+( 40pt,40pt);
	}\]

\subsubsection{Some Convention}

	Although it is theoretically possible for a tree to have
	multiple, nested $T\fek$, each with different parameters $k$,
	we limit our interest to two small classes of trees.
	They are
	\begin{itemize}
		\item trees consisting of one transformation $T$ (that is not $T\fek$);
			or
		\item trees consisting of three transformations $T\rat,T\fek,T\err$,
			wherein
		\begin{itemize}
			\item all $T\fek$ correspond to the same parameter $k$,
			\item every root-to-leaf path passes exactly one $T\fek$,
			\item every root-to-$T\fek$ path passes only $T\rat$, and
			\item every $T\fek$-to-leaf path passes only $T\err$.
		\end{itemize}
	\end{itemize}
	
	That said, in the second case, $T\fek$  might not locate at the same depth.
	It turns out allowing $T\fek$ to be at different depths
	bursts the performance,
	theoretically and practically.
	
	Denote by $𝒯$ a tree of channels with root channel $W$.

\subsection{\Bha{} Parameter and Process}

	The \emph{\Bha{} parameter} $Z(w)$ of a channel $w$
	measures the unreliability, the badness, of the channel.
	For instance for BDMC
	\begin{gather}
		I(w)+Z(w)≥1, \\
		I(w)+Z(w)^2≤1,\rlap{ and} \\
		I(w)\log2+Z(w)≤1,
	\end{gather}
	by \cite[Corollary~5]{JA18}.
	That is to say, this pair of parameters $\(I(w),Z(w)\)$ lies in
	\[\tikz[Tri]\draw
		(0,1)pic{dot}node[<<]{$(0,1)$}--
		(1,0)pic{dot}node[>>]{$(1,0)$}
		plot[domain=0:1](1-\x^2,\x)
		plot[domain=0:1](\x,1-\x*.693)
		(0.804021,.442695)pic{dot}node[^>]{$(.80,.44)$}
	;.\]
	
	That said, an explicit definition of \Bha{} parameters is not presented here
	since all we need in this work is the following two properties
	playing as axioms:
	\begin{itemize}
		\item (Regarding transformations) \cite[Lemma~33]{MT14}
			For any transformation $T$ we are interested in,
			it has an operator norm $\absT$ such that
			for any channel $w$ we are interested in
			and any outcome $v$ of $T(w)$,
			the multiple $\absT Z(w)$ bounds $Z(v)$ from above.
		\item (Regarding error probability) \cite[Lemma~22]{MT14}
			For any $q$-ary channel $w$ we are interested in,
			the multiple $qZ(w)$ bounds from above the probability that
			a decoder fails to decode a single symbol transmitted through $w$.
	\end{itemize}
	
	Given the nice properties, the general strategy is
	to fully control $Z(w)$ for as many $w$ as possible in a tree,
	and then rewrite the resulting inequalities in terms of error probabilities.
	During this process, it is not important anymore
	what the \Bha{} parameter is.
	In theory, it could be replaced by any function
	that satisfies the aforementioned two axioms.
	Starting from Section~\ref{sec:zpp}, we will call it $Z$-parameter instead.
	
	Given a channel tree $𝒯$ with root channel $W$,
	define two discrete-time stochastic processes $W_i,Z_i$
	and a stopping time $τ$ as follows:
	Start from the root channel $W_0≔W$; and let $Z_0≔Z(W_0)$.
	For any $i≥0$, if $W_i$ is a leaf channel, let $τ≔i$.
	If, otherwise, $W_i$ has children, choose a child uniformly at random
	as $W_{i+1}$; and let $Z_{i+1}≔Z(W_{i+1})$.
	(c.f.\ \cite[Section~IV, third paragraph]{Arikan09}.)
	
	In case of \Arikan's polar codes,
	$Z_i$ is a martingale over BEC and is a super-martingale over other BDMC
	\cite[Proposition~9]{Arikan09}.
	For other binary kernels over general BDMC,
	\cite[Remark~5]{KSU10} claims that it is difficult to characterize,
	but they manage to prove a useful statement \cite[Lemma~10]{KSU10}.
	For larger alphabets,
	\cite[Lemma~33]{MT14} claims that it is very similar to the binary case.
	We provide our generalization in Lemma~\ref{lem:bounded}.
	
	For a tree $𝒯$ as in Formula~(\ref{eq:tre5}),
	a possible instance of the process is
	
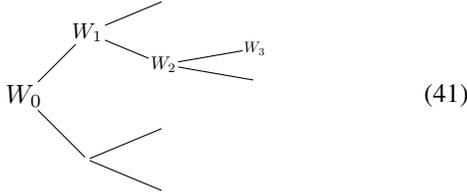
\[\tikz[ut]{
		\utnode$W_{0}$${
			\utnode$W_{1}$${
				\utnode$$${}{} }{
				\utnode$W_{2}$${
					\utnode$W_{3}$${}{} }{
					\utnode$$${}{} } } }{
			\utnode$$${
				\utnode$$${}{} }{
				\utnode$$${}{} } }
	}\]
	with $τ=3$ and $W_τ=W_3$.
	The probability measure of this path is $1/8$.
	For another instance
	
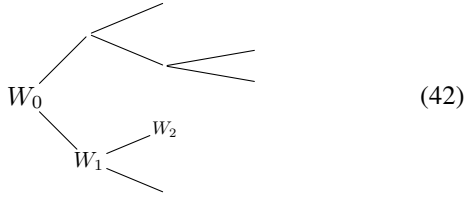
\[\tikz[ut]{
		\utnode$W_{0}$${
			\utnode$$${
				\utnode$$${}{} }{
				\utnode$$${
					\utnode$$${}{} }{
					\utnode$$${}{} } } }{
			\utnode$W_{1}$${
				\utnode$W_{2}$${}{} }{
				\utnode$$${}{} } }
	}\]
	with $τ=2$ and $W_τ=W_2$, the probability measure is $1/4$.
	
\subsection{Construct Code and Communicate}

	In a given tree $𝒯$, non-leaf vertices represent
	channels that are consumed to obtain their children.
	They are not available to users.
	Leaves of $𝒯$, however, represent channels that are available to users.
	
	A person who wants to send messages can
	(a) choose a subset $𝒜$ of leaves,
	(b) transmit uncoded messages through leaf channels in $𝒜$, and
	(c) transmit predictable symbols through the remaining leaf channels.
	
	This tree-leaves pair $(𝒯,𝒜)$ determines a block code.
	A block code has block length~$N$, code rate~$R$, and error probability~$P$.
	The following is how to read-off these parameters from the pair $(𝒯,𝒜)$.
	
	For every leaf channel $w$ in $𝒯$,
	the probability $ℙ\{W_τ=w\}$ is the reciprocal of an integer.
	This integer is the product of the ``$ℓ$s'' of
	$W_0,W_1,\dotsc,W_{τ-1}$ when $W_τ=w$.
	
	The \emph{block length} $N$ of $𝒯$ is the least positive integer
	such that $Nℙ(W_τ=w)$ is an integer for every leaf channel $w$, i.e.,
	\[N≔\lcm_{w:\text{leaf}}\frac1{ℙ\{W_τ=w\}},\]
	when $T\fek$ does not present.
	When $T\fek$ does present,
	\[N≔\lcm_{w:\text{leaf}}\frac k{ℙ\{W_τ=w\}}.\]
	
	The \emph{code rate} $R$ of $(𝒯,𝒜)$
	is the probability that $W_τ$ ends up in $𝒜$.
	\[R≔ℙ\{W_τ∈𝒜\}.\]
	
	The \emph{error probability} $P$ is the probability that
	any leaf channel  in $𝒜$ fails to transmit the message.
	For \Arikan's polar codes,
	this quantity is less than the weighted sum
	\[∑_{w\in𝒜}Nℙ\{W_τ=w\}Z(w)\]
	by \cite[Proposition~2]{Arikan09}.
	For other binary kernels, \cite[Formula~(12)]{KSU10} claims the same.
	For larger alphabets, it is still true that the error probability
	is less than a multiple of the weighted sum \cite[Lemma~22]{MT14}.
	Later in Section~\ref{sec:zpp}, we will \emph{define}
	the error probability to be the sum.

\subsection{The Three Regimes}

	To investigate the tradeoff among
	block length $N$, code rate $R$, and error probability $P$,
	researchers have developed three general directions:
	\begin{itemize}
		\item error exponent regime (varying $N,P$);
		\item scaling exponent regime (varying $N,R$); and
		\item moderate deviations regime (varying $N,P,R$ at once).
	\end{itemize}
	See \cite[Abstract and Section~1]{MHU16} for an alternative introduction.

\subsubsection{Error Exponent Regime}

	The error exponent regime studies
	the tradeoff between $N,P$ when $R$ is bounded from below.
	That is, if we want to communicate at a certain rate $R\lb$
	and ask for longer and longer codes,
	what is the gain of $P$ in exchange for $N$?
	
	For a series of block codes (including random codes), the number
	\[\liminf_{N→∞}\frac{-\log P}N\]
	measures how fast $P$ decays to zero
	and is called the error exponent \cite{Gallager65}.
	Hence the name error exponent regime.
	For random codes with $R\lb$ fixed, the error exponent is positive,
	and it is an interesting s to approximate the error exponent.
	
	However, for other codes such as polar codes
	or random codes with ``fast growing $R$'' (will explain soon),
	$-\log P$ is sub-linear in $N$ so the error exponent vanishes.
	In such case, the second best thing is the quantity
	\[β'≔\liminf_{N→∞}\frac{\log(-\log P)}{\log N}\]
	being positive.
	
	The best possible $β'$ a coding scheme can obtain
	is denoted by $β$ in some literature.
	For codes with positive error exponent, $β=1$.
	(And being $1$ is optimal.)
	For \Arikan's polar codes with $R\lb$ fixed, $β=1/2$ \cite{AT09}.
	For polar codes with arbitrary kernels with $R\lb$ fixed,
	$β$ is the average of logarithms of the \emph{partial distances}
	\cite{KSU10}.
	Chances are that some deliberately selected kernels
	produce polar codes with $β$ arbitrarily close to $1$, but not exactly $1$.
	
	However, for polar codes with ``fast growing $R$'' (will explain soon),
	$β'$ is strictly less than $β$,
	and how much $β'$ is less than $β$
	depends largely on how fast $R$ is approaching the capacity.
	This dependency is the main interest of this work.
	
	In Section~\ref{sec:padic}, we will define the \emph{$∂$-dice}
	which generalizes the usual partial distances.
	We pretend this extra level of abstraction
	makes possible application in other paradigms, e.g.\ LDPC.
	Readers are invited to read ``partial distance''
	every time they see ``\texttt{\string\partial}-dice''.
	See Appendix~\ref{app:eer} for a note on error exponent regime.
	
\subsubsection{Scaling Exponent Regime}

	The scaling exponent regime studies
	the tradeoff between $N,R$ when $P$ is bounded from above.
	That is, if we want to communicate at a certain error probability $P\ub$
	and ask for longer and longer codes,
	what is the gain of $R$ in exchange for $N$?
	
	The number
	\[μ'≔\liminf_{N\to\infty}\frac{\log N}{-\log\(I(W)-R\)}\]
	measures how fast $R$ approaches the capacity
	and is sometimes called the scaling exponent.
	Hence the name scaling exponent regime.
	
	The best possible $μ'$ a coding scheme can obtain
	is denoted by $μ$ in some literature.
	For random codes with $P\ub$ fixed, $μ=2$.
	(And being $2$ is optimal.)
	For \Arikan's polar codes with $P\ub$ fixed,
	$μ=3.627$ on BEC \cite{FV14} and $μ≤4.714$ on other channels \cite{MHU16}.
	For polar codes with arbitrary kernels,
	it is difficult to approximate but researchers tried to bound \cite{MHU16}.
	Chances are that some randomly selected kernels produce polar codes
	with $μ$ arbitrarily close to $2$, but not exactly $2$ \cite{FHMV17}.
	
	However, for random codes and polar codes
	with ``fast decaying $P$'' (will explain soon),
	$μ'$ will be strictly more than $μ$,
	and how much $μ'$ is more than $μ$
	depends largely on how fast $P$ is decaying to zero.
	This dependency is the main interest of this work.
	
	In Section~\ref{sec:muexp} we will define the \emph{$μ^*$-exponent}
	which is a variant of $μ$.
	The definition of $μ^*$ is made so that, say,
	proving $μ^*≤5$ is much easier than proving $μ≤5$,
	and then our analysis nonsense (as opposite to abstract nonsense)
	will complete the rest of proof.
	See Appendix~\ref{app:ser} for a note on scaling exponent regime.

\subsubsection{Moderate Deviations Regime}

	We mentioned above that $β'≤1$ and $1$ can be achieved.
	We also mentioned that $1/μ'≤1/2$ and $1/2$ can be achieved.
	These poses new questions:
	Are those all restrictions?
	Can, in particular, a family of codes achieve $(β',1/μ')=(1,1/2)$?
	
	The moderate deviations regime studies $N,R,P$ as a whole
	to answer these questions.
	The answer turns out to be NO.
	There are more fundamental restrictions on the pair $(β',1/μ')$, i.e., on
	\[（
		\liminf_{N\to\infty}\frac{\log(-\log P)}{\log N},
		\liminf_{N\to\infty}\frac{-\log\(I(W)-R\)}{\log N}
	）,\]
	that stop a family of codes from achieving $(1,1/2)$.
	
	The restrictions can be seen in the following way:
	That $0≤1/μ'≤1/2$ is illustrated by this vertical segment
	\[\tikz[Tri]\draw
		(0,.5)pic{dot}node[<<]{$(0,1/2)$}--
		(0,0 )pic{dot}node[<<]{$(0,0)$}
	;.\]
	That $0≤β'≤1$ is illustrated by this horizontal segment
	\[\tikz[Tri]\draw
		(0,0 )pic{dot}node[<<]{$(0,0)$}--
		(1,0 )pic{dot}node[>>]{$(1,0)$}
	;.\]
	The moderate deviations regime then shows that the pair $(β',1/μ')$
	lies in, or on the boundary of, the following right triangle
	\[\tikz[Tri]\draw
		(0,.5)pic{dot}node[<<]{$(0,1/2)$}--
		(0,0 )pic{dot}node[<<]{$(0,0)$}--
		(1,0 )pic{dot}node[>>]{$(1,0)$}--cycle
	;.\]
	It also shows that every point inside or on the boundary
	is achievable by random codes
	\[\tikz[Tri]\fill
		(0,.5)pic{dot}node[<<]{$(0,1/2)$}--
		(0,0 )pic{dot}node[<<]{$(0,0)$}--
		(1,0 )pic{dot}node[>>]{$(1,0)$}--cycle
	;.\]
	So far polar codes achieve
	\[\tikz[Tri]{
		\fill(0,0)--plot file{AriRS2.txt}--cycle;
		\draw
			(0,.5)pic{dot}node[<<]{$(0,1/2)$}--
			(0,0 )pic{dot}node[<<]{$(0,0)$}--
			(1,0 )pic{dot}node[>>]{$(1,0)$};
	}\]
	on BEC.
	We will expand it to
	\[\tikz[Tri]\fill
		(0,0 )pic{dot}node[<<]{$(0,0)$}--
		(0,.5)pic{dot}node[<<]{$(0,1/2)$}--
		(1,0 )pic{dot}node[>>]{$(1,0)$}--cycle
	;\]
	on BEC.
	
	See Appendix~\ref{app:mdr} for a note on moderate deviations regime.

\subsection{Large Deviations Theory}

	Assume $Y$ is a discrete, bounded random variable.
	Let $Y_1,Y_2,\dotsc$ be i.i.d.\ copies of $Y$.
	Let $S_n≔Y_1+Y_2+\dotsb+Y_n$ be the partial sum.
	Let $y$ be a number that is about, but smaller than, $𝔼Y$.
	We want to control the probability
	\[ℙ｛\frac{S_n}n≤y｝\]
	in terms of $y$ and the distribution of $Y$.
	
	The canonical argument goes as follows:
	For every $λ<0$,
	\begin{align}
		ℙ｛\frac{S_n}n≤y｝ &= ℙ\{\exp(λS_n)≥\exp(λny)\} \\
		&≤ 𝔼[\exp(λS_n)]\exp(-λny) \\
		&= 𝔼[\exp(λY)]^n\exp(-λy)^n
	\end{align}
	by the Chernoff bound and independency.
	Take logarithms and divide by $-n$:
	\[\frac{-1}n\logℙ｛\frac{S_n}n≤y｝≥λy-\log𝔼[\exp(λY)].\]
	
	Since the right hand side of the inequality contains a free parameter $λ<0$,
	it makes sense to take the supremum and treat it as a function of $y$
	\[\frac{-1}n\logℙ｛\frac{S_n}n≤y｝≥\sup_{λ<0}λy-\log𝔼[\exp(λY)].
		\label{eq:suped}\]
	That motivates the definition of the \emph{\Cramer{} function}
	\[\CF(y)≔\sup_{λ<0}λy-\log𝔼[\exp(λY)].\]
	
	Two non-obvious comments:
	(a) Take the supremum over $λ∈ℝ$ still gives the same result for $y<𝔼Y$.
		Doing so makes it 
		the Legendre--Fenchel transformation
		of the cumulant generating function of $Y$.
	(b) $\CF$ as defined above is the largest possible function
		such that Formula~(\ref{eq:suped}) holds.
	
	See \cite[Theorem~2.1.24 and 2.2.3]{DZ10} for more on this topic.


\section{Preliminary}

	In this section, we consolidate the notations
	that will be useful to state and prove theorems.

\subsection{Channel Transformation}

	A \emph{communication channel} is a triple $(𝒳,𝒴,W)$
	of a finite input alphabet $𝒳$, a finite output alphabet $𝒴$,
	and an one-step Markov process
	\[W:𝒳⟶𝒴.\]
	To abuse notation, write $W$ to mean the full triple.
	The cardinality of $𝒳$ is called
	the input size of $W$, or the \emph{arity} of $W$ for short.
	
	Let $𝒞$ be the set of channels we are interested in.
	A \emph{channel transformation} is a triple $(𝒟,ℓ,T)$
	of a domain $𝒟⊂𝒞$, a length $ℓ∈ℕ$, and a map
	\[T:𝒟⟶𝒞^ℓ.\]
	To abuse notation, write $T$ to mean the full triple.
	
	In this work, every $𝒟$ consists of channels of the same arity.
	We refer to this number as the arity of $𝒟$, or the arity of $T$ for short.
	For instance, $T_\Ari$ works on channels of binary input,
	so $T_\Ari$ has arity $2$, or it is a binary ($2$-ary) transformation.
	
	Unless stated otherwise, transformations in this work are such that $ℓ≥2$ and
	\[T:𝒟⟶𝒟^ℓ.\]
	Therefore, it is well-defined
	when the same transformation is applied iteratively.
	For instance, Formula~(\ref{eq:tre4}) begins with $T_\Ari(W)=(W^♭,W^♯)$
	and then $T_\Ari(W^♭)=((W^♭)^♭,(W^♭)^♯)$ and $T_\Ari(W^♯)=((W^♯)^♭,(W^♯)^♯)$.
	They all are of binary input.
	
	We also define an exceptional transformation $(𝒟,1,T\fek)$ where
	\[T\fek:𝒟⟶𝒞\]
	transforms $q$-ary channels to $q^k$-ary channels,
	for some integer parameter $k$.
	This corresponds to the fact that
	$k$ instances of $q$-ary channels can be seen as a $q^k$-ary channel.
	Or dually, a $k$-tuple of $q$-bits can be seen as a $q^k$-bit.

\subsection{Channel Tree}

	A \emph{channel tree} $𝒯$ is a rooted tree
	where each vertex is a channel in $𝒞$,
	and each non-leaf vertex $w$ corresponds to a transformation $T$
	such that $T(w)$ are children of $w$.
	In this work, channel trees are generated by
	\begin{itemize}
		\item Begin with a channel $W$ as the root of a new tree.
		\item For each leaf channel $w$, run a deterministic algorithm
			that observes the current tree
			and decides wether to apply a certain transformation or not.
		\item If $T$ is applied, append synthetic channels $T(w)$
			as children of $w$.
	\end{itemize}
	Most channel trees in this work are finite.
	In fact, a good algorithm will stop applying transformations
	once the depth reaches some prescribed number $n$.
	
	For instance, let $𝒯\pf(W,T,n)$ be the channel tree generated as follows:
	\begin{itemize}
		\item Begin with $W$ as the root of a new tree.
		\item For each leaf channel $w$,
			apply $T$ if the depth of $w$ is not yet $n$.
			(The algorithm merely checks the depth.)
		\item By applying $T$, we mean to
			append synthetic channels $T(w)$ as children of $w$.
	\end{itemize}
	
	Convention: the root has depth $0$;
	the tree $𝒯\pf(W,T,n)$ has $ℓ^n$ leaves,
	where $ℓ$ is the length of $T$.
	Some examples are
	Formula~(\ref{eq:tre2}) being $𝒯\pf(W,T_\Ari,1)$;
	Formula~(\ref{eq:tre4}) being $𝒯\pf(W,T_\Ari,2)$;
	Formula~(\ref{eq:tre8}) being $𝒯\pf(W,T_\Ari,3)$; and
	Formula~(\ref{eq:trel}) being $𝒯\pf(W,T_\KSU,1)$.
	
	More involved example:
	Formula~(\ref{eq:tre6}) is $𝒯\pf(W,T_\Ari,1)$
	except that a leaf $W^♭$ is merged with $𝒯\pf(W^♭,T_\GBLB,1)$,
	and the other leaf $W^♯$ is merged with $𝒯\pf(W^♯,T_\GBLB,1)$.
	
	Let $𝒯\pf(W,T,∞)$ be the infinite tree.
	This is useful when arguing about the process $Z_i$ (defined below)
	without having to worry about whether $i≤n$ or not.

\subsection{\TP$Z${Z}-Parameter and Processes} \label{sec:zpp}

	A $Z$-parameter will be a function $Z:𝒞→[0,1]$
	measuring the unreliability, the badness, of a given channel.
	It does not have to be exactly the \Bha{} parameter,
	but could be any function such that a multiple of $Z(w)$ bounds,
	from above, the probability that
	a decoder fails to decode a single symbol transmitted through $w$.
	
	Given a channel tree with root channel $W$, define
	a discrete-time stochastic process $W_i$ and a stopping time $τ$ as follows:
	Start from the root channel $W_0≔W$.
	For any $i≥0$, if $W_i$ is a leaf channel, let $τ≔i$.
	If, otherwise, $W_i$ has $ℓ$ children,
	choose an integer $X_{i+1}$ from $1,2,\dotsc,ℓ$ uniformly at random,
	and let $W_{i+1}$ be the $X_{i+1}$-th child of $W_i$.
	
	Be careful that \latin{a priori}
	$X_i$ are neither independent nor identical.
	This is because $X_1$ controls the number of children of $W_1$,
	which affects the distribution of $X_2$.
	However, they are i.i.d.\ in $𝒯\pf(W,T,∞)$.
	
	Let $Z_i$ be $Z(W_i)$.
	Let $\Y_i$ be $\log(\log Z_i/\log Z_{i-1})$;
	this is the ``empirical increment'' of
	$\log(-\log Z_i)$.
	Let $T_{i-1}$ be the transformation applied to $W_{i-1}$.
	Then the empirical increment can also be written as
	\[\Y_i=\log\frac{\log Z\(X_i\text{-th component of }T_{i-1}(W_{i-1})\)}
		{\log Z(W_{i-1})}. \label{eq:empinc}\]
	This motivates the definition of the ``theoretical increment''
	(without underline)
	\[Y_i≔\liminf_{\substack{w∈𝒟\\\mathclap{Z(w)→0}}}\log\frac
		{\log Z\(X_i\text{-th component of }T_{i-1}(w)\)}
		{\log Z(w)}.\! \label{eq:theinc}\]
	The purpose of defining two types of ``increments'' is that
	$Y_i$ are i.i.d.\ in $𝒯\pf(W,T,∞)$ and
	approximate $\Y_i$ in a certain context.
	It is easy to study $Y_i$ and then predict $\Y_i$ accordingly.

\subsection{Root-to-Leaf Path as Sample, Vertex as Event}

	The process $W_i$ implicitly assumes a sample space:
	the set of all root-to-leaf paths of $𝒯$.
	Each vertex lies on a subset of root-to-leaf paths,
	which form an event.
	Thus we can talk about the probability measure of a vertex.
	It is the probability that
	the trajectory $W_0,W_1,\dotsc,W_τ$ passes that vertex.
	
	Furthermore, for any two vertices,
	their corresponding events are disjoint if and only if
	neither of them is a descendant of the other one.
	Thus it makes sense to say two vertices are disjoint or not.
	For a subset of pairwise-disjoint vertices,
	its probability measure is the sum of probability measures of these vertices.
	It is also the probability that
	the trajectory $W_0,W_1,\dotsc,W_τ$ passes any of these vertices.
	
	Let $w$ be a synthetic channel at depth $j$.
	When $W_j$ happens to be $w$,
	the trajectory $W_0,W_1,\dotsc,W_{j-1}$ is uniquely determined.
	(In entropy notation, $H(W_i|W_j)=0$ for $0≤i<j$.)
	It also determines $T_0,X_0,\Y_0,Y_0,Z_0$
	and their successors up to $T_j,X_j,\Y_j,Y_j,Z_j$.

\subsection{Construct Code and Communicate}

	Let $𝒯$ be a channel tree and $𝒜$ be a subset of leaves of $𝒯$.
	The pair $(𝒯,𝒜)$ defines a block code.
	
	The \emph{block length} $N$ of $(𝒯,𝒜)$ is
	\[N≔\lcm_{w:\text{leaf}}\frac1{ℙ(w)}\]
	when $T\fek$ does not present.
	When $T\fek$ does present,
	\[N≔\lcm_{w:\text{leaf}}\frac k{ℙ(w)}.\]
	
	The \emph{code rate} $R$ of $(𝒯,𝒜)$ is
	\[R≔ℙ(𝒜).\]
	
	The \emph{error probability} $P$ of $(𝒯,𝒜)$ is \emph{defined} as
	\[P≔∑_{w\in𝒜}Nℙ(w)Z(w).\]

\subsection{The \TP$∂${∂}-Dice of a Transformation} \label{sec:padic}

	Let $T$, or formally $(𝒟,ℓ,T)$, be a length-$ℓ$ transformation.
	Let $X$ be a random integer chosen uniformly from $1,2,\dotsc,ℓ$.
	Define the \emph{$∂$-dice of $T$}:
	\[Y≔\liminf_{\substack{w∈𝒟\\Z(w)→0}}\log
		\frac{\log Z\(X\text{-th component of }T(w)\)}{\log Z(w)}.\]
	Compare this to Formulae (\ref{eq:empinc}) and~(\ref{eq:theinc}):
	$Y$ is the ``prototype'' of $Y_i$ in $𝒯\pf(W,T,∞)$,
	i.e., $Y_i$ are i.i.d.\ copies of $Y$.
	
	Call $T$ \emph{bounded} if there exists a number,
	denoted by $\absT$, such that, for all $w∈𝒟$,
	\[\frac{Z(\text{every component of }T(w))}{Z(w)}<\absT.\]
	Call $T$ \emph{powerful} if
	\[ℙ\{Y>0\}>0.\]
	The following lemma motivates a necessary condition for our main theorems.
	\begin{lem} \label{lem:bounded}
		Consider $𝒯\pf(W,T,∞)$ for any $W∈𝒟$.
		If $T$ is bounded, then
		\[Y≥0.\]
		If $T$ is bounded and powerful, and $ϵ>0$ is small enough,
		then there exists $δ>0$ such that
		\[(Z_i∧δ)^ϵ\text{ is a super-martingale}.\]
		Here $Z_i∧δ$ is a shorthand for $\min(Z_i,δ)$.
	\end{lem}
	\begin{IEEEproof}
		For the first statement,
		\begin{align}
			Y
			&≥ \liminf_{\substack{w∈𝒟\\Z(w)→0}}\log
				\frac{\log(Z(w)\absT)}{\log Z(w)} \\
			&= \liminf_{\substack{w∈𝒟\\Z(w)→0}}\log
				（1+\frac{\log\absT}{\log Z(w)}） \\
			&= \log(1+0).
		\end{align}
		
		For the second statement,
		start from
		\[ℙ\{Y≤0\}=1-ℙ\{Y>0\}<1.\]
		Pick a smaller $ϵ>0$ such that
		\[ℙ\{Y<2ϵ\}<1.\]
		Pick a smaller $ϵ>0$ such that
		\[\absT^ϵℙ\{Y<2ϵ\}<1.\]
		Pick a number $δ>0$ such that
		\[\absT^ϵℙ\{Y<2ϵ\}+δ^{ϵ·ϵ}ℙ\{Y≥2ϵ\}≤1. \label{eq:atmostone}\]
		Pick a smaller $δ>0$ such that
		\[\inf_{\substack{w∈𝒟\\Z(w)<δ}}
			\log\frac{\log Z\(X\text{-th component of }T(w)\)}{\log Z(w)}>Y-ϵ.\]
		Note that this is saying
		\[Z_{i-1}<δ\text{ implies }\Y_i>Y_i-ϵ.\]
		
		Now bound $𝔼［(Z_i∧δ)^ϵ｜Z_0,\dotsc,Z_{i-1}］$
		by considering one plus two cases:
		(a) If $Z_{i-1}≥δ$, then it is automatically true that
		\begin{align}
			𝔼［(Z_i∧δ)^ϵ｜Z_0,\dotsc,Z_{i-1}］
			&≤ 𝔼［δ^ϵ｜Z_0,\dotsc,Z_{i-1}］ \\
			&= δ^ϵ \\
			&= (Z_{i-1}∧δ)^ϵ.
		\end{align}
		(b-i) If $Z_{i-1}<δ$ and $Y_i<2ϵ$, then
		\[(Z_i∧δ)^ϵ≤Z_i^ϵ≤Z_{i-1}^ϵ\absT^ϵ.\]
		(b-ii) If $Z_{i-1}<δ$ and $Y_i≥2ϵ$, then
		\[Z_i=Z_{i-1}^{\exp\Y_i}≤Z_{i-1}^{1+\Y_i}
			<Z_{i-1}^{1+Y_i-ϵ}≤Z_{i-1}^{1+ϵ}<Z_{i-1}δ^ϵ\]
		and hence
		\[(Z_i∧δ)^ϵ≤Z_i^ϵ≤Z_{i-1}^ϵδ^{ϵ·ϵ}.\]
		(b) Combine (b-i) and (b-ii) to get, when $Z_{i-1}<δ$,
		\begin{align}
			&{} 𝔼［(Z_i∧δ)^ϵ｜Z_1,\dotsc,Z_{i-1}］ \\
			&≤ Z_{i-1}^ϵ\absT^ϵℙ\{Y_i<2ϵ\}+Z_{i-1}^ϵδ^{ϵ·ϵ}ℙ\{Y_i≥2ϵ\} \\
			&= Z_{i-1}^ϵ\(\absT^ϵℙ\{Y_i<2ϵ\}+δ^{ϵ·ϵ}ℙ\{Y_i≥2ϵ\}\) \\
			&≤ Z_{i-1}^ϵ·1 \\
			&= (Z_{i-1}∧δ)^ϵ
		\end{align}
		where the last inequality is by Formula~(\ref{eq:atmostone}).
		Combine (a) and (b) to get
		\[𝔼［(Z_i∧δ)^ϵ｜Z_1,\dotsc,Z_{i-1}］≤(Z_{i-1}∧δ)^ϵ.\]
		This proves
		\[(Z_i∧δ)^ϵ\text{ is a super-martingale}.\]
	\end{IEEEproof}
	
	For $T_\Ari$, the lemma does not imply, but it is true,
	that $Z_i$ is a super-martingale \cite[Proposition~9]{Arikan09}.

\subsubsection{The \TP$β^*${β*}-exponent of \TP$T${T}} \label{sec:beexp}

	Define the $β^*$-exponent:
	\[β^*≔\frac{𝔼Y}{\logℓ}.\]

\subsection{The \TP$μ^*${μ*}-Exponent of a Transformation} \label{sec:muexp}

	Let $T$, or formally $(𝒟,ℓ,T)$, be a transformation and let $W∈𝒟$.
	Let $𝒜_n$ be the subset of leaves $w$ in $𝒯\pf(W,T,n)$ such that%
	\footnote{Please be informed that Formula~(\ref{eq:quasipoly})
		is not an \latin{ad hoc} definition.
		We merely choose a handy instance of quasi-polynomial
		to avoid being flooded with Big-$O$ notations.}
	\[Z(w)<\exp(-n^{2/3}). \label{eq:quasipoly}\]
	Define the \emph{$μ^*$-exponent} of $T$:
	\[μ^*≔\sup_{W∈𝒟}\limsup_{n→∞}\frac{\log N_n}{-\log\(I(W)-ℙ(𝒜_n)\)}.\]
	This definition is not perfect
	because $I(W)-ℙ(𝒜_n)$ is not necessary positive.
	(We can always specify a code whose code rate exceeds the Shannon capacity.)
	Of course we know that $I(W)-ℙ(𝒜_n)≤0$, or even $I(W)-ℙ(𝒜_n)≤O(N_n^{-.49})$,
	is too good to be true.
	So we alter the definition a little bit
	\[μ^*≔\sup_{\!W∈𝒟\!}\limsup_{n→∞}
		\frac{-\log N_n}{\log\max\(I(W)-ℙ(𝒜_n),N_n^{-1/2}\)}\]
	so that $μ^*$ is at least $2$.
	
	We will make use of the definition of $μ^*$ in the following manner.
	\begin{lem} \label{lem:remainder}
		Assume $E_0^{m-√n}$ is
		an arbitrary subset of disjoint vertices in $𝒯\pf(W,T,m)$.
		Let $A_m$ be the set of leaves $w$
		that satisfy $Z(w)<\exp(-m^{2/3})$ but have no ancestor in $E_0^{m-√n}$.
		Then
		\[I(W)-ℙ(E_0^{m-√n}∪A_m)≤N_m^{-1/μ^*+o(1)}.\]
	\end{lem}
	\begin{IEEEproof}
		Every leaf in $𝒜_m-A_m$ has some ancestor in $E_0^{m-√n}$,
		so $ℙ(𝒜_m-A_m)≤ℙ(E_0^{m-√n})$.
		This implies $ℙ(𝒜_m)≤ℙ(E_0^{m-√n}∪A_m)$ and 
		\begin{align}
			I(W)-ℙ(E_0^{m-√n}∪A_m) &≤ \\
			I(W)-ℙ(𝒜_m) &≤ N_m^{-1/μ^*+o(1)}.
		\end{align}
		The last inequality is
		a simple consequence of $\limsup$ in the definition of $μ^*$.
	\end{IEEEproof}
	
	In general, we have the following.
	
	\begin{lem} \label{lem:predicate}
		Assume $A_0^{m-√n}$ is
		an arbitrary subset of disjoint vertices in $𝒯\pf(W,T,m)$.
		Let $φ$ be a predicate of channels.
		Let $A_m$ be the set of leaves $w$
		that satisfy $φ$ but have no ancestor in $A_0^{m-√n}$.
		Then
		\[I(W)-ℙ(A_0^{m-√n}∪A_m)≤I(W)-ℙ\{φ(W_m)\}.\]
	\end{lem}
	\begin{IEEEproof}
		Same logic as Lemma~\ref{lem:remainder}.
		By the way, when this lemma is applied,
		$φ(w)$ will be $Z(w)<\exp\(-\exp(m^{1/3})\)$.
	\end{IEEEproof}

\subsection{The \Cramer{} Function}

	Assume $Y$ is a discrete, bounded random variable.
	Let $Y_1,Y_2,\dotsc$ be i.i.d.\ copies of $Y$.
	Define the \emph{\Cramer{} function} of $Y$:
	\[\CF(y)≔\sup_{λ<0}λy-\log𝔼[\exp(λY)].\]
	It is such that
	\[ℙ｛\frac{Y_1+Y_2+\dotsb+Y_n}n≤y｝≤\exp\(-n\CF(y)\). \label{eq:CFaxiom}\]


\section{The Recruit-Train-Retain Template}

	The recruit-train-retain template helps us
	understand the distribution of $Z_n$
	by first understanding the distribution of $Z_m$ for some $m<n$.
	
	An over-simplified template is as follows:
	\begin{rtr}
		\item[Recruit]
			Sometimes $Z_m$ is quite small.
			Calculate $ℙ\{Z_m\text{ is quite small}\}$.
		\item[Train]
			When $Z_i$ is quite small,
			there is a positive chance that $Z_{i+1}$ gets smaller.
			Repeat this for $i=m,m+1,\dotsc,n-1$;
			it is very unlikely not to get smaller at all.
		\item[Retain] 
			By syllogism, most of the $Z_n$ will be extremely small.
			Keep these extremely small $Z_n$,
			and freeze those $Z_n$ that are not extremely small enough.
	\end{rtr}
	In terms of Sankey diagram:
	
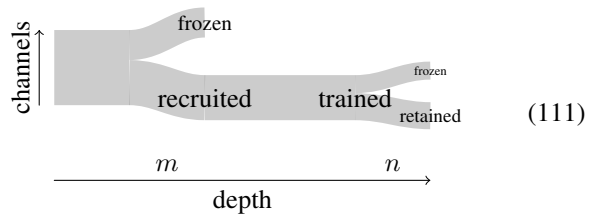
\[\tikz{
		\draw[->](0,-\sdwidth) +(1.5,0)node[^]{$m$}
		                       +(4.5,0)node[^]{$n$}
		                 +(0,0)--+(5,0)node[_,pos=.5]{depth};
		\draw[->](-.2,0)--+(0,\sdwidth)node[^,pos=.5,rotate=90]{channels};
		\tikzset{sd}
		\sddive(1);
		\sdcutby.4{
			\sddive(1)node{frozen};
		}{
			\sddive(1)node{recruited};
			\sddive(2)node{trained};
			\sdcutby.4{
				\sddive(1)node{frozen};
			}
			{
				\sddive(1)node{retained};
			}
		}
	}\]
	See Formula~(\ref{eq:sdT}) in Appendix~\ref{app:bigsankey}
	for the big diagram.
	
	This diagram records the fact that synthetic channels at depth $m$
	are classified into two groups based on their $Z$-parameters.
	The upper group consists of bad channels (large $Z_m$) and is frozen.
	The lower group consists of good channels (small $Z_m$) and
	is recruited and trained in the sense that we want to investigate their
	children $Z_{m+1}$, grandchildren $Z_{m+2}$, and all the way up to $Z_n$.
	Then these $Z_n$ are further classified into two groups.
	Those that are mediocrely small are frozen;
	those that are extremely small are retained --- they go to $𝒜_n$.
	
	We will see the template and Sankey diagrams multiple times.

\subsection{A Brief History}

	The first appearance dated back to \cite{AT09} with $m≔n^{3/4}$.
	The argument goes as follows:
	\begin{rtr}
		\item[Recruit]
			Control $ℙ\{Z_m<.875^m\}$; this is almost $I(W)$.
		\item[Train]
			Condition on the event $\{Z_m<.875^m\}$.
			For $i=m,m+1,\dotsc,n-1$,
			\[Z_{i+1}≈\begin{cases*}
				Z_i & with probability $1/2$ \\
				Z_i^2 & with probability $1/2$ \\
				\end{cases*}.\]
			That is, it gets squared with probability $1/2$.
		\item[Retain]
			By syllogism, conditioning on the event $\{Z_m<.875^m\}$,
			the quantity (how many times it is squared)
			$\log_2(\log Z_n/\log Z_m)$ is about $(n-m)/2$ with high probability.
	\end{rtr}
	That is to say, with probability $I(W)-o(1)$ it holds that
	\[\log_2(-\log Z_n)=(n-o(n))/2.\]
	Hence the $β'$-exponent
	\[\frac{\log(-\log P)}{\log N}≈\frac{\log_2(-\log Z_n)}{\log_22^n}
		⟶\frac12.\]
	The argument can be summarized by the Sankey diagram:
	
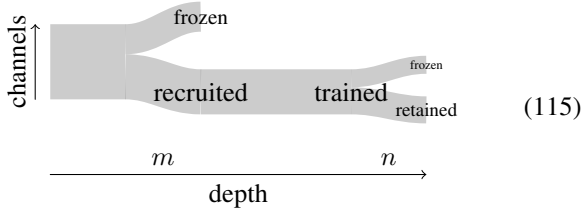
\[\tikz{
		\draw[->](0,-\sdwidth) +(1.5,0)node[^]{$m$}
		                       +(4.5,0)node[^]{$n$}
		                 +(0,0)--+(5,0)node[_,pos=.5]{depth};
		\draw[->](-.2,0)--+(0,\sdwidth)node[^,pos=.5,rotate=90]{channels};
		\tikzset{sd}
		\sddive(1);
		\sdcutby.4{
			\sddive(1)node{frozen};
		}{
			\sddive(1)node{recruited};
			\sddive(2)node{trained};
			\sdcutby.4{
				\sddive(1)node{frozen};
			}
			{
				\sddive(1)node{retained};
			}
		}
	}\]
	
	\cite{KSU10} claims to generalize the argument
	to handle cases like the following.
	\begin{rtr}
		\item[Recruit] Control $ℙ\{Z_m<.ρ^m\}$ for some magic choice of $ρ$;
			this it almost $I(W)$.
		\item[Train] Condition on the event $\{Z_m<ρ^m\}$.
			For $i=m,m+1,\dotsc,n-1$,
			\[Z_{i+1}≈\begin{cases*}
				Z_i^4 & with probability $1/2$ \\
				Z_i^5 & with probability $1/3$ \\
				Z_i^7 & with probability $1/6$
				\end{cases*}.\]
		\item[Retain] By syllogism, condition on the event $\{Z_m<ρ^m\}$,
			the quantity $\log_2(\log Z_n/\log Z_m)$
			is about, with high probability,
			\[(n-m)·（\frac12\log4+\frac13\log5+\frac16\log7）.\]
	\end{rtr}
	But part of the proof of \cite[Theorem~11]{KSU10}
	is omitted in the original paper.
	However, the idea is the same Sankey diagram:
	
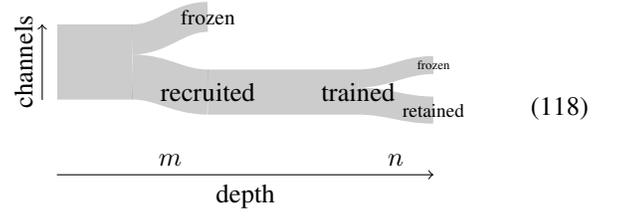
\[\tikz{
		\draw[->](0,-\sdwidth) +(1.5,0)node[^]{$m$}
		                       +(4.5,0)node[^]{$n$}
		                 +(0,0)--+(5,0)node[_,pos=.5]{depth};
		\draw[->](-.2,0)--+(0,\sdwidth)node[^,pos=.5,rotate=90]{channels};
		\tikzset{sd}
		\sddive(1);
		\sdcutby.4{
			\sddive(1)node{frozen};
		}{
			\sddive(1)node{recruited};
			\sddive(2)node{trained};
			\sdcutby.4{
				\sddive(1)node{frozen};
			}
			{
				\sddive(1)node{retained};
			}
		}
	}\]
	
	Another argument appears in \cite{MHU16}.
	\begin{rtr}
		\item[Recruit]
			Control $ℙ\{Z_m<.5^m\}$, where $m=γn$ for some fixed ratio $0<γ<1$.
		\item[Train]
			Condition on the event $\{Z_m<.5^m\}$.
			Track the process $Z_m,Z_{m+1},\dotsc,Z_n$.
		\item[Retain] Control $\log_2(\log Z_n/\log Z_m)$ and $\log(-\log Z_n)$.
	\end{rtr}
	The unchanging part of \cite{MHU16} is the Sankey diagram:
	
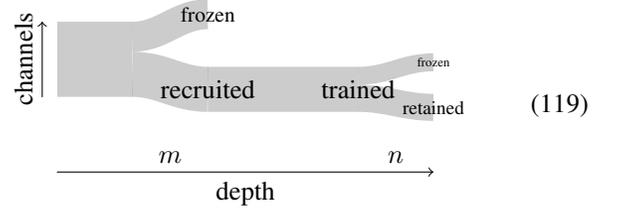
\[\tikz{
		\draw[->](0,-\sdwidth) +(1.5,0)node[^]{$m$}
		                       +(4.5,0)node[^]{$n$}
		                 +(0,0)--+(5,0)node[_,pos=.5]{depth};
		\draw[->](-.2,0)--+(0,\sdwidth)node[^,pos=.5,rotate=90]{channels};
		\tikzset{sd}
		\sddive(1);
		\sdcutby.4{
			\sddive(1)node{frozen};
		}{
			\sddive(1)node{recruited};
			\sddive(2)node{trained};
			\sdcutby.4{
				\sddive(1)node{frozen};
			}
			{
				\sddive(1)node{retained};
			}
		}
	}\]
	
	The innovative part of \cite{MHU16} is that $m$ is parameterized by $γ$.
	That is, they are free to choose $γ$ before spending
	$γn$ steps in the recruit phase and $(1-γ)n$ steps in the train phase.
	A rule of thumb is, a longer recruit phase makes $R$ better;
	and a longer train phase makes $P$ better.
	Thus they obtain a tradeoff between $R$ and $P$.
	The following plot shows pairs of $(β',1/μ')$ they achieve.
	
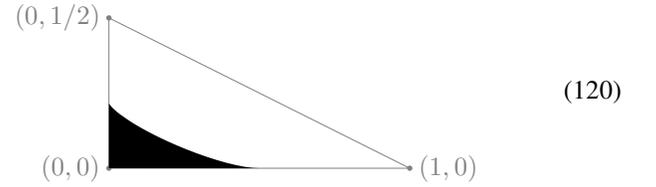
\[\tikz[TRi]{
		\draw[hl]
			(0,.5)pic{dot}node[<<]{$(0,1/2)$}--
			(0,0 )pic{dot}node[<<]{$(0,0)$}--
			(1,0 )pic{dot}node[>>]{$(1,0)$}--cycle;
		\fill(0,0)--plot[domain=.001:.5]({G(\x)*\x},{(1-G(\x))/mu});
	}\]

\subsection{Disposing Bad Synthetic Channels}

	Our contribution in the last work \cite{WD18} is
	what we now called the \emph{disposable recruit-train-retain template}.
	The idea is as follows.
	\begin{rtr}
		\item[Recruit]
			Control $ℙ\{Z_m<.5^m\}$ for $m=√n,2√n,\dotsc,n\rat$.
		\item[Train]
			Condition on the event
			$\{\text{$Z_m<.5^m$ but $Z_i≥.5^i$ for $i=0,\dotsc,m-√n$}\}$.
			Track the process $Z_m,Z_{m+1},\dotsc,Z_n$.
		\item[Retain] Control $\log_2(\log Z_n/\log Z_m)$ and $\log(-\log Z_n)$.
	\end{rtr}
	In terms of Sankey diagram:
	
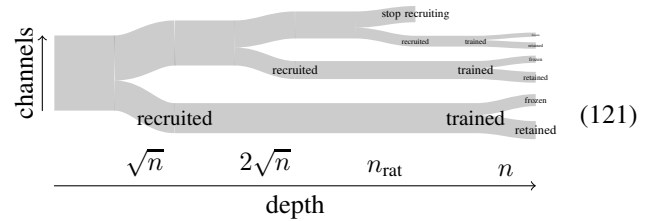
\[\tikz[xscale=.8]{
		\draw[->](0,-\sdwidth) +(1.5,0)node[^]{$√n$}
		                       +(3.5,0)node[^]{$2√n$}
		                       +(5.5,0)node[^]{$n\rat$}
		                       +(7.5,0)node[^]{$n$}
		                 +(0,0)--+(8,0)node[_,pos=.5]{depth};
		\draw[->](-.2,0)--+(0,\sdwidth)node[^,pos=.5,rotate=90]{channels};
		\tikzset{sd}
		\sddive(1);
		\sdcutby.6{
			\sddive(1);
			\sddive(1);
			\sdcutby.6{
				\sddive(1);
				\sddive(1);
				\sdcutby.6{
					\sddive(1)node{stop recruiting};
				}{
					\sddive(1)node{recruited};
					\sddive(1)node{trained};
					\sdcutby.4{
						\sddive(1)node{frozen};
					}
					{
						\sddive(1)node{retained};
					}
				}
			}{
				\sddive(1)node{recruited};
				\sddive(3)node{trained};
				\sdcutby.4{
					\sddive(1)node{frozen};
				}
				{
					\sddive(1)node{retained};
				}
			}
		}{
			\sddive(1)node{recruited};
			\sddive(5)node{trained};
			\sdcutby.4{
				\sddive(1)node{frozen};
			}
			{
				\sddive(1)node{retained};
			}
		}
	} \label{eq:disposable}\]
	See Formula~(\ref{eq:sdD1}) in Appendix~\ref{app:bigsankey}
	for the big diagram.
	
	This approach recruits $Z_m$ in several rounds to maximize the rate.
	Plus it trains $Z_i$ for almost the same depth as \cite{MHU16} does.
	Thus it should outperform the latter.
	Our final result, besides what \cite{MHU16} achieved before in gray,
	is the dark region below.
	
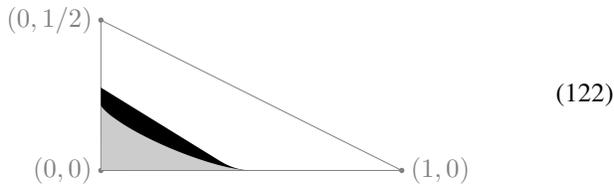
\[\tikz[TRi]{
		\draw[hl]
			(0,.5)pic{dot}node[<<]{$(0,1/2)$}--
			(0,0 )pic{dot}node[<<]{$(0,0)$}--
			(1,0 )pic{dot}node[>>]{$(1,0)$}--cycle;
		\fill(0,0)--plot file{AriRS2.txt};
		\fill[black!20](0,0)--plot[domain=.001:.5]({G(\x)*\x},{(1-G(\x))/mu});
		;
	}\]
	
	We personally believe
	this is the maximum region \Arikan's polar codes can achieve.
	We do not see any obvious way to improve our inner bound in \cite{WD18}.
	Nonetheless, there is a hope that 
	since polar codes generalize to other kernels,
	they might achieve a larger region.
	
	In fact,
	Theorem~\ref{thm:interior} shows any point inside the triangle is achievable.
	And Corollary~\ref{cor:hypotenuse} extends the conclusion to the hypotenuse.
	Both Theorem~\ref{thm:interior} and Corollary~\ref{cor:hypotenuse}
	rely on Theorem~\ref{thm:collaborate}, which heavily relies on this
	disposable recruit-train-retain template.

\subsection{Recycling Bad Synthetic Channels}

	In Formula~(\ref{eq:disposable}),
	a recruited synthetic channel is trained until depth $n$.
	As mentioned above, training this much reduces the error probability a lot.
	But it either requires very fine control on how good $Z_m$ are to begin with
	or we will have to freeze a lot of innocent $Z_n$
	(i.e., the $Z_n$ that we believe are good but are not able to prove),
	which hurts the rate.
	
	In case of \cite{WD18}, which considers only Arikan's polar codes,
	we do have fine control on $Z_m$ provided by \cite{FV14}.
	No innocent $Z_n$ is frozen.
	However a result like \cite{FV14} is missing for general polar codes.
	So we came up with a workaround.
	
	In the following version, synthetic channels will
	be trained for depth $√n$ and immediately be frozen or retained.
	And then the next round of recruit-train-retain starts.
	There will be $√n$ rounds in total.
	Therefore, even if some synthetic channel is frozen,
	there is a chance that it(s descendants)
	will be recruited in the upcoming rounds.
	
	This makes it a \emph{recyclable recruit-train-retain template}.
	\begin{rtr}
		\item[Recruit]
			Control $ℙ\{Z_m<.5^m\}$ for $m=√n,2√n,\dotsc,n-√n$.
		\item[Train]
			Condition on the event
			$\bigl\{\exp\(-\exp(n^{1/3})\)≤Z_m<\exp(-m^{2/3})\bigr\}$.
			Track the process $Z_m,Z_{m+1},\dotsc,Z_{m+√n}$.
		\item[Retain] Control $\log_2(\log Z_{m+√n}/\log Z_m)$
			and $\log(-\log Z_n)$.
	\end{rtr}
	In terms of Sankey diagram:
	
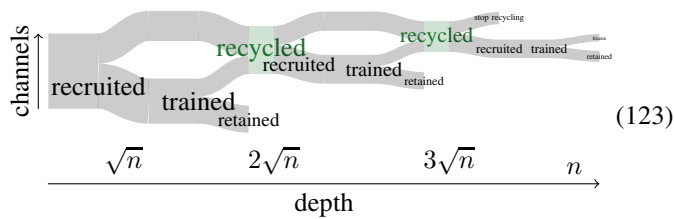
\[\tikz[xscale=.666]{
		\draw[->](1.5,-\sdwidth)node[^]{$√n$}
		         (4.5,-\sdwidth)node[^]{$2√n$}
		         (8.0,-\sdwidth)node[^]{$3√n$}
		        (10.5,-\sdwidth)node[^]{$n$}
		  (0,-\sdwidth)--+(11,0)node[_,pos=.5]{depth};
		\draw[->](-.2,0)--+(0,1)node[^,pos=.5,rotate=90]{channels};
		\tikzset{sd}
		\sddive(1);
		\sdcutby.4{
			\sddive(1);
			\sddive(1);
			\sdsave[a]
		}{
			\sddive(1)node[pos=0]{recruited};
			\sddive(1)node{trained};
			\sdcutby.4{
				\sdload[a](1);
				\sddive(.5)[recycol]node[pos=.5]{recycled};
				\sdcutby.4{
					\sddive(1);
					\sddive(1);
					\sdsave[b]
				}{
					\sddive(1)node[pos=.5]{recruited};
					\sddive(1)node{trained};
					\sdcutby.4{
						\sdload[b](1);
						\sddive(.5)[recycol]node[pos=.5]{recycled};
						\sdcutby.4{
							\sddive(1)node{stop recycling};
						}{
							\sddive(1)node{recruited};
							\sddive(1)node{trained};
							\sdcutby.4{
								\sddive(1)node{frozen};
							}{
								\sddive(1)node{retained};
							}
						}
					}{
						\sddive(1)node{retained};
					}
				}
			}{
				\sddive(1)node{retained};
			}
		}
		\pgf@picminx-5mm\global\pgf@picminx\pgf@pt@aa\pgf@picminx
		\pgf@picmaxx11cm\global\pgf@picmaxx\pgf@pt@aa\pgf@picmaxx
	}\]
	See Formula~(\ref{eq:sdR1}) in Appendix~\ref{app:bigsankey}
	for the big diagram.
	
	This approach does not minimize $Z_n$ to a satisfactory, finalized level.
	But it reduces $Z_n$ to somewhere
	that barely makes the disposable version efficient
	without having to worry about innocent $Z_n$.
	We will demonstrate this recyclable recruit-train-retain template
	in the proof of Lemma~\ref{lem:recyclable},
	which is the key to Theorems \ref{thm:disposable} and~\ref{thm:collaborate}.


\section{Main Results: To Interpolate \TP$β^*${β*} and \TP$μ^*${μ*}}

	We present three statements at once
	so readers immediately see the similarity.
	In fact, Theorem~\ref{thm:collaborate} can be proven by combining 
	the proofs of Lemma~\ref{lem:recyclable} and Theorem~\ref{thm:disposable}.
	
	\begin{lem} \label{lem:recyclable}
		Let $T$ be a length-$ℓ$, bounded transformation
		with $μ^*$-exponent $μ^*$ and $∂$-dice $Y$.
		If
		\[ℙ\{Y=0\}<ℓ^{-1/μ^*},\]
		then $T$ produces block codes $(𝒯_n,𝒜_n)$ such that
		\begin{gather}
			N_n=ℓ^n, \\
			R_n>I(W)-N_n^{-1/μ^*+o(1)},\rlap{ and} \\
			P_n<\exp\(-\exp(n^{1/3})\).
		\end{gather}
		For $n$ large enough.
	\end{lem}
	\begin{IEEEproof}
		First-time reader may skim Section~\ref{sec:recyclertr}.
		Second-time may skim Section~\ref{pf:recyclable}
		with white lies in mind:
		lie number one: $\Y_i=Y_i$;
		lie number two: $a_m$ and $b_m$ are about $ℓ^{-(m-√n)/μ^*}$;
		lie number three: $c_m/b_m$ and $d_m/b_m$ are about $ℓ^{√n/μ^*}$;
		lie number four: $g_m-b_m$ is about $ℓ^{m/μ^*}$;
		lie number five: $g_m$ is about $2^{m/√n}ℓ^{-m/μ^*}$.
		Third-time reader may realize that the whole proof,
		Section~\ref{pf:recyclable}, is an attempt to prove those lies
		but ends up barely proving the lemma by something weaker.
	\end{IEEEproof}
	
	\begin{thm} \label{thm:disposable}
		Let $T$ be a length-$ℓ$, bounded transformation
		with $μ^*$-exponent $μ^*$ and $∂$-dice $Y$.
		Let $\CF$ be the \Cramer{} function of $Y$.
		If
		\begin{gather}
			ℙ\{Y=0\}<ℓ^{-1/μ^*}
			\shortintertext{and, for $π∈[0,1]$,}
			\frac{(1-π)\logℓ}{μ'-πμ^*}<\CF（\frac{β'μ'\logℓ}{μ'-πμ^*}）,
		\end{gather}
		then $T$ produces block codes $(𝒯_n,𝒜_n)$ such that
		\begin{gather}
			N_n=ℓ^n, \\
			R_n>I(W)-N_n^{-1/μ'},\rlap{ and} \\
			P_n<\exp(-N_n^{β'})
		\end{gather}
		for $n$ large enough.
	\end{thm}
	\begin{IEEEproof}
		First-time reader may skim Section~\ref{sec:disposertr}.
		Se\-cond-time reader may skim Section~\ref{pf:disposable}
		with white lies in mind:
		lie number one: $\Y_i=Y_i$;
		lie number two: $a_m$ and $b_m$ are about $ℓ^{-(m-√n)/μ^*}$;
		lie number three: $c_m/b_m$ and $d_m/b_m$ are about $ℓ^{-n/μ'+m/μ^*}$;
		lie number four: $f_m$ is about $ℓ^{-m/μ^*}$;
		lie number five: $g_m-f_m$ is about $2mℓ^{-n/μ'+√n/μ^*}$;
		lie number six: $g_m$ is about $ℓ^{-m/μ^*}$.
		Third-time reader may realize that the whole proof,
		Section~\ref{pf:disposable}, is an attempt to prove those lies
		but ends up barely proving the theorem by something weaker.
	\end{IEEEproof}
	
	\begin{thm} \label{thm:collaborate}
		Let $T\rat$ be a $q$-ary, length-$ℓ$, bounded transformation
		with $μ^*$-exponent $μ^*\rat$ and $∂$-dice $Y\rat$.
		Let $T\fek$ be transforming $q$-ary channels to $q^k$-ary channels.
		Let $T\err$ be a $q^k$-ary, length-$ℓ$, bounded transformation
		with $∂$-dice $Y\err$.
		Let $\CF\err$ be the \Cramer{} function of $Y\err$.
		If
		\begin{gather}
			ℙ\{Y\rat=0\}<ℓ^{-1/μ\rat^*}
			\shortintertext{and, for $π∈[0,1]$,}
			\frac{(1-π)\logℓ}{μ'-πμ^*\rat}<
				\CF\err（\frac{β'μ'\logℓ}{μ'-πμ^*\rat}）,
		\end{gather}
		then $T\rat,T\fek,T\err$ produce block codes $(𝒯_n,𝒜_n)$ such that
		\begin{gather}
			N_n=kℓ^n, \\
			R_n>I(W)-N_n^{-1/μ'},\rlap{ and} \\
			P_n<\exp(-N_n^{β'})
		\end{gather}
		for $n$ large enough.
	\end{thm}
	\begin{IEEEproof}
		The first-time reader may believe the white lie that
		this is an easy implication of
		Lemma~\ref{lem:recyclable} and Theorem~\ref{thm:disposable}.
		The second-time reader may realize that it is not an easy implication,
		but Section~\ref{pf:collaborate} tries to explain it is an implication.
	\end{IEEEproof}
	Readers may notice that in Theorem~\ref{thm:collaborate},
	$(β',1/μ')$ is highly related to $μ^*\rat$ and $Y\err$
	instead of $μ^*\err$ or $Y\rat$.
	That is, the code rate is controled by $T\rat$ and
	the error probability is controled by $T\err$.
	This explains why and how we should mix two kernels.


\section{Prove Lemma~\ref{lem:recyclable}
	by Recyclable Recruit-Train-Retain Template} \label{pf:recyclable}

	Consider $𝒯\pf(W,T,n)$.
	We are going to choose a subset of leaf channels $𝒜_n$.

\subsection{First choose some constants}

	By Lemma~\ref{lem:bounded}, $Y≥0$.
	Start from
	\[\lim_{Υ→+∞}𝔼[Υ^{-Y}]=ℙ\{Y=0\}<ℓ^{-1/μ^*}.\]
	Pick a number $Υ\gg\exp(1)$ such that
	\[𝔼[Υ^{-Y}]<ℓ^{-1/μ^*}.\]
	Pick a number $ϵ>0$ such that
	\[𝔼[Υ^{-Y}]Υ^{2ϵ}<ℓ^{-1/μ^*}. \label{eq:atmostgap}\]
	Pick a smaller $ϵ>0$ and a number $δ>0$ such that
	\[(Z_i∧δ)^ϵ\text{ is a super-martingale}\]
	as in Lemma~\ref{lem:bounded}.
	Recall from the proof of Lemma~\ref{lem:bounded},
	\[\inf_{\substack{w∈𝒟\\Z(w)<δ}}
		\log\frac{\log Z\(X\text{-th component of }T(w)\)}{\log Z(w)}>Y-ϵ.\]
	Note that this is saying
	\[Z_{i-1}<δ\text{ implies }\Y_i>Y_i-ϵ.\]

\subsection{Second fill in the recyclable template} \label{sec:recyclertr}

	Let $E_0^0$ be the empty set.
	For $m=√n,2√n,\dotsc,n-√n$, define helically 
	$A_m$, $B_m$, $C_m$, $D_m$, $E_m$, $E_0^m$ as follows:
	\begin{rtr}
		\item[Recruit]
			Let $A_m$ be the set of synthetic channels $w$ at depth $m$
			that satisfy $Z(w)<\exp(-m^{2/3})$
			but have no ancestor in $E_0^{m-√n}$.
		\item[Train]
			Let $B_m$ be the set of synthetic channels at depth $m+√n$
			that are descendants of synthetic channels in $A_m$.
		\item[Retain]
			Let $C_m$ be the set of synthetic channels $w$ in $B_m$
			such that $Z(v)≥δ$
			for some ancestor $v$ of $w$ at depth $m,m+1,\dotsc,m+√n$.
			Let $D_m$ be the set of synthetic channels $w$ in $B_m-C_m$
			such that
			\[\frac{y_{m+1}+y_{m+2}+\dotsb+y_{m+√n}}{√n}≤2ϵ \label{eq:y2e}\]
			where $y_{m+i}$ are the values that $Y_{m+i}$ take
			when $W_{m+√n}=w$ happens.
			Let $E_m$ be $B_m-C_m-D_m$.
			Let $E_0^m$ be $E_0^{m-√n}∪E_m$.
	\end{rtr}
	In terms of Sankey diagram:
	
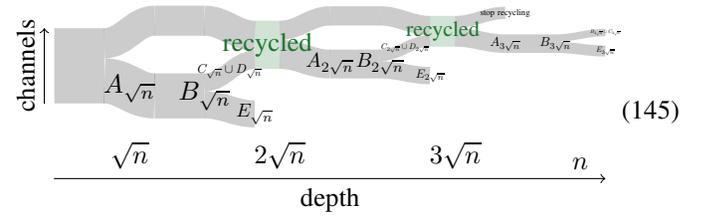
\[\tikz[xscale=.666]{
		\draw[->](1.5,-\sdwidth)node[^]{$√n$}
		         (4.5,-\sdwidth)node[^]{$2√n$}
		         (8.0,-\sdwidth)node[^]{$3√n$}
		        (10.5,-\sdwidth)node[^]{$n$}
		  (0,-\sdwidth)--+(11,0)node[_,pos=.5]{depth};
		\draw[->](-.2,0)--+(0,1)node[^,pos=.5,rotate=90]{channels};
		\tikzset{sd}
		\sddive(1);
		\sdcutby.4{
			\sddive(1);
			\sddive(1);
			\sdsave[a]
		}{
			\sddive(1)node[pos=.5]{$A_{√n}$};
			\sddive(1)node{$B_{√n}$};
			\sdcutby.4{
				\sdload[a](1)node[pos=.5]{$C_{√n}∪D_{√n}$};
				\sddive(.5)[recycol]node[pos=.5](recycled){};
				\onforeground\node[recycol]at(recycled){recycled};
				\sdcutby.4{
					\sddive(1);
					\sddive(1);
					\sdsave[b]
				}{
					\sddive(1)node{$A_{2√n}$};
					\sddive(1)node{$B_{2√n}$};
					\sdcutby.4{
						\sdload[b](1)node[pos=.5]{$C_{2√n}∪D_{2√n}$};
						\sddive(.5)[recycol]node[pos=.5](recycled){};
						\onforeground\node[recycol]at(recycled){recycled};
						\sdcutby.4{
							\sddive(1)node{stop recycling};
						}{
							\sddive(1)node{$A_{3√n}$};
							\sddive(1)node{$B_{3√n}$};
							\sdcutby.4{
								\sddive(1)node{$B_{3√n}∪C_{3√n}$};
							}{
								\sddive(1)node{$E_{3√n}$};
							}
						}
					}{
						\sddive(1)node{$E_{2√n}$};
					}
				}
			}{
				\sddive(1)node{$E_{√n}$};
			}
		}
		\pgf@picminx-5mm\global\pgf@picminx\pgf@pt@aa\pgf@picminx
		\pgf@picmaxx11cm\global\pgf@picmaxx\pgf@pt@aa\pgf@picmaxx
	} \label{eq:sdr2}\]
	See Formula~(\ref{eq:sdR2}) in Appendix~\ref{app:bigsankey}
	for the big diagram.
	
	Let $a_m$, $b_m$, $c_m$, $d_m$, $e_m$, $e_0^m$
	be the probability measures of the corresponding capital-letter events.
	Let $g_m$ be $I(W)-e_0^m$.
	
		Readers are encouraged to compare this subsection
		(\ref{sec:recyclertr}) with Section~\ref{sec:disposertr}
		and to figure out what in the template makes
		Formula~(\ref{eq:sdr2}) different from Formula~(\ref{eq:sdd2}).
		More generally, all subsections in this section (\ref{pf:recyclable})
		are parallel to those in Section~\ref{pf:disposable}.

\subsection{Third estimate \TP$c_m/b_m${cm/bm}}

	It is not hard to see from the definitions that $C_m$ is a subset of $B_m$,
	so the target quantity
	\[\frac{c_m}{b_m}=\frac{ℙ(C_m)}{ℙ(B_m)}=ℙ(C_m|B_m)\]
	is a conditional probability.
	It is also not hard to see that $B_m$ and $A_m$ refer to the same event,
	so
	\[\frac{c_m}{b_m}=ℙ(C_m|B_m)=ℙ(C_m|A_m).\]
	
	The event defined by $C_m$ is equal to
	\[\{Z_{m+i}≥δ\text{ for some }0≤i≤√n\}.\]
	Let $σ$ be the stopping time
	\[\min\(\{0≤s≤√n|Z_{m+s}≥δ\}∪\{√n\}\).\]
	That is, the first index that makes up the inequality, or the largest index.
	Then $C_m$ is also equal to
	\[\{Z_{m+σ}≥δ\}=\{(Z_{m+σ}∧δ)^ϵ≥δ^ϵ\}.\]
	By how $δ,ϵ$ are chosen,
	$(Z_{m+i}∧δ)^ϵ$ for $i=0,1,\dotsc,√n$ is a super-martingale.
	Thus there is a Doob's inequality-flavor bound
	(c.f.\ \cite[Theorem~5.4.1]{Durrett10})
	\begin{align}
		\frac{c_m}{b_m}
		&= ℙ(C_m|A_m) \\
		&≤ 𝔼［(Z_{m+σ}∧δ)^ϵ｜A_m］δ^{-ϵ} \\
		&≤ 𝔼［(Z_m∧δ)^ϵ｜A_m］δ^{-ϵ}
	\end{align}
	
	On the other hand, the event defined by $A_m$ is equal to
	\[\bigl\{Z_m<\exp(-m^{2/3})\bigr\}.\]
	Thus
	\begin{align}
		\frac{c_m}{b_m}
		&≤ 𝔼［(Z_m∧δ)^ϵ｜A_m］δ^{-ϵ} \\
		&< \exp(-m^{2/3})^ϵδ^{-ϵ} \\
		&= \exp(-m^{2/3}ϵ-ϵ\logδ).
	\end{align}
	And this is an upper bound on $c_m/b_m$.

\subsection{Forth estimate \TP$d_m/b_m${dm/bm}}

	Since we are in $𝒯(W,T,n)$,
	the $∂$-dices $Y_{m+i}$ are independent of the event $A_m$.
	As a consequence, the condition imposed by Formula~(\ref{eq:y2e})
	is independent of $A_m$, which we know from the previous subsection
	refers to the same event as $B_m$ does.
	Thus $d_m/b_m=ℙ(D_m)/ℙ(A_m)$ is at most the probability that
	\[\frac{Y_{m+1}+Y_{m+2}+\dotsb+Y_{m+√n}}{√n}≤2ϵ.\]
	To bound the probability measure of this event,
	it suffices to bound the probability measure of the event
	\[\{Y_{m+1}+Y_{m+2}+\dotsb+Y_{m+√n}≤2ϵ√n\}.\]
	This is equivalent to the probability measure of
	\[\{Υ^{-Y_{m+1}-Y_{m+2}-\dotsb-Y_{m+√n}}≥Υ^{-2ϵ√n}\}.\]
	By the Chernoff bound, it is less than
	\begin{align}
		&{} 𝔼[Υ^{-Y_{m+1}-Y_{m+2}-\dotsb-Y_{m+√n}}]Υ^{2ϵ√n} \\
		&= 𝔼[Υ^{-Y}]^{√n}Υ^{2ϵ√n} \\
		&= \(𝔼[Υ^{-Y}]Υ^{2ϵ}\)^{√n} \\
		&< (ℓ^{-1/μ^*})^{√n} \\
		&= ℓ^{-√n/μ^*}
	\end{align}
	where the last inequality is by Formula~(\ref{eq:atmostgap}).
	And this is an upper bound on $d_m/b_m$.

\subsection{Fifth estimate \TP$e_0^{n-√n}${e0n-√n}}

	Notice that $E_m$ is a subset of $B_m$, so
	\[0≤\frac{e_m}{b_m}≤1.\]
	Notice also that
	\begin{align}
		g_{m-√n}-b_m
		&= I(W)-e_0^{m-√n}-a_m \\
		&= I(W)-ℙ(E_0^{m-√n}∪A_m) \\
		&≤ N_m^{-1/μ^*+o(1)}
	\end{align}
	where the last inequality is by Lemma~\ref{lem:remainder}.
	So
	\[(g_{m-√n}-b_m)^+≤N_m^{-1/μ^*+o(1)}=ℓ^{-m/(μ^*+o(1))}.\]
	Here $(g_{m-√n}-b_m)^+$ is $\max(g_{m-√n}-b_m,0)$.
	Similarly let $g_m^+$ be $\max(g_m,0)$.
	
	Now we calculate $g_m$
	\begin{align}
		&= g_{m-√n}-e_m \\
		&= g_{m-√n}（1-\frac{e_m}{b_m}）+(g_{m-√n}-b_m)\frac{e_m}{b_m} \\
		&≤ g_{m-√n}^+（1-\frac{e_m}{b_m}）+(g_{m-√n}-b_m)^+\frac{e_m}{b_m} \\
		&≤ g_{m-√n}^+（1-\frac{e_m}{b_m}）+(g_{m-√n}-b_m)^+ \\
		&= g_{m-√n}^+（\frac{c_m}{b_m}+\frac{d_m}{b_m}）+(g_{m-√n}-b_m)^+ \\
		&≤ g_{m-√n}^+\(ℓ^{-√n/μ^*}+\exp(-m^{2/3}ϵ-ϵ\logδ)\)\\
		&{\mkern183mu} +ℓ^{-m/(μ^*+o(1))}.
	\end{align}
	
	Starting from $m≥O(n^{3/4})$
	the term $ℓ^{-√n/μ^*}$ dominates the term $\exp(-m^{2/3}ϵ-ϵ\logδ)$.
	Thus it suffices to solve the recurrence relation
	\[\begin{cases}
		g_{O(n^{3/4})}≤1; \\
		g_m≤2g_{m-√n}^+ℓ^{-√n/μ^*}+ℓ^{-m/(μ^*+o(1))}.
	\end{cases}\]
	The result is
	\[g_{n-√n}≤ℓ^{-n/(μ^*+o(1))}=N_n^{-1/μ^*+o(1)}.\]
	By algebra
	\[e_0^{n-√n}=I(W)-g_{n-√n}≥I(W)-N_n^{-1/μ^*+o(1)}.\]

\subsection{Sixth estimate how good
	synthetic channels in \TP$E_0^{n-√n}${E0n-√n} are}

	They are synthetic channels such that during the time they are being trained,
	$Z(W_{m+i})$ is never larger than $δ$,
	so $\Y_{m+i}>Y_{m+i}-ϵ$.
	They are also synthetic channels such that
	\[\frac{Y_{m+1}+Y_{m+2}+\dotsb+Y_{m+√n}}{√n}>2ϵ\]
	so
	\[\Y_{m+1}+\Y_{m+2}+\dotsb+\Y_{m+√n}>ϵ√n.\]
	Therefore for every $w∈E_m$ and $v$ its ancestor at depth $m$,
	by telescoping
	\[\log(-\log Z(w))-\log(-\log Z(v))>ϵ√n.\]
	But $v∈A_m$ are such that $Z(v)≤\exp(-m^{2/3})$,
	so
	\[Z(w)<\exp\(-\exp(ϵ√n)m^{2/3}\).\]
	Sum over $E_0^{n-√n}$:
	\[∑_{w∈E_0^{n-√n}}Z(w)<N_n\exp\(-\exp(ϵ√n)m^{2/3}\).\]
	Let $𝒜_n$ be the set of synthetic channels at depth $n$
	that are descendants of synthetic channels in $E_0^{n-√n}$.
	Then the inequality lifts
	\[∑_{w∈𝒜_n}Z(w)<\absT^nN_n\exp\(-\exp(ϵ√n)m^{2/3}\).\]
	Eventually, as $n→∞$, replacing $√n$ by $n^{1/3}$ eats up other minor terms:
	\[∑_{w∈𝒜_n}Z(w)<\exp\(-\exp(n^{1/3})\).\]

\subsection{Seventh we announce the code}

	\[\(𝒯\pf(W,T,n),𝒜_n\)\]
	has block length
	\[N_n=ℓ^n,\]
	code rate
	\[R_n=ℙ(𝒜_n)=e_0^{n-√n}≥I(W)-N_n^{1/μ^*+o(1)},\]
	and error probability
	\[P_n=∑_{w∈𝒜_n}Z(w)<\exp\(-\exp(n^{1/3})\).\]
	This finishes the proof of Lemma~\ref{lem:recyclable}.


\section{Prove Theorem~\ref{thm:disposable}
	by Disposable Recruit-Train-Retain Template} \label{pf:disposable}

	Consider $𝒯\pf(W,T,n)$.
	We are going to choose a subset of leaf channels $𝒜_n$.

\subsection{First choose some constants}

	Pick a number $ϵ>0$ such that, for all $π∈[0,1]$,
	\[\frac{(1-π)\logℓ}{μ'-πμ^*}
		<\CF（\frac{β'μ'\logℓ}{μ'-πμ^*}+ϵ）. \label{eq:lessthanCF}\]
	Pick a smaller $ϵ>0$ such that
	if all $μ'$ are replaced by $μ'-ϵ$ in this ineqaulity,
	then it still holds for all $π∈[0,1]$.
	Pick a smaller $ϵ>0$ and a number $δ>0$ such that
	\[(Z_i∧δ)^ϵ\text{ is a super-martingale}\]
	as in Lemma~\ref{lem:bounded}.
	Recall from the proof of Lemma~\ref{lem:bounded},
	\[\inf_{\substack{w∈𝒟\\Z(w)<δ}}
		\log\frac{\log Z\(X\text{-th component of }T(w)\)}{\log Z(w)}>Y-ϵ.\]
	Note that this is saying
	\[Z_{i-1}<δ\text{ implies }\Y_i>Y_i-ϵ.\]

\subsection{Second fill in the disposable template} \label{sec:disposertr}

	Let $n\rat$ be $nμ^*/μ'$.
	Let both $A_0^0$ and $E_0^0$ be the empty set.
	For $m=√n,2√n,\dotsc,n\rat$, define helically 
	$A_m$, $A_0^m$, $B_m$, $C_m$, $D_m$, $E_m$, $E_0^m$ as follows:
	\begin{rtr}
		\item[Recruit]
			Let $A_m$ be the set of synthetic channels $w$ at depth $m$ that
			satisfy $Z(w)<\exp\(-\exp(m^{1/3})\)$
			but have no ancestor in $A_0^{m-√n}$.
			Let $A_0^m$ be $A_0^{m-√n}∪A_m$.
		\item[Train]
			Let $B_m$ be the set of synthetic channels at depth $n$
			that are descendants of synthetic channels in $A_m$.
		\item[Retain]
			Let $C_m$ be the set of synthetic channels $w$ in $B_m$ such that
			$Z(v)≥δ$ for some ancestor $v$ of $w$ at depth $m,m+1,\dotsc,n$.
			Let $D_m$ be the set of synthetic channels $w$ in $B_m-C_m$
			such that
			\[\frac{y_{m+1}+y_{m+2}+\dotsb+y_n}{n-m}≤\frac{β'\logℓ}{1-m/n}+ϵ.
				\label{eq:yfrace}\]
			where $y_{m+i}$ are the values that $Y_{m+i}$ take
			when $W_n=w$ happens.
			Let $E_m$ be $B_m-C_m-D_m$.
			Let $E_0^m$ be $E_0^{m-√n}∪E_m$.
	\end{rtr}
	In terms of Sankey diagram:
	
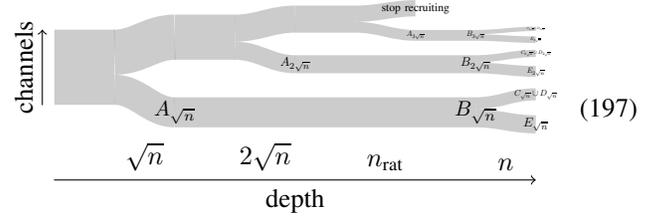
\[\tikz[xscale=.8]{
		\draw[->](0,-\sdwidth) +(1.5,0)node[^]{$√n$}
		                       +(3.5,0)node[^]{$2√n$}
		                       +(5.5,0)node[^]{$n\rat$}
		                       +(7.5,0)node[^]{$n$}
		                 +(0,0)--+(8,0)node[_,pos=.5]{depth};
		\draw[->](-.2,0)--+(0,\sdwidth)node[^,pos=.5,rotate=90]{channels};
		\tikzset{sd}
		\sddive(1);
		\sdcutby.6{
			\sddive(1);
			\sddive(1);
			\sdcutby.6{
				\sddive(1);
				\sddive(1);
				\sdcutby.6{
					\sddive(1)node{stop recruiting};
				}{
					\sddive(1)node{$A_{3√n}$};
					\sddive(1)node{$B_{3√n}$};
					\sdcutby.4{
						\sddive(1)node{$C_{3√n}∪D_{3√n}$};
					}
					{
						\sddive(1)node{$E_{3√n}$};
					}
				}
			}{
				\sddive(1)node{$A_{2√n}$};
				\sddive(3)node{$B_{2√n}$};
				\sdcutby.4{
					\sddive(1)node{$C_{2√n}∪D_{2√n}$};
				}
				{
					\sddive(1)node{$E_{2√n}$};
				}
			}
		}{
			\sddive(1)node{$A_{√n}$};
			\sddive(5)node{$B_{√n}$};
			\sdcutby.4{
				\sddive(1)node{$C_{√n}∪D_{√n}$};
			}
			{
				\sddive(1)node{$E_{√n}$};
			}
		}
	} \label{eq:sdd2}\]
	See Formula~(\ref{eq:sdD2}) in Appendix~\ref{app:bigsankey}
	for the big diagram.
	
	Let $a_m$, $b_m$, $c_m$, $d_m$, $e_m$, $e_0^m$
	be the probability measures of the corresponding capital-letter events.
	Let $f_m$ be $I(W)-a_0^m$.
	Let $g_m$ be $I(W)-e_0^m$.
	Let $π$ be $m/n\rat$.
	
		Readers are encouraged to compare this subsection
		(\ref{sec:disposertr}) with Section~\ref{sec:recyclertr}
		and to figure out what in the template makes
		Formula~(\ref{eq:sdd2}) different from Formula~(\ref{eq:sdr2}).
		More generally, all subsections in this section (\ref{pf:disposable})
		are parallel to those in Section~\ref{pf:recyclable}.

\subsection{Third estimate \TP$c_m/b_m${cm/bm}}

	It is not hard to see from the definitions that $C_m$ is a subset of $B_m$,
	so the target quantity
	\[\frac{c_m}{b_m}=\frac{ℙ(C_m)}{ℙ(B_m)}=ℙ(C_m|B_m)\]
	is a conditional probability.
	It is also not hard to see that $B_m$ and $A_m$ refer to the same event,
	so
	\[\frac{c_m}{b_m}=ℙ(C_m|B_m)=ℙ(C_m|A_m).\]
	
	The event defined by $C_m$ is equal to
	\[\{Z_{m+i}≥δ\text{ for some $0≤i≤n-m$}\}.\]
	Let $σ$ be the stopping time
	\[\min\(\{0≤s≤n-m|Z_{m+s}≥δ\}∪\{n\}\).\]
	That is, the first index that makes up the inequality, or the largest index.
	Then $C_m$ is also equal to
	\[\{Z_{m+σ}≥δ\}=\{(Z_{m+σ}∧δ)^ϵ≥δ^ϵ\}\]
	By how $δ,ϵ$ are chosen,
	$(Z_{m+i}∧δ)^ϵ$ for $i=0,1,\dotsc,n-m$ is a super-martingale.
	Thus there is a Doob's inequality-flavor bound
	(c.f.\ \cite[Theorem~5.4.1]{Durrett10})
	\begin{align}
		\frac{c_m}{b_m}
		&= ℙ(C_m|A_m) \\
		&≤ 𝔼［(Z_{m+σ}∧δ)^ϵ｜A_m］δ^{-ϵ} \\
		&≤ 𝔼［(Z_m∧δ)^ϵ｜A_m］δ^{-ϵ}
	\end{align}
	
	On the other hand, the event defined by $A_m$ is equal to
	\[\bigl\{Z_m<\exp\(-\exp(m^{1/3})\)\bigr\}.\]
	Thus
	\begin{align}
		\frac{c_m}{b_m}
		&≤ 𝔼［(Z_m∧δ)^ϵ｜A_m］δ^{-ϵ} \\
		&< \exp\(-\exp(m^{1/3})\)^ϵδ^{-ϵ} \\
		&= \exp\(-\exp(m^{1/3})ϵ-ϵ\logδ\).
	\end{align}
	And this is an upper bound on $c_m/b_m$.

\subsection{Forth estimate \TP$d_m/b_m${dm/bm}}

	Since we are in $𝒯(W,T,n)$,
	the $∂$-dices $Y_{m+i}$ are independent of the event $A_m$.
	As a consequence, the condition imposed by Formula~(\ref{eq:yfrace})
	is independent of $A_m$, which we know from the previous subsection
	refers to the same event as $B_m$ does.
	Thus $d_m/b_m=ℙ(D_m)/ℙ(A_m)$ is at most probability that
	\[\frac{Y_{m+1}+Y_{m+2}+\dotsb+Y_n}{n-m}≤\frac{β'\logℓ}{1-m/n}+ϵ.\]
	Here $m/n=πn\rat/n=πμ^*/μ'$,
	so the right hand side of the inequality is
	\[\frac{β'\logℓ}{1-πμ^*/μ'}+ϵ=\frac{β'μ'\logℓ}{μ'-πμ^*}+ϵ.\]
	By Formula~\ref{eq:CFaxiom}, the probability that
	\[\frac{Y_{m+1}+Y_{m+2}+\dotsb+Y_n}{n-m}≤\frac{β'μ'\logℓ}{μ'-πμ^*}+ϵ\]
	is bounded from above by
	\[\exp（-(n-m)·\CF（\frac{β'μ'\logℓ}{μ'-πμ^*}+ϵ））\]
	And Formula~(\ref{eq:lessthanCF}) helps bound this from above by
	\[\exp（-(n-m)\frac{(1-π)\logℓ}{μ'-πμ^*}）,\]
	where the argument of $\exp$ is
	\[-(n-m)\frac{(1-π)\logℓ}{μ'-πμ^*}=-（\frac n{μ'}-\frac m{μ^*}）\logℓ.\]
	Put $\exp$ back; it becomes
	\[ℓ^{-n/μ'+m/μ^*}.\]
	And this is an upper bound on $d_m/b_m$.

\subsection{Fifth estimate \TP$e_0^{n\rat}${e0nrat}}

	Notice that $E_m$ is a subset of $B_m$, so
	\[0≤\frac{e_m}{b_m}≤1.\]
	Notice also that
	\begin{align}
		f_m &= I(W)-a_0^{m-√n}-a_m \\
		&= I(W)-ℙ(A_0^{m-√n}∪A_m) \\
		&≤ N_m^{-1/μ^*+o(1)}
	\end{align}
	where the last inequality is by
	Lemma \ref{lem:recyclable} and~\ref{lem:predicate}.
	So
	\[f_m^+≤N_m^{-1/μ^*+o(1)}=ℓ^{-m/(μ^*+o(1))}.\]
	Here $f_m^+$ is $\max(f_m,0)$.
	Similarly let $(g_{m-√n}-b_m)^+$ be $\max(g_{m-√n}-b_m,0)$.
	
	Now we calculate $g_m-f_m^+$
	\begin{align}
		&= g_{m-√n}-e_m-(f_{m-√n}-b_m)^+ \\
		&≤ g_{m-√n}-e_m-(f_{m-√n}-b_m)^+\frac{e_m}{b_m} \\
		&≤ g_{m-√n}-e_m-(f_{m-√n}^+-b_m)\frac{e_m}{b_m} \\
		&= g_{m-√n}-f_{m-√n}^++f_{m-√n}^+（1-\frac{e_m}{b_m}） \\
		&= g_{m-√n}-f_{m-√n}^++f_{m-√n}^+（\frac{c_m}{b_m}+\frac{d_m}{b_m}） \\
		&≤ g_{m-√n}-f_{m-√n}^++ℓ^{-(m-√n)/μ^*}× \\
		&\quad （\exp\(-\exp(m^{1/3})ϵ-ϵ\logδ\)+ℓ^{-n/μ'+m/μ^*}）
	\end{align}
	
	In the last line, the term $ℓ^{-n/μ'+m/μ^*}$ dominates
	the other doubly-exponential term as $n→∞$.
	Thus it suffices to solve the recurrence relation
	\[\begin{cases}
		g_0-f_0^+=0; \\
		g_m-f_m^+≤g_{m-√n}-f_{m-√n}^++2ℓ^{-n/μ'+√n/μ^*}.
	\end{cases}\]
	The result is
	\[g_{n\rat}-f_{n\rat}^+≤ℓ^{-n/(μ'+o(1))}.\]
	In other words
	\[g_{n\rat}≤f_{n\rat}^++ℓ^{-n/(μ'+o(1))}=ℓ^{-n/(μ'+o(1))}.\]
	Since right after Formula~(\ref{eq:lessthanCF}) we replaced $μ'$ by $μ'-ϵ$,
	this $ϵ$ cancels $o(1)$ as $n→∞$.
	Hence we can really say that
	\[g_{n\rat}≤ℓ^{-n/μ'}=N_n^{-1/μ'}\]
	and that
	\[e_0^{n\rat}=I(W)-g_{n\rat}≥I(W)-N_n^{-1/μ'}.\]

\subsection{Sixth estimate how good
	synthetic channels in \TP$E_0^{n\rat}${E0nrat} are}

	They are synthetic channels such that during the time they are being trained,
	$Z(W_{m+i})$ is never larger than $δ$.
	Therefore $\Y_{m+i}>Y_{m+i}-ϵ$ holds.
	They are also synthetic channels such that
	\[\frac{Y_{m+1}+Y_{m+2}+\dotsb+Y_n}{n-m}>\frac{β'\logℓ}{1-m/n}+ϵ\]
	so
	\[\Y_{m+1}+\Y_{m+2}+\dotsb+\Y_n>β'n\logℓ.\]
	Therefore for every $w∈E_m$ and $v$ its ancestor at depth $m$,
	by telescoping
	\[\log(-\log Z(w))-\log(-\log Z(v))>β'n\logℓ.\]
	But $v$ are such that $Z(v)≤\exp\(-\exp(m^{1/3})\)$,
	so
	\[Z(w)<\exp\(-\exp(β'n\logℓ+m^{1/3})\).\]
	Sum over $E_0^{n\rat}$:
	\[∑_{w∈E_0^{n\rat}}Z(w)<N_n\exp\(-\exp(β'n\logℓ+m^{1/3})\).\]
	Let $𝒜_n$ be $E_0^{n\rat}$.
	Eventually, as $n→∞$,
	the term $m^{1/3}$ eats up the term $N_n$ in front of $\exp$:
	\[∑_{w∈𝒜_n}Z(w)<\exp\(-\exp(β'n\logℓ)\)=\exp(-N_n^{β'}).\]

\subsection{Seventh we announce the code}

	\[\(𝒯\pf(W,T,n),𝒜_n\)\]
	has block length
	\[N_n=ℓ^n,\]
	code rate
	\[R_n=ℙ(𝒜_n)=e_0^{n\rat}≥I(W)-N_n^{1/μ'},\]
	and error probability
	\[P_n=∑_{w∈𝒜_n}Z(w)<\exp(-N_n^{β'}).\]
	This finishes the proof of Theorem~\ref{thm:disposable}.


\section{Prove Theorem~\ref{thm:collaborate}
	by Combining Lemma~\ref{lem:recyclable}~and~Theorem~\ref{thm:disposable}}
	\label{pf:collaborate}

\subsection{First apply Lemma~\ref{lem:recyclable} to \TP$T\rat${Trat}}

	The conditions posed in Lemma~\ref{lem:recyclable}
	coincide with conditions posed on $T\rat$.
	Therefore $T\rat$ produces block codes such that
	\begin{gather}
		N_m=ℓ^m, \\
		R_m>I(W)-N_m^{-1/μ^*\rat+o(1)},\rlap{ and} \label{eq:goodrat}\\
		P_m<\exp\(-\exp(m^{1/3})\).
	\end{gather}

\subsection{Second grow a special channel tree}

	\[𝒯\pf(W,T\rat,n\rat,T\err,n).\]
	Here $n$ is a positive integer and $n\rat$ is $nμ^*/μ'$.
	\begin{rtr}
		\item[Stock] Begin with $𝒯\pf(W,T\rat,n)$.
		\item[Prune] Let $A_m$ and $A_0^m$ be defined
			as in Section~\ref{sec:disposertr}.
			For every synthetic channel in $A_0^{n\rat}$, detach its descendants.
		\item[Graft] To every leaf channel $w$ in the remaining channel tree,
			append $𝒯\pf(w^k,T\err,n-\depth(w))$.
			Here $w^k=T\fek(w)$ is the $k$-th power of $w$.
	\end{rtr}
	In terms of Sankey diagram:
	
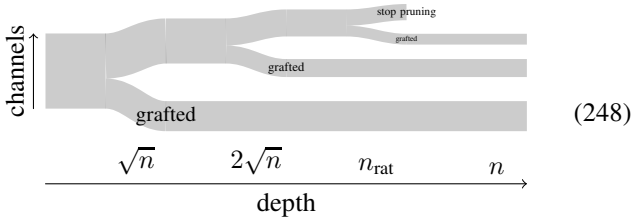
\[\tikz[xscale=.8]{
		\draw[->](0,-\sdwidth) +(1.5,0)node[^]{$√n$}
		                       +(3.5,0)node[^]{$2√n$}
		                       +(5.5,0)node[^]{$n\rat$}
		                       +(7.5,0)node[^]{$n$}
		                 +(0,0)--+(8,0)node[_,pos=.5]{depth};
		\draw[->](-.2,0)--+(0,\sdwidth)node[^,pos=.5,rotate=90]{channels};
		\tikzset{sd}
		\sddive(1);
		\sdcutby.6{
			\sddive(1);
			\sddive(1);
			\sdcutby.6{
				\sddive(1);
				\sddive(1);
				\sdcutby.6{
					\sddive(1)node{stop pruning};
				}{
					\sddive(1)node{grafted};
					\sddive(2);
				}
			}{
				\sddive(1)node{grafted};
				\sddive(4);
			}
		}{
			\sddive(1)node{grafted};
			\sddive(6);
		}
	}\]
	See Formula~(\ref{eq:sdG}) in Appendix~\ref{app:bigsankey}
	for the big diagram.
	
	Here is a small, but illustrative, example:
	Stock: choose $  T_\Ari$ to be $T\rat$ and
	prepare $𝒯\pf(W,T_\Ari,3)$ to begin with
	(Unlike Formula~(\ref{eq:tre8}), we omit labeling $T_\Ari$.)
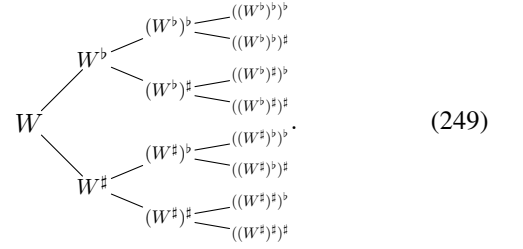
	\[\tikz[ut]{
		\vtnode$W$2{
			\vtnode$W^♭$2{
				\vtnode$(W^♭)^♭$2{
					\vtnode$((W^♭)^♭)^♭$1{}
				}{
					\vtnode$((W^♭)^♭)^♯$1{}
				}
			}{
				\vtnode$(W^♭)^♯$2{
					\vtnode$((W^♭)^♯)^♭$1{}
				}{
					\vtnode$((W^♭)^♯)^♯$1{}
				}
			}
		}{
			\vtnode$W^♯$2{
				\vtnode$(W^♯)^♭$2{
					\vtnode$((W^♯)^♭)^♭$1{}
				}{
					\vtnode$((W^♯)^♭)^♯$1{}
				}
			}{
				\vtnode$(W^♯)^♯$2{
					\vtnode$((W^♯)^♯)^♭$1{}
				}{
					\vtnode$((W^♯)^♯)^♯$1{}
				}
			}
		}
	}.\]
	Prune: if it happens that $A_0^m$ contains $W^♯,(W^♭)^♯,((W^♭)^♭)^♯$
	(highlighted in yellow background), remove their descendants.
	
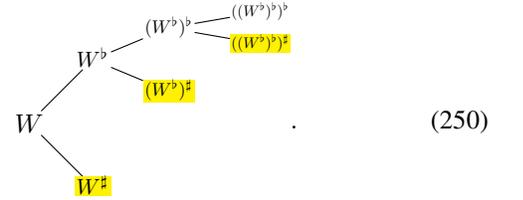
\[\tikz[ut]{
		\vtnode$W$2{
			\vtnode$W^♭$2{
				\vtnode$(W^♭)^♭$2{
					\vtnode$((W^♭)^♭)^♭$1{}
				}{
					\vtnode[h]$((W^♭)^♭)^♯$1{}
				}
			}{
				\vtnode[h]$(W^♭)^♯$1{}
			}
		}{
			\vtnode[h]$W^♯$1{}
		}
	}.\]
	Graft: let $k=1$ (so $T\fe1$ does nothing) and
	choose $T_\Ari$ again as $T\err$; attach three trees $𝒯\pf((W^♯)^1,T_\Ari,2)$
	and $𝒯\pf(((W^♭)^♯)^1,T_\Ari,1)$ and $𝒯\pf((((W^♭)^♭)^♯)^1,T_\Ari,0)$
	to the corresponding leaves.
	
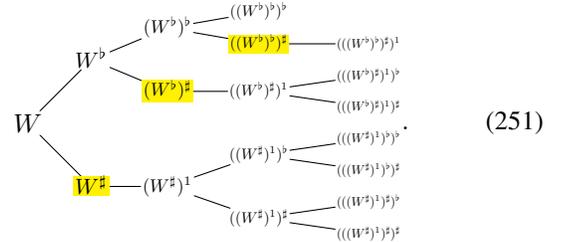
\[\tikz[ut]{
		\vtnode$W$2{
			\vtnode$W^♭$2{
				\vtnode$(W^♭)^♭$2{
					\vtnode$((W^♭)^♭)^♭$1{}
				}{
					\vtnode[h]$((W^♭)^♭)^♯$1{
						\vtnode$(((W^♭)^♭)^♯)^1$1{}
					}
				}
			}{
				\vtnode[h]$(W^♭)^♯$1{
					\vtnode$((W^♭)^♯)^1$2{
						\vtnode$(((W^♭)^♯)^1)^♭$1{}
					}{
						\vtnode$(((W^♭)^♯)^1)^♯$1{}
					}
				}
			}
		}{
			\vtnode[h]$W^♯$1{
				\vtnode$(W^♯)^1$2{
					\vtnode$((W^♯)^1)^♭$2{
						\vtnode$(((W^♯)^1)^♭)^♭$1{}
					}{
						\vtnode$(((W^♯)^1)^♭)^♯$1{}
					}
				}{
					\vtnode$((W^♯)^1)^♯$2{
						\vtnode$(((W^♯)^1)^♯)^♭$1{}
					}{
						\vtnode$(((W^♯)^1)^♯)^♯$1{}
					}
				}
			}
		}
	}.\]
	The depth of the attached subtrees are chosen
	such that the resulting tree is balanced.
	It is practically pointless, but legal and coherent,
	to have $T\fek$ at the bottom of a channel tree.
	
	Here is another example.
	This time $k=2$ so $T\rat$ and $T\err$ are of different arities.
	Stock
	\[\utupper2\utupper
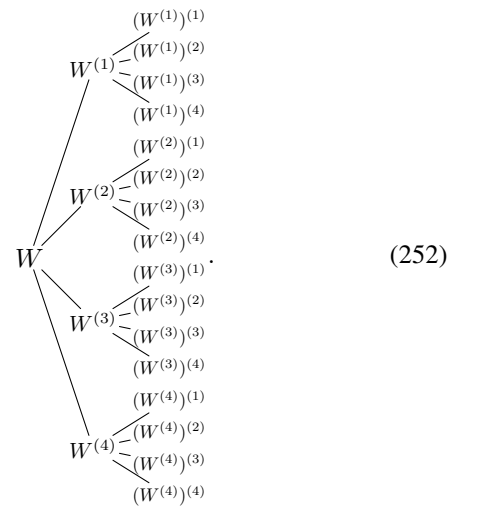
\tikz[ut]{
		\vtnode$W$4{
			\vtnode$W^{(1)}$4{
					\vtnode$(W^{(1)})^{(1)}$1{}
				}{
					\vtnode$(W^{(1)})^{(2)}$1{}
				}{
					\vtnode$(W^{(1)})^{(3)}$1{}
				}{
					\vtnode$(W^{(1)})^{(4)}$1{}
				}
		}{
			\vtnode$W^{(2)}$4{
					\vtnode$(W^{(2)})^{(1)}$1{}
				}{
					\vtnode$(W^{(2)})^{(2)}$1{}
				}{
					\vtnode$(W^{(2)})^{(3)}$1{}
				}{
					\vtnode$(W^{(2)})^{(4)}$1{}
				}
		}{
			\vtnode$W^{(3)}$4{
				\vtnode$(W^{(3)})^{(1)}$1{}
			}{
				\vtnode$(W^{(3)})^{(2)}$1{}
			}{
				\vtnode$(W^{(3)})^{(3)}$1{}
			}{
				\vtnode$(W^{(3)})^{(4)}$1{}
			}
		}{
			\vtnode$W^{(4)}$4{
				\vtnode$(W^{(4)})^{(1)}$1{}
			}{
				\vtnode$(W^{(4)})^{(2)}$1{}
			}{
				\vtnode$(W^{(4)})^{(3)}$1{}
			}{
				\vtnode$(W^{(4)})^{(4)}$1{}
			}
		}
	}.\]
	Prune
	\[\utupper2\utupper
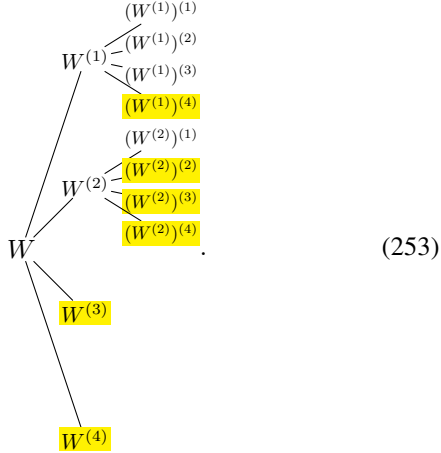
\tikz[ut]{
		\vtnode$W$4{
			\vtnode$W^{(1)}$4{
					\vtnode$(W^{(1)})^{(1)}$1{}
				}{
					\vtnode$(W^{(1)})^{(2)}$1{}
				}{
					\vtnode$(W^{(1)})^{(3)}$1{}
				}{
					\vtnode[h]$(W^{(1)})^{(4)}$1{}
				}
		}{
			\vtnode$W^{(2)}$4{
					\vtnode$(W^{(2)})^{(1)}$1{}
				}{
					\vtnode[h]$(W^{(2)})^{(2)}$1{}
				}{
					\vtnode[h]$(W^{(2)})^{(3)}$1{}
				}{
					\vtnode[h]$(W^{(2)})^{(4)}$1{}
				}
		}{
			\vtnode[h]$W^{(3)}$1{}
		}{
			\vtnode[h]$W^{(4)}$1{}
		}
	}.\]
	Graft
	\[\utupper2\utupper
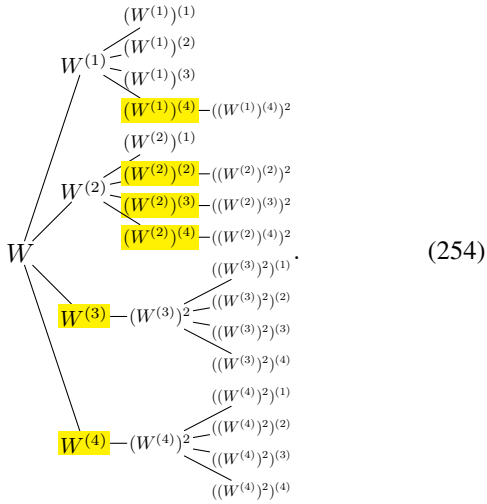
\tikz[ut]{
		\vtnode$W$4{
			\vtnode$W^{(1)}$4{
					\vtnode$(W^{(1)})^{(1)}$1{}
				}{
					\vtnode$(W^{(1)})^{(2)}$1{}
				}{
					\vtnode$(W^{(1)})^{(3)}$1{}
				}{
					\vtnode[h]$(W^{(1)})^{(4)}$1{
						\vtnode$((W^{(1)})^{(4)})^2$1{}
					}
				}
		}{
			\vtnode$W^{(2)}$4{
					\vtnode$(W^{(2)})^{(1)}$1{}
				}{
					\vtnode[h]$(W^{(2)})^{(2)}$1{
						\vtnode$((W^{(2)})^{(2)})^2$1{}
					}
				}{
					\vtnode[h]$(W^{(2)})^{(3)}$1{
						\vtnode$((W^{(2)})^{(3)})^2$1{}
					}
				}{
					\vtnode[h]$(W^{(2)})^{(4)}$1{
						\vtnode$((W^{(2)})^{(4)})^2$1{}
					}
				}
		}{
			\vtnode[h]$W^{(3)}$1{
				\vtnode$(W^{(3)})^2$4{
					\vtnode$((W^{(3)})^2)^{(1)}$1{}
				}{
					\vtnode$((W^{(3)})^2)^{(2)}$1{}
				}{
					\vtnode$((W^{(3)})^2)^{(3)}$1{}
				}{
					\vtnode$((W^{(3)})^2)^{(4)}$1{}
				}
			}
		}{
			\vtnode[h]$W^{(4)}$1{
				\vtnode$(W^{(4)})^2$4{
					\vtnode$((W^{(4)})^2)^{(1)}$1{}
				}{
					\vtnode$((W^{(4)})^2)^{(2)}$1{}
				}{
					\vtnode$((W^{(4)})^2)^{(3)}$1{}
				}{
					\vtnode$((W^{(4)})^2)^{(4)}$1{}
				}
			}
		}
	}.\]

\subsection{Third look at \TP$T\fek${T⊂k}}

	Applying $T\fek$ increases the error probability $k$ times.
	But since we are dealing with error probabilities
	that are doubly exponential in $n$,
	A $k$-fold increase is easily eaten up by other minor terms.
	
	Similarly, $T\fek$ increases the block length $k$ times,
	which is negligible by fluctuating $β',μ'$ a little bit.

\subsection{Forth apply Theorem~\ref{thm:disposable} to \TP$T\err${Terr}}

	The proof of Theorem~\ref{thm:disposable}
	presented in Section~\ref{pf:disposable}
	reasons on the channel tree $𝒯\pf(W,T,n)$, which is different from
	what we have here, namely $𝒯\pf(W,T\rat,n\rat,T\err,n)$.
	However, we claim that this is not a mismatch.
	
	Imagine we copy-and-paste the proof here and replace
	all $μ^*$ by $μ^*\rat$, all $T$ by $T\err$, and all $Y$ by $Y\err$.
	Then the proof relies on three, and only these three facts:
	\begin{itemize}
		\item The subset $A_m$ is a collection of synthetic channels $w$
			at depth $m$ such that $Z(w)<\exp\(-\exp(m^{1/3})\)$.
		\item The subsets $A_0^m$ satisfy $I(W)-ℙ(A_0^m)≤N_m^{-1/μ^*\rat+o(1)}$.
		\item Subtrees rooted at synthetic channels in $A_m$
			are generated by applying $T\err$ till depth $n$.
	\end{itemize}
	Any other information, such as the transformation applied to $W_0$,
	is irrelevant to the proof.
	In fact, the argument does not care at all what happens before $A_0^{n\rat}$.

	We now verify that these three facts hold
	in case of $𝒯\pf(W,T\rat,n\rat,T\err,n)$:
	The first fact is the definition of $A_m$,
	which we inherit in growing the channel tree.
	The second fact is by
	Formula~(\ref{eq:goodrat}) and Lemma~\ref{lem:predicate}.
	The third fact is by
	how we grow the channel tree $𝒯\pf(W,T\rat,n\rat,T\err,n)$.
	Now we are sure that all three facts hold.
	
	Let $𝒜_n$ be defined as in Section~\ref{pf:disposable},
	the proof of Theorem~\ref{thm:disposable}.

\subsection{Fifth we announce the code}

	\[\(𝒯\pf(W,T\rat,n\rat,T\err,n),𝒜_n\)\]
	has block length
	\[N_n=kℓ^n,\]
	code rate
	\[R_n=I(W)-N_n^{1/μ'},\]
	and error probability
	\[P_n<\exp(N_n^{β'}).\]
	This finishes the proof of Theorem~\ref{thm:collaborate}.


\section{Application: To Approach the Hypotenuse}

	In this section, fix the relation $ℓ=2^k$.
	
	\begin{lem} \label{lem:randomgl}
		Assume BEC.
		There exist binary, length-$ℓ$, bounded transformations $T\rat$
		with $μ^*$-exponents $μ^*\rat$ and $∂$-dices $Y\rat$ such that
		\begin{gather}
			ℙ\{Y\rat=0\}<ℓ^{-1/μ^*\rat}
			\shortintertext{and, as $ℓ→∞$,}
			μ^*\rat⟶2.
		\end{gather}
	\end{lem}
	\begin{IEEEproof}
		That $μ^*\rat→2$ is by \cite[Theorem 2 and~3]{FHMV17}.
		On BEC, $Z$-parameters form a martingale,
		so transformations are bounded.
		The condition on $ℙ\{Y\rat=0\}$ is a consequence of the fact that
		an $[n,n-√n]$-random code has minimal distance at least $2$
		with high probability
		or the fact that an $[n,√n]$-random code has no all-zero column.
	\end{IEEEproof}
	
	\begin{lem} \label{lem:reedsolomon}
		There exist $ℓ$-ary, length-$ℓ$, bounded transformations $T\err$
		with $∂$-dices $Y\err$ following
		the uniform distribution on $\log1,\log2,\dotsc,\logℓ$ for all $ℓ≔2^k$.
	\end{lem}
	\begin{IEEEproof}
		\cite{MT10f,MT10h,MT14}.
	\end{IEEEproof}
	
	\begin{thm} \label{thm:interior}
		Assume BEC.
		For every point $(β',1/μ')$ inside the right triangle
		\[\tikz[Tri]\draw
			(0,0 )pic{dot}node[<<]{$(0,0)$}--
			(0,.5)pic{dot}node[<<]{$(0,1/2)$}--
			(1,0 )pic{dot}node[>>]{$(1,0)$}--cycle
		;,\]
		there exist $k,ℓ$ and transformations $T\rat,T\fek,T\err$
		that produce block codes $(𝒯_n,𝒜_n)$ such that
		\begin{gather}
			N_n=kℓ^n, \\
			R_n>I(W)-N_n^{-1/μ'},\rlap{ and} \\
			P_n<\exp(-N_n^{β'})
		\end{gather}
		for $n$ large enough.
	\end{thm}
	\begin{IEEEproof}
		See Section~\ref{pf:interior} right after the corollary below.
	\end{IEEEproof}
	
	\begin{cor} \label{cor:hypotenuse}
		Assume BEC.
		For every point $(β',1/μ')$ on the hypotenuse of the right triangle
		\[\tikz[Tri]\draw
			(0,0 )pic{dot}node[<<]{$(0,0)$}--
			(0,.5)pic{dot}node[<<]{$(0,1/2)$}--
			(1,0 )pic{dot}node[>>]{$(1,0)$}--cycle
		;\]
		and every monotonically increasing, unbounded function $h$,
		there exist a series of polar-like codes $(𝒯_n,𝒜_n)$ such that
		\begin{gather}
			N_n=kℓ^n, \\
			R_n>I(W)-N_n^{-1/μ'+o(1)}, \\
			P_n<\exp(-N_n^{β'-o(1)}),
			\shortintertext{and}
			\text{complexity}<h(N)N\log N\vphantom{N_n^{β'}}
		\end{gather}
		for $n$ large enough.

	\end{cor}
	\begin{IEEEproof}
		Approximate the point on the hypotenuse
		by points inside the right triangle.
		Apply Theorem~\ref{thm:interior} to each point
		and then apply the diagonal argument
		(as in the proof of Arzel\`a--Ascoli theorem).
	\end{IEEEproof}

\subsection{Proof of Theorem~\ref{thm:interior}} \label{pf:interior}

	Fix a point $(β',1/μ')$ inside the right triangle.
	Since we have Theorem~\ref{thm:collaborate}
	and Lemma \ref{lem:randomgl} and~\ref{lem:reedsolomon},
	it suffices to find an $ℓ≔2^k$,
	which determines $μ^*\rat$ (probabilistically) and $Y\err$,
	such that, for all $π∈[0,1]$,
	\[\frac{(1-π)\logℓ}{μ'-πμ^*\rat}<
		\CF\err（\frac{β'μ'\logℓ}{μ'-πμ^*\rat}）.\]
	
	Start from the fact that
	$Y\err$ follows the uniform distribution on $\log1,\log2,\dotsc,\logℓ$.
	The cumulant generating function satisfies
	\[\log𝔼[\exp(λY\err)]=\log𝔼[X\err^λ]\]
	where $X\err$ follows the uniform distribution on $1,2,\dotsc,ℓ$.
	For $-1<λ<0$, the $λ$-moment is
	\[𝔼X\err^λ=\frac1{ℓ}∑_{X=1}^ℓX^λ
		<\frac1{ℓ}∫_0^ℓX^λ\,\mathrm dX=\frac{ℓ^λ}{λ+1}.\]
	This leads to an approximation
	\[\log𝔼[\exp(λY\err)]<λ\logℓ-\log(λ+1).\]
	The \Cramer{} function is then bounded by
	\[\CF\err(y)≥\sup_{λ<0}λy-λ\logℓ+\log(λ+1).\]
	Redeem the supremum at $\logℓ-y=1/(λ+1)$ to obtain
	\begin{align}
		&{} \CF\err(y) \\
		&>（\frac1{\logℓ-y}-1）(y-\logℓ)+\log\frac1{\logℓ-y} \\
		&= -1+\logℓ-y-\log(\logℓ-y) \\
		&≥ -1+\logℓ-y-\log\logℓ+\frac y{\logℓ} \\
		&= (\logℓ-y)（1-\frac1{\logℓ}）-\log\logℓ. \label{eq:CFline}
	\end{align}
	The last line is linear in $y$.
	It is $\logℓ-1-\log\logℓ≈\logℓ$ when $y=0$
	and is $0$ when
	\[y=y^*≔\logℓ-\frac{\log\logℓ}{1-1/\logℓ}.\]
	
	Back to the fact that $(β',1/μ')$ is inside the right triangle
	\[\tikz[TRi]{
		\draw[hl]
			(0,.5)pic{dot}node[<<]{$(0,1/2)$}--
			(0,0 )pic{dot}node[<<]{$(0,0)$}--
			(1,0 )pic{dot}node[>>]{$(1,0)$}--cycle;
		\draw(.25,.25)pic{dot}coordinate[pin=45:{$(β',1/μ')$}]
	}.\]
	There exist a $μ^*\rat>2$ (by letting $ℓ→∞$)
	\[\tikz[TRi]{
		\draw[hl]
			(0,.5)pic{dot}node[<<]{$(0,1/2)$}--
			(0,0 )pic{dot}node[<<]{$(0,0)$}--
			(1,0 )pic{dot}node[>>]{$(1,0)$}--cycle;
		\draw(0,.35)pic{dot}node[<<]{$(0,1/μ^*\rat)$}
	}\]
	and a $y^*/\logℓ<1$ (by letting $ℓ→∞$)
	\[\tikz[TRi]{
		\draw[hl]
			(0,.5)pic{dot}node[<<]{$(0,1/2)$}--
			(0,0 )pic{dot}node[<<]{$(0,0)$}--
			(1,0 )pic{dot}node[>>]{$(1,0)$}--cycle;
		\draw(.875,0)pic{dot}node[_]{$(y^*/\logℓ,0)\;$}
	}\]
	such that these three points are collinear
	\[\tikz[TRi]{
		\draw[hl]
			(0,.5)pic{dot}node[<<]{$(0,1/2)$}--
			(0,0 )pic{dot}node[<<]{$(0,0)$}--
			(1,0 )pic{dot}node[>>]{$(1,0)$}--cycle;
		\draw
			(0,.35)pic{dot}node[<<]{$(0,1/μ^*\rat)$}--
			(.875,0)pic{dot}node[_]{$(y^*/\logℓ,0)\;$}
			(.25,.25)pic{dot}coordinate[pin=45:{$(β',1/μ')$}]
	}.\]
	
	Fix $ℓ,μ^*\rat,y^*$ as above.
	The term
	\[\frac{y^*-y}{y^*μ^*\rat}\logℓ\]
	is also linear in $y$.
	It is less than $\logℓ/2$ when $y=0$ and is $0$ when $y=y^*$.
	Thus, for all $0≤y≤y^*$,
	\[\frac{y^*-y}{y^*μ^*\rat}\logℓ≤(\logℓ-y)（1-\frac1{\logℓ}）-\log\logℓ\]
	because the inequality holds for endpoints and both sides are linear in $y$.
	Concatenate with Formula~(\ref{eq:CFline}) to obtain, for all $0≤y≤y^*$,
	\[\frac{y^*-y}{y^*μ^*\rat}\logℓ<\CF\err(y).\]
	On each side of the inequality, replace $y$
	with the corresponding side of the equality due to collinearity below
	\[\frac{y^*(μ'-μ^*\rat)}{μ'-πμ^*\rat}
		=\frac{β'μ'\logℓ}{μ'-πμ^*\rat} \label{eq:ysub} \]
	to get
	\[\frac{(1-π)\logℓ}{μ'-πμ^*\rat}<\CF\err（\frac{β'μ'\logℓ}{μ'-πμ^*\rat}）.\]
	This is exactly what we need to apply Theorem~\ref{thm:disposable}.
	
	This proof is very similar to \cite[Corollary~8]{WD18}.


\section{Further Implications} \label{sec:bigtriangle}

	There is another way to state Theorem~\ref{thm:disposable}.
	We put this as a claim since we omit the details of the proof.
	\begin{conj} \label{conj:convexhull}
		Let $T$ be a length-$ℓ$, bounded transformation
		with $μ^*$-exponent $μ^*$ and $∂$-dice $Y$.
		Let $\CF$ be the \Cramer{} function of $Y$.
		If $(β',1/μ')$ does not lie in the convex hull
		of the point $(0,1/μ^*)$ union the epigraph of the function
		\[β⟼\frac{\CF(β\logℓ)}{\logℓ},\]
		then $(β',1/μ')$ is possible.
	\end{conj}
	\begin{IEEEproof}[Sketch]
		As a function of $π$, consider points
		\[Q(π)≔（\frac{β'μ'}{μ'-πμ^*},\frac{1-π}{μ'-πμ^*}）.\]
		Here is the trace of $Q(π)$ when $π=0,.1,\dotsc,1$:
		for $π=0$, $Q(0)$ coincides with $(β',1/μ')$;
		for $π=1$, $Q(1)$ is on the horizontal axis;
		for intermediate $π$, the $Q(π)$ moves along the ray
		starting at $(0,1/μ^*)$ through $(β',1/μ')$.
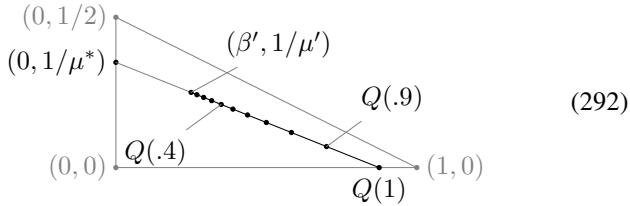
		\[\tikz[TRi]{
			\draw[hl]
				(0,.5)pic{dot}node[<<]{$(0,1/2)$}--
				(0,0 )pic{dot}node[<<]{$(0,0)$}--
				(1,0 )pic{dot}node[>>]{$(1,0)$}--cycle;
			\draw
				(0,.35)pic{dot}node[<<]{$(0,1/μ^*)$}
				(0,.35)edge[hl](.25,.25)
				(.25,.25)pic{dot}coordinate[pin=45:{$(β',1/μ')$}]--
				(.875,0)pic{dot}node[_]{$Q(1)$}
				foreach\p in{1,...,9}{
					({1/(4-.\p/.35)},{(1-.\p)/(4-.\p/.35)})pic{dot}
					\ifnum\p=4 coordinate[pin=225:{$Q(.\p)$}]\fi
					\ifnum\p=9 coordinate[pin=45:{$Q(.\p)$}]\fi
				};
		}\]
		
		From the graph, we learn that:
		$(β',1/μ')$ does not lie in the convex hull
		iff $Q(π)$ is not in the epigraph for all $π∈[0,1]$;
		The later happens iff $μ<\CF(β\logℓ)/\logℓ$ for all $π∈[0,1]$;
		iff the criteria of Theorem~\ref{thm:disposable} are met.
	\end{IEEEproof}
	
	Here is a running example:
	For $T_\Ari$, the rescaled \Cramer\ function $β↦\CF(β\log2)/\log2$
	coincides with the relative entropy
	\[β⟼1+β\log_2β+(1-β)\log_2(1-β)\]
	for $0≤β≤1/2$.
	For $1/2≤β≤1$, the ``classical definition'' of the \Cramer\ function
	still coincides with the relative entropy.
	In our definition, however, we insist that
	the supremum is taken over negative $λ$ so $\CF$ vanishes.
	In the following graph, the curve is the relative entropy and
	the shaded area is the epigraph of $β↦\CF(β\log2)/\log2$
	
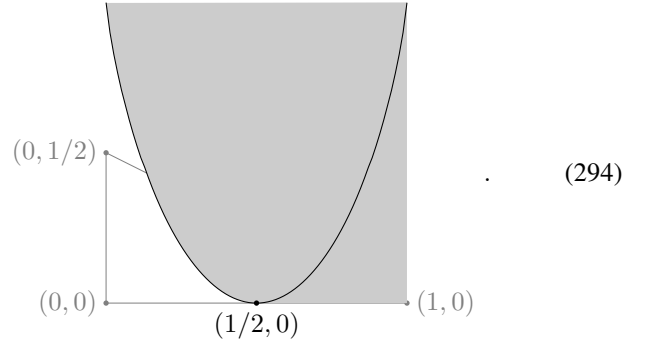
\[\tikz[TRi]{
		\draw[hl]
			(0,.5)pic{dot}node[<<]{$(0,1/2)$}--
			(0,0 )pic{dot}node[<<]{$(0,0)$}--
			(1,0 )pic{dot}node[>>]{$(1,0)$}--cycle;
		\fill[black!20]plot[domain=.0001:.5](\x,{1-H(\x)})-|(1,1);
		\draw plot[domain=.0001:.9999,samples=79](\x,{1-H(\x)});
		\draw(.5,0)pic{dot}node[_]{$(1/2,0)$};
	}.\]
	Together with $(0,1/μ^*)$ they form a convex hull
	
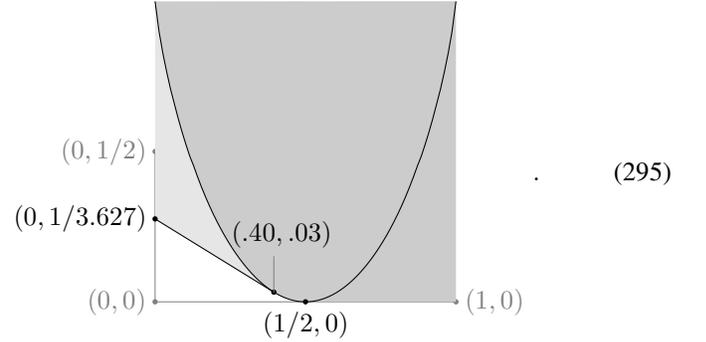
\[\tikz[TRi]{
		\draw[hl]
			(0,.5)pic{dot}node[<<]{$(0,1/2)$}--
			(0,0 )pic{dot}node[<<]{$(0,0)$}--
			(1,0 )pic{dot}node[>>]{$(1,0)$}--cycle;
		\fill[black!10]plot file{AriRS2.txt}|-(0,1);
		\fill[black!20]plot[domain=.0001:.5](\x,{1-H(\x)})-|(1,1);
		\draw plot[domain=.0001:.9999,samples=79](\x,{1-H(\x)});
		\draw(0,1/mu)pic{dot}node[<<]{$(0,1/3.627)$}--
			(0.3947,0.0322)pic{dot}coordinate[pin=90:{$\;\;(.40,.03)$}];
		\draw(.5,0)pic{dot}node[_]{$(1/2,0)$};
	}.\]
	
	Back to Claim~\ref{conj:convexhull}.
	If $(β',1/μ')$ is here
	
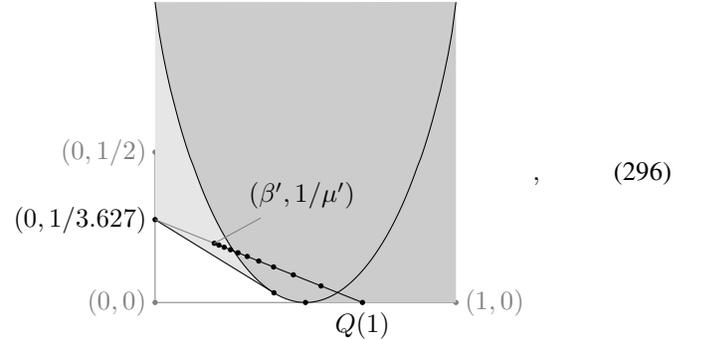
\[\tikz[TRi]{
		\draw[hl]
			(0,.5)pic{dot}node[<<]{$(0,1/2)$}--
			(0,0 )pic{dot}node[<<]{$(0,0)$}--
			(1,0 )pic{dot}node[>>]{$(1,0)$}--cycle;
		\fill[black!10]plot file{AriRS2.txt}|-(0,1);
		\fill[black!20]plot[domain=.0001:.5](\x,{1-H(\x)})-|(1,1);
		\draw plot[domain=.0001:.9999,samples=79](\x,{1-H(\x)});
		\draw(0,1/mu)pic{dot}node[<<]{$(0,1/3.627)$}--(0.395,0.0322)pic{dot};
		\draw(.5,0)pic{dot};
		\draw
			(0,1/mu)edge[hl](5/7/mu,5/7/mu)
			(5/7/mu,5/7/mu)pic{dot}coordinate[pin=45:{$(β',1/μ')$}]--
			(2.5/mu,0)pic{dot}node[_]{$Q(1)$}
			foreach\p in{1,...,9}{
				({1/(1.4*mu-.\p*mu)},{(1-.\p)/(1.4*mu-.\p*mu)})pic{dot}
			}
	},\]
	then some $Q(π)$ is in the epigraph
	and the criteria of Theorem~\ref{thm:disposable} fail.
	On the other hand, if $(β',1/μ')$ is here
	
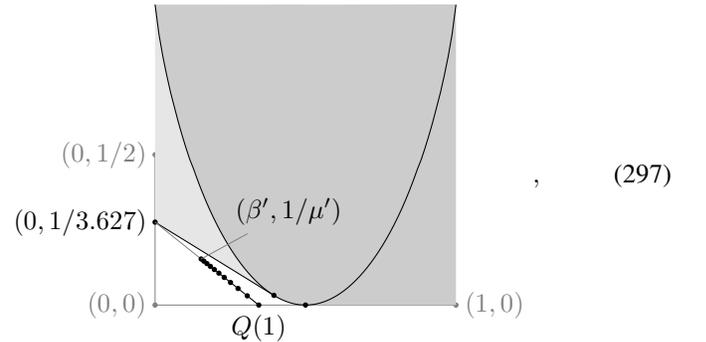
\[\tikz[TRi]{
		\draw[hl]
			(0,.5)pic{dot}node[<<]{$(0,1/2)$}--
			(0,0 )pic{dot}node[<<]{$(0,0)$}--
			(1,0 )pic{dot}node[>>]{$(1,0)$}--cycle;
		\fill[black!10]plot file{AriRS2.txt}|-(0,1);
		\fill[black!20]plot[domain=.0001:.5](\x,{1-H(\x)})-|(1,1);
		\draw plot[domain=.0001:.9999,samples=79](\x,{1-H(\x)});
		\draw(0,1/mu)pic{dot}node[<<]{$(0,1/3.627)$}--(0.395,0.0322)pic{dot};
		\draw(.5,0)pic{dot};
		\draw
			(0,1/mu)edge[hl](5/9/mu,5/9/mu)
			(5/9/mu,5/9/mu)pic{dot}coordinate[pin=45:{$(β',1/μ')$}]--
			(5/4/mu,0)pic{dot}node[_]{$Q(1)$}
			foreach\p in{1,...,9}{
				({1/(1.8*mu-.\p*mu)},{(1-.\p)/(1.8*mu-.\p*mu)})pic{dot}
			}
	},\]
	then all $Q(π)$ are outside the epigraph
	and Theorem~\ref{thm:disposable} applies.
	Another interesting case is when $(β',1/μ')$ is in
	the tiny tip area at the bottom.
	Therein all $Q(π)$ are outside the epigraph
	and Theorem~\ref{thm:disposable} applies.

\subsection{Moderate Deviations Regime Recovers
	Error Exponent Regime as a Special Case}

	The following is a consequence of the Claim~\ref{conj:convexhull}
	plus the fact that $\CF(y)$ reaches zero at $y=𝔼Y$.
	\begin{prop}
		Let $T$ be a length-$ℓ$, bounded transformation
		with $μ^*$-exponent $μ^*<∞$ and $β^*$-exponent $β^*>0$.
		For any $β'<β^*$, there exists $1/μ'>0$ such that
		$(β',1/μ')$ is possible.
	\end{prop}
	See also \cite[Theorem~2.16]{BGS18}.

\subsection{Moderate Deviations Regime Recovers
	Scaling Exponent Regime as a Special Case}

	The following is another consequence of the Claim~\ref{conj:convexhull}.
	\begin{prop}
		Let $T$ be a length-$ℓ$, bounded transformation
		with $μ^*$-exponent $μ^*<∞$ and $β^*$-exponent $β^*>0$.
		For any $1/μ'<1/μ^*$, there exists $β'>0$ such that
		$(β',1/μ')$ is possible.
	\end{prop}
	This is a generalization of \cite[Corollary~8]{WD18}.

\subsection{\Arikan's Polar Codes Attacking on BEC}

	The three corner dots are $(0,.5)$, $(0,0)$, and $(1,0)$.
	\cite{GX13} proves that it is possible to achieve $(β',1/μ')=(.49,O(1))$.
	It is represented as a point very close to $(.5,0)$.
	\cite{MHU16} proves an interpolating result.
	Their curve connects $(0,1/4.627)$ and $(.5,0)$ and is drawn below.
	Theorem~\ref{thm:disposable} (and also \cite{WD18}) implies a better curve.
	This curve connects $(0,1/3.627)$ and $(.5,0)$.
	Notice that in this scenario, $μ^*=3.627$ is given by \cite{FV14}.
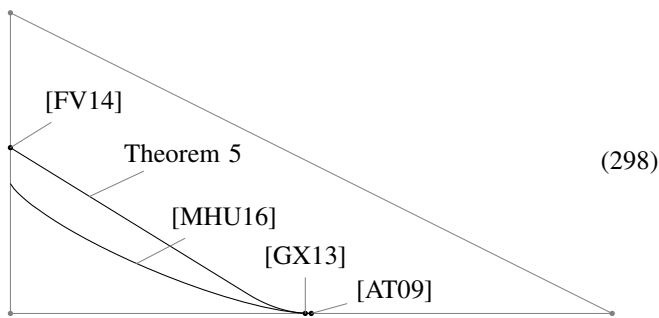
	\[\tikz[TRI]{
		\draw[hl](0,.5)pic{dot}--(0,0)pic{dot}--(1,0)pic{dot}--cycle;
		\draw plot file{AriRS2.txt}
			(.1316,.1946)coordinate[pin=45:Theorem~\ref{thm:disposable}];
		\draw plot[domain=.001:.5]({G(\x)*\x},{(1-G(\x))/mu})
			({G(.3)*.3},{(1-G(.3))/mu})coordinate[pin=45:\cite{MHU16}];
		\draw(0,1/mu)pic{dot}coordinate[pin=45:\cite{FV14}];
		\draw(.49,.0001)pic{dot}coordinate[pin=90:\cite{GX13}];
		\draw(.5,0)pic{dot}coordinate[pin=4:\cite{AT09}];
	}\]

\subsection{\Arikan's Polar Codes Attacking on BDMC}

	BDMC is not far from BEC in the sense that almost all treatments
	are the same except $μ^*=4.714$ instead of $3.627$.
	In particular, the curves are drawn using the same formulae
	with the new $μ^*$.
	So this time the Theorem~\ref{thm:disposable} curve
	connects $(0,1/4.714)$ and $(.5,0)$.
	And the \cite{MHU16} curve connects $(0,1/5.714)$ and $(.5,0)$.
	Notice that in this scenario, $μ^*=4.714$ is given by \cite{MHU16}.
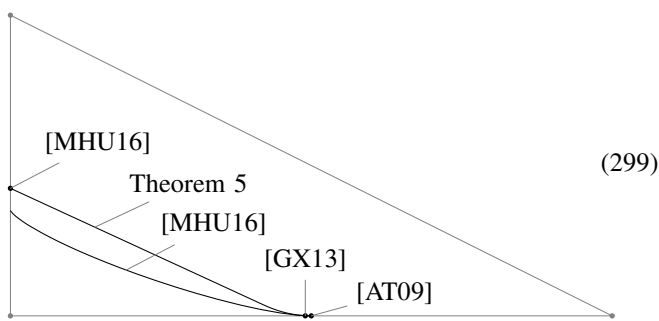
	\[\tikz[TRI]{
		\draw[hl](0,.5)pic{dot}--(0,0)pic{dot}--(1,0)pic{dot}--cycle;
		\pgfmathdeclarefunction*{mu}{0}{\pgfmathparse{4.714}}
		\draw plot file{AriRS2AWGN.txt}
			(.1403,.1475)coordinate[pin=45:Theorem~\ref{thm:disposable}];
		\draw plot[domain=.001:.5]({G(\x)*\x},{(1-G(\x))/mu})
			({G(.3)*.3},{(1-G(.3))/mu})coordinate[pin=45:\cite{MHU16}];
		\draw(0,1/mu)pic{dot}coordinate[pin=45:\cite{MHU16}];
		\draw (.49,.0001)pic{dot}coordinate[pin=90:\cite{GX13}];
		\draw(.5,0)pic{dot}coordinate[pin=4:\cite{AT09}];
	}\]

\subsection{\Arikan's Polar Codes Attacking on AWGN}

	\cite{FT17} analyzes the AWGN channel and mimic \cite{MHU16}.
	They end up with the same curve as the bottom one
	in the previous plot that connects $(0,1/5.714)$ and $(.5,0)$.
	Theorem~\ref{thm:disposable} implies the same curve as the top one
	in the previous plot that connects $(0,1/4.714)$ and $(.5,0)$.
	Notice that in this scenario, $μ^*=4.714$ is given by \cite{FT17}.
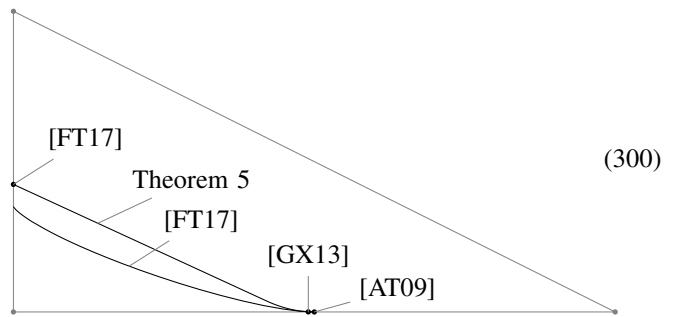
	\[\tikz[TRI]{
		\draw[hl](0,.5)pic{dot}--(0,0)pic{dot}--(1,0)pic{dot}--cycle;
		\pgfmathdeclarefunction*{mu}{0}{\pgfmathparse{4.714}}
		\draw plot file{AriRS2AWGN.txt}
			(.1403,.1475)coordinate[pin=45:Theorem~\ref{thm:disposable}];
		\draw plot[domain=.001:.5]({G(\x)*\x},{(1-G(\x))/mu})
			({G(.3)*.3},{(1-G(.3))/mu})coordinate[pin=45:\cite{FT17}];
		\draw(0,1/mu)pic{dot}coordinate[pin=45:\cite{FT17}];
		\draw (.49,.0001)pic{dot}coordinate[pin=90:\cite{GX13}];
		\draw(.5,0)pic{dot}coordinate[pin=4:\cite{AT09}];
	}\]

\subsection{Polar Codes with Larger Kernels Attacking on BEC}

\subsubsection{Pessimistic Case}

	We present two fake curves that illustrates the fact that
	Theorem~\ref{thm:disposable} can be used
	to connect $(0,1/μ^*)$ and $(β^*,0)$.
	The left curve with \cite{FHMV17} as an endpoint shows that
	there are kernels such that $1/μ^*$ are arbitrarily close to $1/2$;
	while the $β^*$-exponents of these kernels are unknown.
	The bottom curve with \cite{KSU10} as an endpoint shows that
	there are kernels such that $β^*$ are arbitrarily close to $1$;
	while the $μ^*$-exponents of these kernels are unknown.
	Besides the two curves, \cite{BGS18} shows that
	it is possible to approach where \cite{KSU10} is with positive $1/μ'$-value.
	
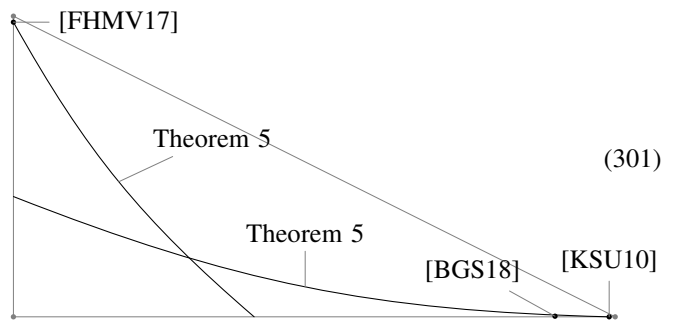
\[\tikz[TRI]{
		\draw[hl](0,.5)pic{dot}--(0,0)pic{dot}--(1,0)pic{dot}--cycle;
		\draw(0,.49)to[bend right=10]node[pos=.5](X){}(.4,0)
			(X.center)coordinate[pin=45:Theorem~\ref{thm:disposable}];
		\draw(0,.2)to[bend right=10]node[pos=.5](X){}(.99,0)
			(X.center)coordinate[pin=90:Theorem~\ref{thm:disposable}];
		\draw(0,.49)pic{dot}coordinate[pin=0:\cite{FHMV17}];
		\draw(.9,.001)pic{dot}coordinate[pin=135:\cite{BGS18}];
		\draw(.99,0)pic{dot}coordinate[pin=90:\cite{KSU10}];
	}\]
	(It seems \cite{BGS18} is a distance away from \cite{KSU10}
	and that is because we do not want labels to overlap.)

\subsubsection{Optimistic Case}

	Moreover, if there are kernels such that
	$(β^*,1/μ^*)$ converges to $(1,1/2)$,
	then Theorem~\ref{thm:disposable} will eventually cover the right triangle.
	
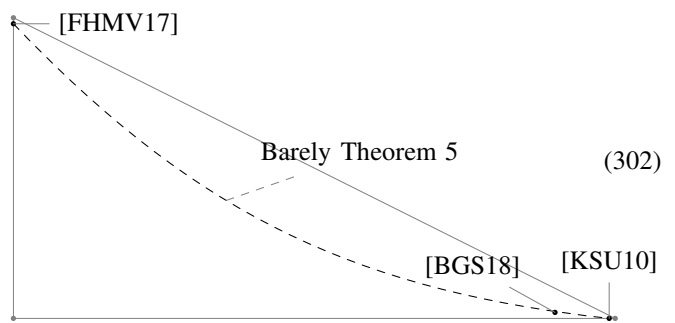
\[\tikz[TRI]{
		\draw[hl](0,.5)pic{dot}--(0,0)pic{dot}--(1,0)pic{dot}--cycle;
		\draw[dashed](0,.49)to[bend right=20]node[pos=.4](X){}(.99,0)
			(X.center)coordinate[pin=45:Barely Theorem~\ref{thm:disposable}];
		\draw(0,.49)pic{dot}coordinate[pin=0:\cite{FHMV17}];
		\draw(.9,.01)pic{dot}coordinate[pin=135:\cite{BGS18}];
		\draw(.99,0)pic{dot}coordinate[pin=90:\cite{KSU10}];
	}\]
	The existence of such kernels is not clear at this stage.
	This is one of the reasons why we develop Theorem~\ref{thm:collaborate} ---
	which is basically saying that we can
	steal the good $μ^*$-exponent from a kernel and
	steal the good $β^*$-exponent from another.
	
	Chances are that random kernels possesses
	good $μ^*$ and good $β^*$-exponents.
	And we can use Hoeffding's inequality
	to control the behavior of \Cramer{} functions.

\subsection{Polar Codes with Larger Kernels Attacking on BDMC}

	For binary channels other than BEC, \cite{FHMV17} does not apply anymore.
	Then \cite{BGNRS18} takes place and proves that
	all kernels, in particular kernels from \cite{KSU10}, have positive $1/μ^*$.
	We draw a fake curve to illustrate that Theorem~\ref{thm:disposable}
	connects the points given by \cite{BGNRS18} and by \cite{KSU10}.
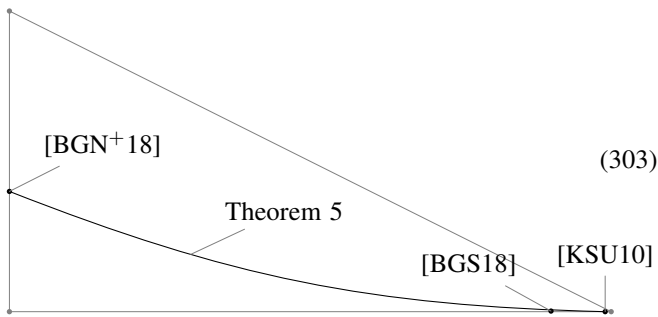
	\[\tikz[TRI]{
		\draw[hl](0,.5)pic{dot}--(0,0)pic{dot}--(1,0)pic{dot}--cycle;
		\draw(0,.2)to[bend right=10]node[pos=.3](X){}(.99,0)
			(X.center)coordinate[pin=45:Theorem~\ref{thm:disposable}];
		\draw(0,.2)pic{dot}coordinate[pin=45:\cite{BGNRS18}];
		\draw(.9,.001)pic{dot}coordinate[pin=135:\cite{BGS18}];
		\draw(.99,0)pic{dot}coordinate[pin=90:\cite{KSU10}];
	}\]

\subsection{Polar Codes with Larger Kernels Attacking on General Channels}

	For channels that are not binary, \cite{KSU10} does not apply anymore.
	Then \cite{BGS18} steps in and comments that
	BCH codes, in general, fill in the blank that
	there are kernels with $β^*$ arbitrarily close to one.
	We again draw a fake curve to illustrate that Theorem~\ref{thm:disposable}
	connects the points representing $1/μ^*$ and $β^*$.
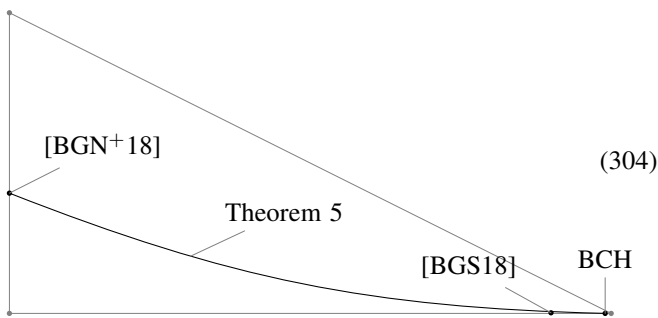
	\[\tikz[TRI]{
		\draw[hl](0,.5)pic{dot}--(0,0)pic{dot}--(1,0)pic{dot}--cycle;
		\draw(0,.2)to[bend right=10]node[pos=.3](X){}(.99,0)
			(X.center)coordinate[pin=45:Theorem~\ref{thm:disposable}];
		\draw(0,.2)pic{dot}coordinate[pin=45:\cite{BGNRS18}];
		\draw(.9,.001)pic{dot}coordinate[pin=135:\cite{BGS18}];
		\draw(.99,0)pic{dot}coordinate[pin=90:BCH];
	}\]

\subsection{Concatenated Polar Codes Attacking on BEC}

	If concatenated polar codes are allowed, then
	Theorem~\ref{thm:interior} shows that
	it is possible to fill the right triangle.
	We draw a fake to illustrate this.
	
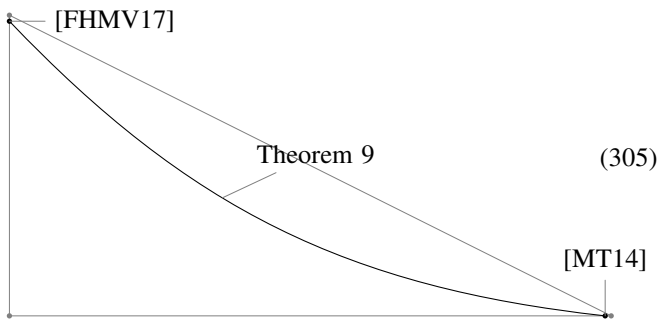
\[\tikz[TRI]{
		\draw[hl](0,.5)pic{dot}--(0,0)pic{dot}--(1,0)pic{dot}--cycle;
		\draw(0,.49)to[bend right=20]node[pos=.4](X){}(.99,0)
			(X.center)coordinate[pin=45:Theorem~\ref{thm:interior}];
		\draw(0,.49)pic{dot}coordinate[pin=0:\cite{FHMV17}];
		\draw(.99,0)pic{dot}coordinate[pin=90:\cite{MT14}];
	}\]

\subsection{Concatenated Polar Codes Attacking on General Channels}

	For general channels other than BEC, \cite{FHMV17} does not apply.
	We may apply Theorem \ref{thm:disposable} or~\ref{thm:collaborate}
	according to whether we want a single kernel or two kernels.
	We draw a fake to illustrate this.
	
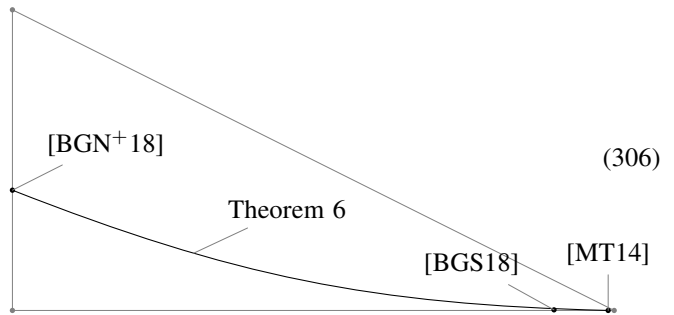
\[\tikz[TRI]{
		\draw[hl](0,.5)pic{dot}--(0,0)pic{dot}--(1,0)pic{dot}--cycle;
		\draw(0,.2)to[bend right=10]node[pos=.3](X){}(.99,0)
			(X.center)coordinate[pin=45:Theorem~\ref{thm:collaborate}];
		\draw(0,.2)pic{dot}coordinate[pin=45:\cite{BGNRS18}];
		\draw(.9,.001)pic{dot}coordinate[pin=135:\cite{BGS18}];
		\draw(.99,0)pic{dot}coordinate[pin=90:\cite{MT14}];
	}\]

\subsection{\Arikan\ and Reed--Solomon Codes Attacking on BEC}

	We consider this a killer application.
	See \cite{BJE10} for a result similar to \cite{HMTU13}.
	See \cite{GB14f} for a result similar to \cite{GX13,BGS18}.
	See \cite{MEKLK13,MEKLK14} for more.
	
	For $k=1$, the transformation $T_\RS2$ is $T_\Ari$.
	There is no concatenation happening.
	
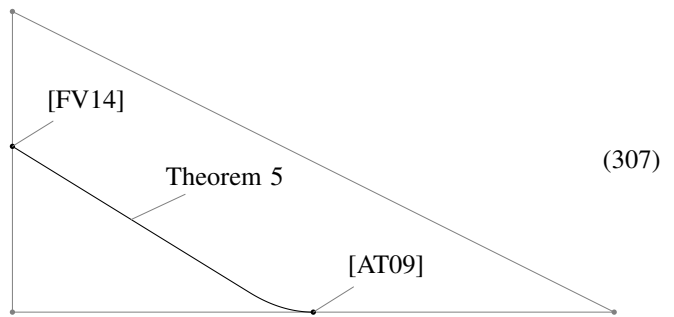
\[\tikz[TRI]{
		\draw[hl](0,.5)pic{dot}--(0,0)pic{dot}--(1,0)pic{dot}--cycle;
		\draw plot file{AriRS2.txt}
			(.1974,.1540)coordinate[pin=45:Theorem~\ref{thm:disposable}];
		\draw(0,1/mu)pic{dot}coordinate[pin=45:\cite{FV14}];
		\draw(.5,0)pic{dot}coordinate[pin=45:\cite{AT09}];
	}\]
	For $k=2$, transformations $T_\Ari^{⊗2},T\fe2,T_\RS4$
	collaboratively beat $T_\Ari$.
	In particular $β^*=(3+\log_23)/8=.57$.
	
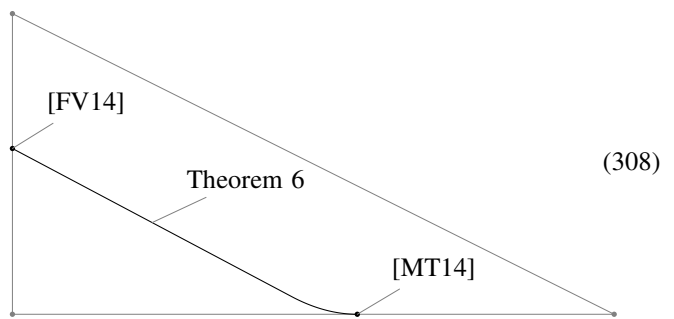
\[\tikz[TRI]{
		\draw[hl](0,.5)pic{dot}--(0,0)pic{dot}--(1,0)pic{dot}--cycle;
		\draw plot file{AriRS4.txt}
			(.2322,.1525)coordinate[pin=45:Theorem~\ref{thm:collaborate}];
		\draw(0,1/mu)pic{dot}coordinate[pin=45:\cite{FV14}];
		\draw(.5731,0)pic{dot}coordinate[pin=45:\cite{MT14}];
	}\]
	For $k=3$, transformations $T_\Ari^{⊗3},T\fe3,T_\RS8$ are even better.
	
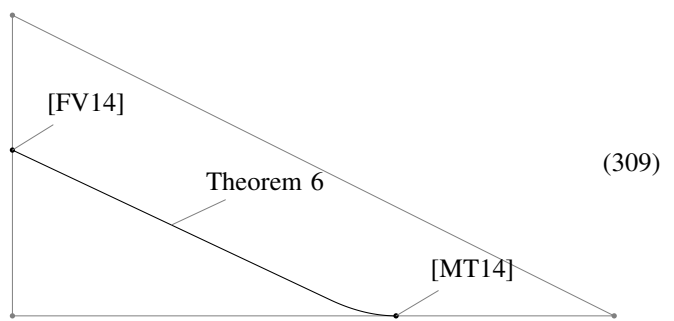
\[\tikz[TRI]{
		\draw[hl](0,.5)pic{dot}--(0,0)pic{dot}--(1,0)pic{dot}--cycle;
		\draw plot file{AriRS8.txt}
			(.2643,.1511)coordinate[pin=45:Theorem~\ref{thm:collaborate}];
		\draw(0,1/mu)pic{dot}coordinate[pin=45:\cite{FV14}];
		\draw(.6375,0)pic{dot}coordinate[pin=45:\cite{MT14}];
	}\]
	We put $k=4$ (and $T_\Ari^{⊗4},T\fe4,T_\RS{16}$) here
	just in case the trend is not clear.
	
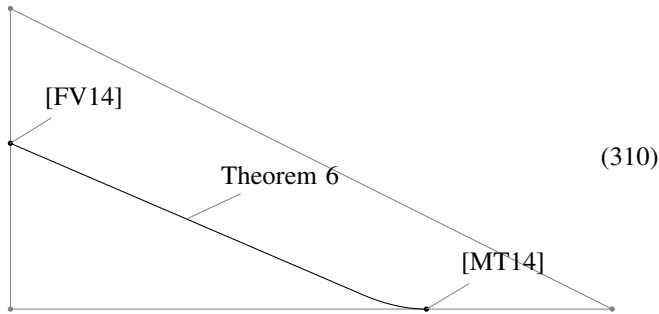
\[\tikz[TRI]{
		\draw[hl](0,.5)pic{dot}--(0,0)pic{dot}--(1,0)pic{dot}--cycle;
		\draw plot file{AriRS16.txt}
			(.2927,.1498)coordinate[pin=45:Theorem~\ref{thm:collaborate}];
		\draw(0,1/mu)pic{dot}coordinate[pin=45:\cite{FV14}];
		\draw(.6914,0)pic{dot}coordinate[pin=45:\cite{MT14}];
	}\]
	It is not hard to see that this series of curves
	eventually converges to a segment
	that connects $(0,1/3.627)$ and $(1,0)$.
	As $k→∞$, points \cite{GB14f,BJE10} also converge to $(1,0)$,
	at a faster pace.
	
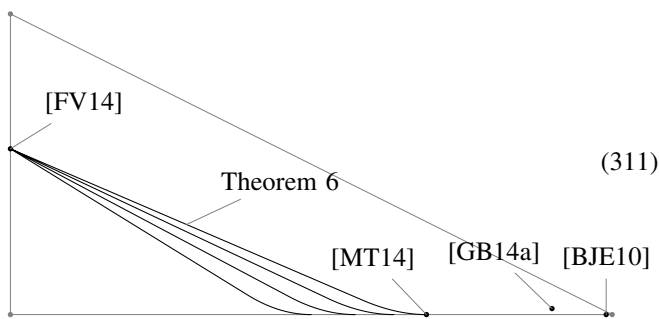
\[\tikz[TRI]{
		\draw[hl](0,.5)pic{dot}--(0,0)pic{dot}--(1,0)pic{dot}--cycle;
		\draw
			plot file{AriRS2.txt}
			plot file{AriRS4.txt}
			plot file{AriRS8.txt}
			plot file{AriRS16.txt}
			(.2927,.1498)coordinate[pin=45:Theorem~\ref{thm:collaborate}];
		\draw(0,1/mu)pic{dot}coordinate[pin=45:\cite{FV14}];
		\draw(.6914,0)pic{dot}coordinate[pin=94:\cite{MT14}];
		\draw(.9,.01)pic{dot}coordinate[pin=94:\cite{GB14f}];
		\draw(.99,0)pic{dot}coordinate[pin=90:\cite{BJE10}];
	}\]
	(It seems \cite{GB14f} is a distance away from \cite{BJE10}
	and that is because we do not want labels to overlap.)

\subsection{\Arikan\ and Reed--Solomon Codes Attacking on BDMC and AWGN}

	For BDMC and AWGN, curves are connecting $(0,1/4.714)$.
	
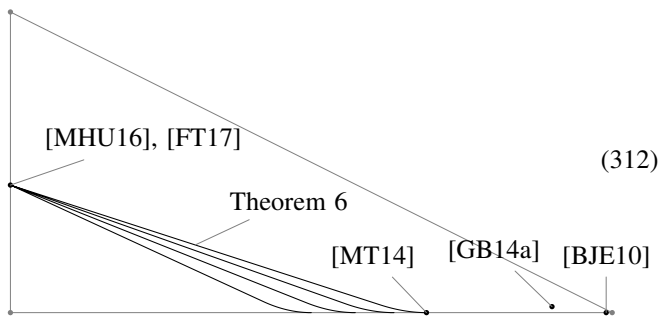
\[\tikz[TRI]{
		\draw[hl](0,.5)pic{dot}--(0,0)pic{dot}--(1,0)pic{dot}--cycle;
		\draw
			plot file{AriRS2AWGN.txt}
			plot file{AriRS4AWGN.txt}
			plot file{AriRS8AWGN.txt}
			plot file{AriRS16AWGN.txt};
		\draw(.3072,.1126)coordinate[pin=45:Theorem~\ref{thm:collaborate}];
		\draw(0,1/4.714)pic{dot}coordinate[pin=45:\cite{MHU16,FT17}];
		\draw(.6914,0)pic{dot}coordinate[pin=94:\cite{MT14}];
		\draw(.9,.01)pic{dot}coordinate[pin=94:\cite{GB14f}];
		\draw(.99,0)pic{dot}coordinate[pin=90:\cite{BJE10}];
	}\]
	
	See Appendix~\ref{app:moreconcat} for more types of concatenations.


\section{Future Works}

	What we do not address in this work is whether
	Theorem \ref{thm:disposable} and~\ref{thm:collaborate} give optimal bounds.
	For one, it is difficult to imagine that a description as simple as
	Claim~\ref{conj:convexhull} is not \emph{the} answer.
	That said, we look forward to a second-order result
	just like \cite{HMTU13} extending \cite{AT09}.
	
	On the other hand, statements and proofs in this work
	heavily rely on the magical value $μ^*$.
	The problem, as of today, is we can bound or approximate $μ^*$
	but do not know if the limit exists.
	Should there be distinct $μ^*$ and $μ_*$
	as limit superior and limit inferior,
	we expect two curves connecting $(0,1/μ^*)$ and $(0,1/μ_*)$ to $(β^*,0)$.
	
	Having Theorem~\ref{thm:interior} and Corollary~\ref{cor:hypotenuse},
	we like to see if they extend to channels other than BEC.
	Particularly, does $μ^*$ achieve $2$ for general channels?
	Furthermore, are there kernels with good $μ^*$ and $β^*$?


\section{Conclusion}

	We provide a merciful generalization of polar codes and
	are able to characterize, for a subclass of polar-like codes,
	the tradeoff among block length, code rate, and error probability
	asymptotically.
	
	We then show that a grafted variant of polar coding
	almost catches up the performance of random codes on BEC,
	if arbitrary kernels are allowed.
	
	If one likes to stick to Reed--Solomon kernels,
	we characterize the performance as well.


\section*{Acknowledgment}

	H.-P.~W.\ thanks Chi-Hua Wang (Purdue University)
	for priceless discussions on probability topics.


{
\hbadness3199
\bibliographystyle{alpha}
\bibliography{LargeDeviations-4}
}


\newpage

\appendix

\subsection{Polar Code Error Exponent Regime} \label{app:eer}

	Inner and outer bounds for usual polar codes \cite{AT09,KSU10,HMTU13,MT14}.
	
	\cite{BBGL17} proposes and solves an interesting question:
	\def\namea{_{\text{Ben}}}
	\def\nameb{_{\text{Bio}}}
	\def\namec{_{\text{Gab}}}
	\def\named{_{\text{Lan}}}
	\def\BBGL{_{\text{BBGL}}}
	We have four matrixes, say $G\namea,G\nameb,G\namec,G\named$.
	They induce four transformations $T\namea,T\nameb,T\namec,T\named$
	and we want to apply them alternately.
	Even more excitingly, we throw dices to decide which transformation to apply.
	
	In this setup, one may argue that the four transformations
	actually form a compound transformation $T\BBGL$ with build-in randomness.
	In particular, the $∂$-dice $Y\BBGL$ follows a compound distribution
	derived from $Y\namea,Y\nameb,Y\namec,Y\named$.
	Not only their result (an $N$-$P$ tradeoff) follows immediately,
	but it also automatically upgrades to an $N$-$R$-$P$ tradeoff.

\subsection{Polar Code Scaling Exponent Regime} \label{app:ser}

	See \cite{FHMV17} for a good review.
	
	Outer bounds
	\cite{Dobrushin61,Strassen64,TZ00,Montanari01,Hayashi09,PPV10}.
	
	Inner bounds
	\cite{KMTU10,HAU14,GB14k,MHU16,FV14,PU16,Hassani13,FHMV17}.
	
	List decoder \cite{MHU15}.

\subsection{Polar Code Moderate Deviations Regime} \label{app:mdr}

	Outer bound \cite{AW10,PV10,AW14,Arikan15,HT15}.
	
	Inner bound \hfill \cite{GX13,MHU16,FT17,BGNRS18,WD18,BGS18}

\subsection{Other Types of Concatenations} \label{app:moreconcat}

	There are a lot of works trying to concatenate polar codes
	with Reed--Solomon codes or RS-polar codes.
	The list includes but is not limited to
	\cite{BJE10,KSH11,MEKLK13,MEKLK14,GB14f,WZLWL17}.
	
	Polar with BCH codes
	\cite{WN14,WNH16}.
	
	Polar with algebraic geometry codes
	\cite{ED13,AM14}.
	
	Polar with LDPC codes \hfill
	\cite{EPN11,EPN13,GQFS14,ZLGYY14,MLZ17,YSLDZR18,ZLHC18}.
	
	Polar with RA codes 
	\cite{YZ16}.
	
	Polar with single parity check code
	\cite{YM17}.
	
	Polar with small, ML-decodable codes
	\cite{SH10,BGZ12}.
	
	Polar with arbitrary outer codes
	\cite{TS11,GB17}.
	
	Polar kernels with various length
	\cite{BBGL17,BCL18,BGLB17,GBLB17}.

\onecolumn

\subsection{Big Sankey Diagram} \label{app:bigsankey}

	\tikzset{
		fit/.style={
			xscale=\textwidth/11cm/.666,
			/utils/exec={
				\sdwidth\pgf@pt@aa\sdwidth\sdlepht.5\sdwidth\sdright.5\sdwidth},
			xscale=#1
		},
		fit/.default=1,
		sd/.append style={
			every node/.style={o,opacity=1}
		}
	}
	
	\[\tikz[fit]{
		\draw[->](0,-\sdwidth) +(1.5,0)node[^]{$m$}
		                       +(4.5,0)node[^]{$n$}
		                 +(0,0)--+(5,0)node[_,pos=.5]{depth};
		\draw[->](-.2,0)--+(0,\sdwidth)node[^,pos=.5,rotate=90]{channels};
		\tikzset{sd}
		\sddive(1);
		\sdcutby.4{
			\sddive(1)node{frozen};
		}{
			\sddive(1)node{recruited};
			\sddive(2)node{trained};
			\sdcutby.4{
				\sddive(1)node{frozen};
			}
			{
				\sddive(1)node{retained};
			}
		}
	} \label{eq:sdT}\]\vfil
	
	\[\tikz[fit=.666]{
		\draw[->](0,-\sdwidth) +(1.5,0)node[^]{$√n$}
		                       +(4.5,0)node[^]{$2√n$}
		                       +(8.0,0)node[^]{$3√n$}
		                      +(10.5,0)node[^]{$n$}
		                +(0,0)--+(11,0)node[_,pos=.5]{depth};
		\draw[->](-.2,0)--+(0,\sdwidth)node[^,pos=.5,rotate=90]{channels};
		\tikzset{sd}
		\sddive(1);
		\sdcutby.4{
			\sddive(1);
			\sddive(1);
			\sdsave[a]
		}{
			\sddive(1)node[pos=0]{recruited};
			\sddive(1)node{trained};
			\sdcutby.4{
				\sdload[a](1);
				\sddive(.5)[recycol]node[pos=.5]{recycled};
				\sdcutby.4{
					\sddive(1);
					\sddive(1);
					\sdsave[b]
				}{
					\sddive(1)node[pos=.5]{recruited};
					\sddive(1)node{trained};
					\sdcutby.4{
						\sdload[b](1);
						\sddive(.5)[recycol]node[pos=.5]{recycled};
						\sdcutby.4{
							\sddive(1)node{stop recycling};
						}{
							\sddive(1)node{recruited};
							\sddive(1)node{trained};
							\sdcutby.4{
								\sddive(1)node{frozen};
							}{
								\sddive(1)node{retained};
							}
						}
					}{
						\sddive(1)node{retained};
					}
				}
			}{
				\sddive(1)node{retained};
			}
		}
		\pgf@picminx-0mm\global\pgf@picminx\pgf@pt@aa\pgf@picminx
		\pgf@picmaxx11cm\global\pgf@picmaxx\pgf@pt@aa\pgf@picmaxx
	} \label{eq:sdR1}\]\vfil
	
	\[\tikz[fit=.666]{
		\draw[->](0,-\sdwidth) +(1.5,0)node[^]{$√n$}
		                       +(4.5,0)node[^]{$2√n$}
		                       +(8.0,0)node[^]{$3√n$}
		                      +(10.5,0)node[^]{$n$}
		                +(0,0)--+(11,0)node[_,pos=.5]{depth};
		\draw[->](-.2,0)--+(0,\sdwidth)node[^,pos=.5,rotate=90]{channels};
		\tikzset{sd}
		\sddive(1);
		\sdcutby.4{
			\sddive(1);
			\sddive(1);
			\sdsave[a]
		}{
			\sddive(1)node[pos=.5]{$A_{√n}$};
			\sddive(1)node{$B_{√n}$};
			\sdcutby.4{
				\sdload[a](1)node[pos=.5]{$C_{√n}∪D_{√n}$};
				\sddive(.5)[recycol]node[pos=.5](recycled){};
				\onforeground\node[recycol]at(recycled){recycled};
				\sdcutby.4{
					\sddive(1);
					\sddive(1);
					\sdsave[b]
				}{
					\sddive(1)node{$A_{2√n}$};
					\sddive(1)node{$B_{2√n}$};
					\sdcutby.4{
						\sdload[b](1)node[pos=.5]{$C_{2√n}∪D_{2√n}$};
						\sddive(.5)[recycol]node[pos=.5](recycled){};
						\onforeground\node[recycol]at(recycled){recycled};
						\sdcutby.4{
							\sddive(1)node{stop recycling};
						}{
							\sddive(1)node{$A_{3√n}$};
							\sddive(1)node{$B_{3√n}$};
							\sdcutby.4{
								\sddive(1)node{$B_{3√n}∪C_{3√n}$};
							}{
								\sddive(1)node{$E_{3√n}$};
							}
						}
					}{
						\sddive(1)node{$E_{2√n}$};
					}
				}
			}{
				\sddive(1)node{$E_{√n}$};
			}
		}
		\pgf@picminx-0mm\global\pgf@picminx\pgf@pt@aa\pgf@picminx
		\pgf@picmaxx11cm\global\pgf@picmaxx\pgf@pt@aa\pgf@picmaxx
	} \label{eq:sdR2}\]
	
	\[\tikz[fit=.8]{
		\draw[->](0,-\sdwidth) +(1.5,0)node[^]{$√n$}
		                       +(3.5,0)node[^]{$2√n$}
		                       +(5.5,0)node[^]{$n\rat$}
		                       +(7.5,0)node[^]{$n$}
		                 +(0,0)--+(8,0)node[_,pos=.5]{depth};
		\draw[->](-.2,0)--+(0,\sdwidth)node[^,pos=.5,rotate=90]{channels};
		\tikzset{sd}
		\sddive(1);
		\sdcutby.6{
			\sddive(1);
			\sddive(1);
			\sdcutby.6{
				\sddive(1);
				\sddive(1);
				\sdcutby.6{
					\sddive(1)node{stop recruiting};
				}{
					\sddive(1)node{recruited};
					\sddive(1)node{trained};
					\sdcutby.4{
						\sddive(1)node{frozen};
					}
					{
						\sddive(1)node{retained};
					}
				}
			}{
				\sddive(1)node{recruited};
				\sddive(3)node{trained};
				\sdcutby.4{
					\sddive(1)node{frozen};
				}
				{
					\sddive(1)node{retained};
				}
			}
		}{
			\sddive(1)node{recruited};
			\sddive(5)node{trained};
			\sdcutby.4{
				\sddive(1)node{frozen};
			}
			{
				\sddive(1)node{retained};
			}
		}
	} \label{eq:sdD1}\]\vfil
	
	\[\tikz[fit=.8]{
		\draw[->](0,-\sdwidth) +(1.5,0)node[^]{$√n$}
		                       +(3.5,0)node[^]{$2√n$}
		                       +(5.5,0)node[^]{$n\rat$}
		                       +(7.5,0)node[^]{$n$}
		                 +(0,0)--+(8,0)node[_,pos=.5]{depth};
		\draw[->](-.2,0)--+(0,\sdwidth)node[^,pos=.5,rotate=90]{channels};
		\tikzset{sd}
		\sddive(1);
		\sdcutby.6{
			\sddive(1);
			\sddive(1);
			\sdcutby.6{
				\sddive(1);
				\sddive(1);
				\sdcutby.6{
					\sddive(1)node{stop recruiting};
				}{
					\sddive(1)node{$A_{3√n}$};
					\sddive(1)node{$B_{3√n}$};
					\sdcutby.4{
						\sddive(1)node{$C_{3√n}∪D_{3√n}$};
					}
					{
						\sddive(1)node{$E_{3√n}$};
					}
				}
			}{
				\sddive(1)node{$A_{2√n}$};
				\sddive(3)node{$B_{2√n}$};
				\sdcutby.4{
					\sddive(1)node{$C_{2√n}∪D_{2√n}$};
				}
				{
					\sddive(1)node{$E_{2√n}$};
				}
			}
		}{
			\sddive(1)node{$A_{√n}$};
			\sddive(5)node{$B_{√n}$};
			\sdcutby.4{
				\sddive(1)node{$C_{√n}∪D_{√n}$};
			}
			{
				\sddive(1)node{$E_{√n}$};
			}
		}
	} \label{eq:sdD2}\]\vfil
	
	\[\tikz[fit=.8]{
		\draw[->](0,-\sdwidth) +(1.5,0)node[^]{$√n$}
		                       +(3.5,0)node[^]{$2√n$}
		                       +(5.5,0)node[^]{$n\rat$}
		                       +(7.5,0)node[^]{$n$}
		                 +(0,0)--+(8,0)node[_,pos=.5]{depth};
		\draw[->](-.2,0)--+(0,\sdwidth)node[^,pos=.5,rotate=90]{channels};
		\tikzset{sd}
		\sddive(1);
		\sdcutby.6{
			\sddive(1);
			\sddive(1);
			\sdcutby.6{
				\sddive(1);
				\sddive(1);
				\sdcutby.6{
					\sddive(1)node{stop pruning};
				}{
					\sddive(1)node{grafted};
					\sddive(2);
				}
			}{
				\sddive(1)node{grafted};
				\sddive(4);
			}
		}{
			\sddive(1)node{grafted};
			\sddive(6);
		}
	} \label{eq:sdG}\]

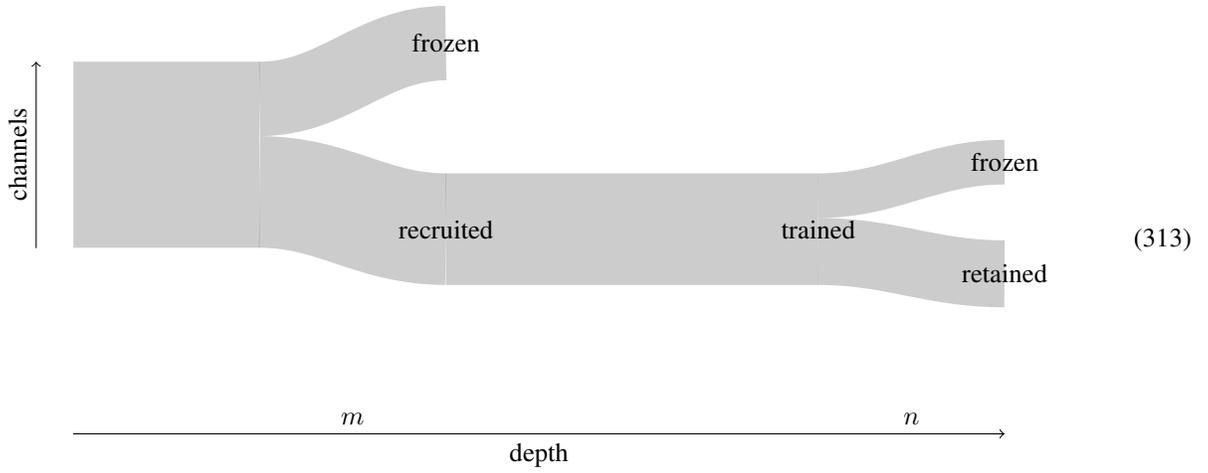
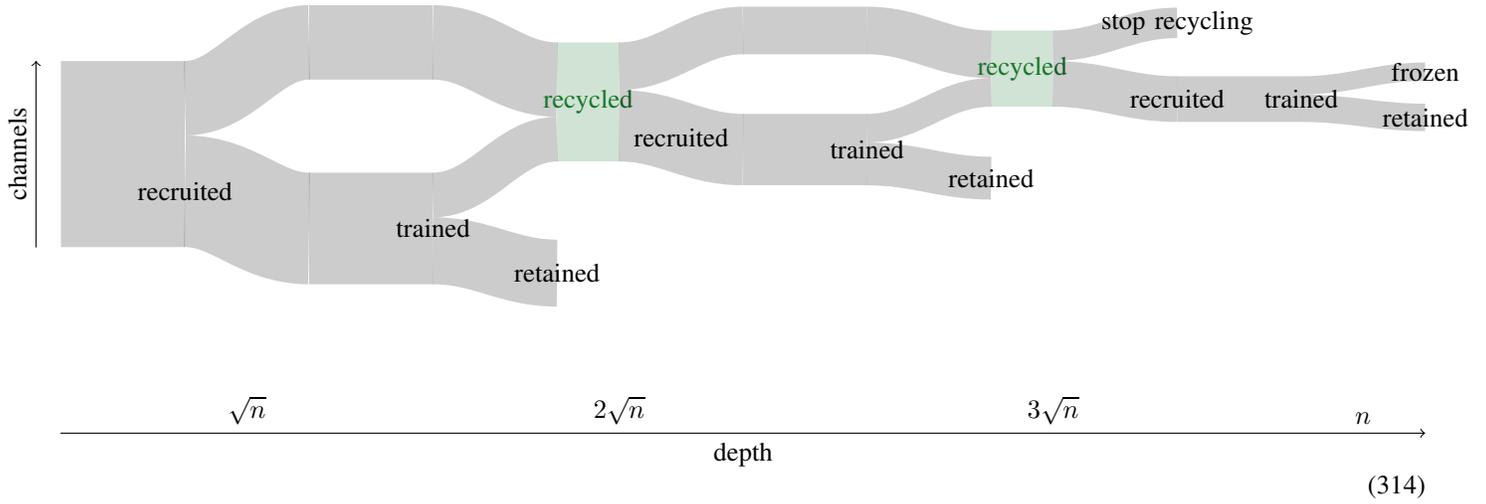
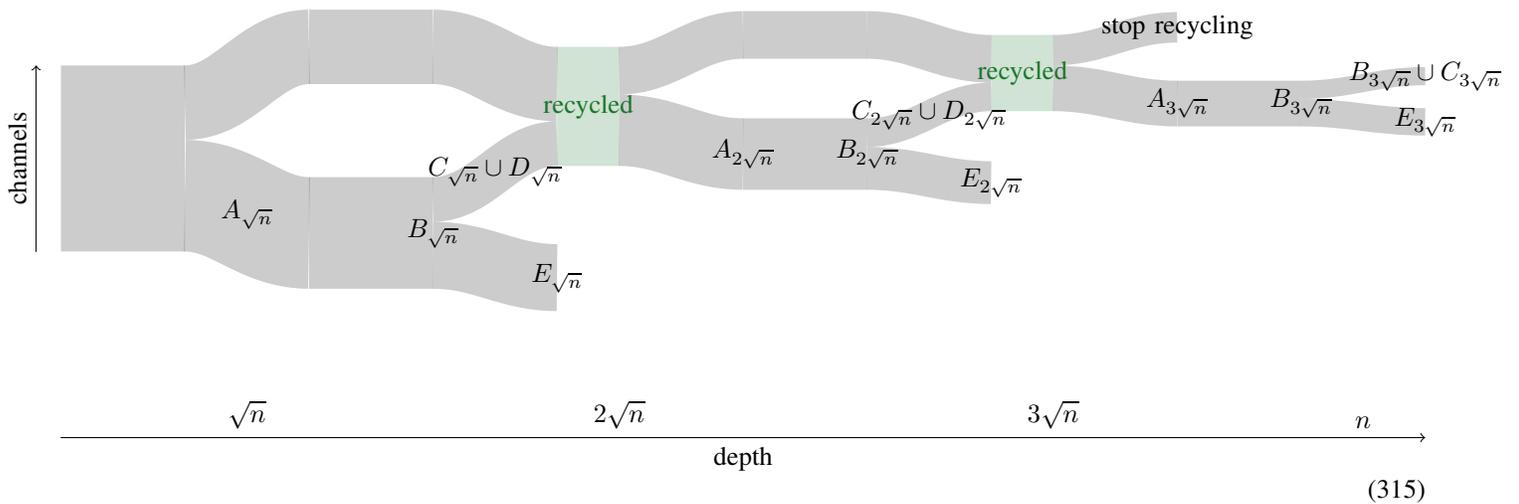
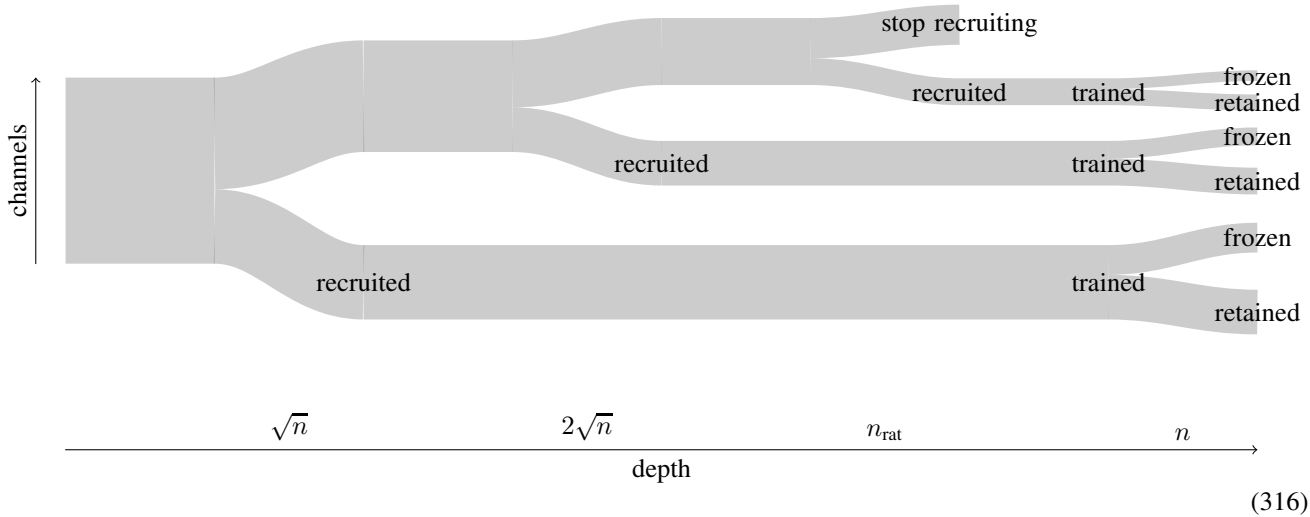
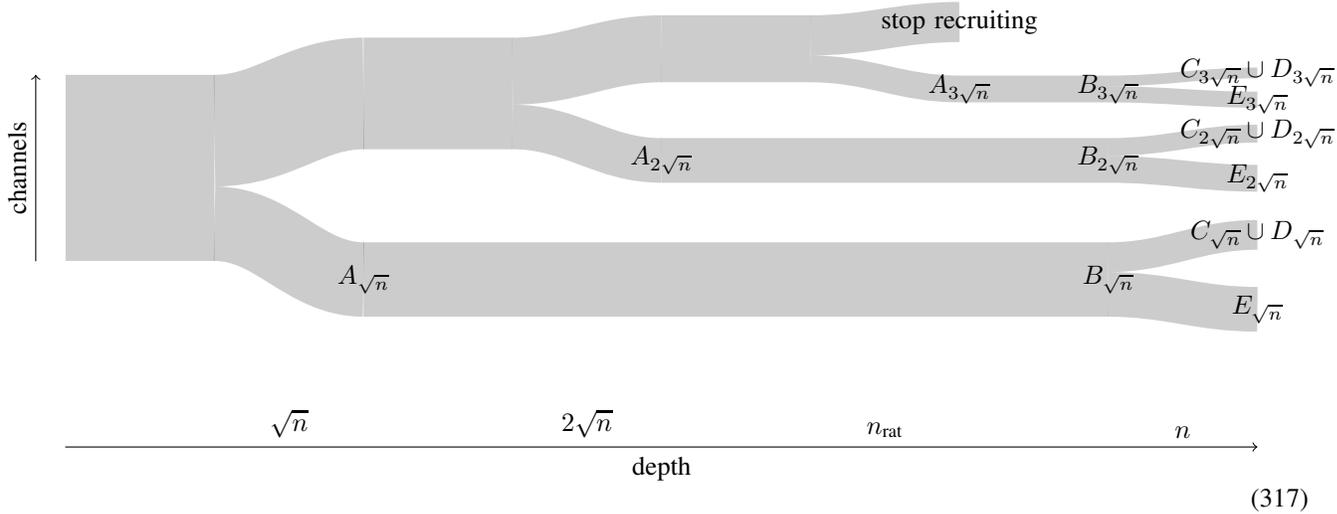
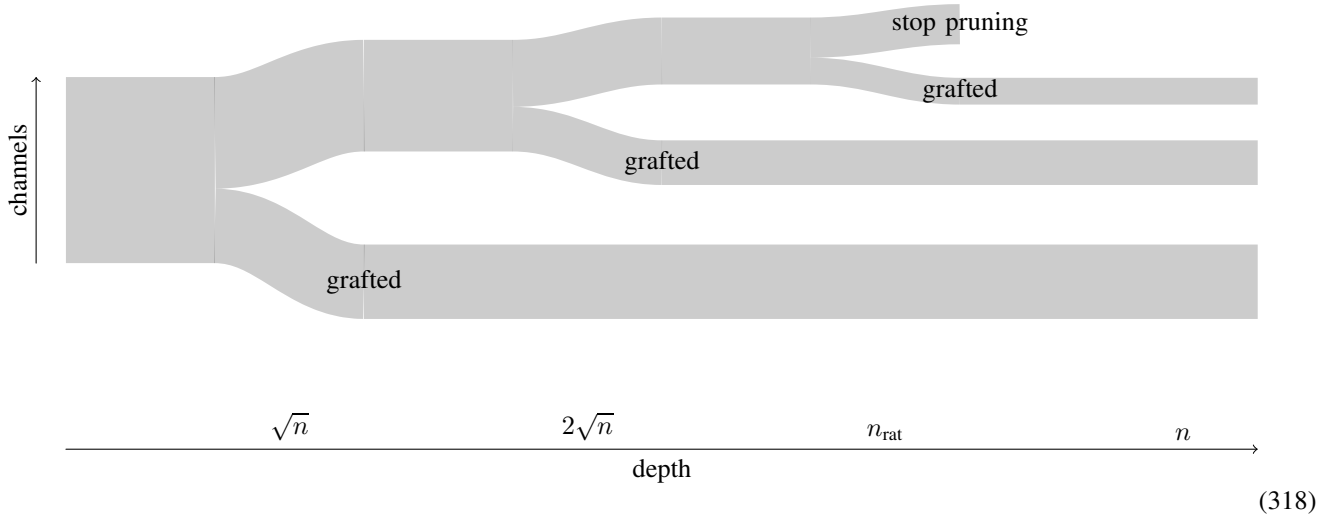
\end{document}